\documentclass[11pt]{article}

\usepackage{graphicx}
\usepackage{float}
\usepackage{amsthm,verbatim,amsmath, epsfig, indentfirst, graphicx, multirow, color, graphics, rotating, amssymb, amsfonts, amscd, xspace, pifont}
\usepackage{hyperref}
\usepackage[mathscr]{euscript}
\usepackage{multirow}
\usepackage{siunitx}
\usepackage[round]{natbib}
\usepackage{amsthm,amsmath, amssymb, amsfonts, amscd, xspace, pifont}
\usepackage{epsfig, psfrag, pstricks}
\usepackage{graphicx}
\usepackage{float, fancybox}

\usepackage{delarray}
\usepackage{hyperref}
\usepackage{url}
\usepackage{anysize}
\usepackage{multirow}
\usepackage{listings}
\usepackage{latexsym,ifthen}
\usepackage{subfigure}
\usepackage{anysize}
\usepackage{natbib}

\newcommand{\bt}{\mathbf{t}}
\newcommand{\bsB}{\boldsymbol{\mathcal{B}}}
\newcommand{\bB}{\boldsymbol{B}}
\newcommand{\bY}{\boldsymbol{Y}}
\newcommand{\bX}{\boldsymbol{X}}
\newcommand{\bZ}{\boldsymbol{Z}}
\newcommand{\bU}{\boldsymbol{U}}
\newcommand{\bE}{\boldsymbol{E}}
\newcommand{\bW}{\boldsymbol{W}}
\newcommand{\bb}{\boldsymbol{b}}
\newcommand{\bu}{\boldsymbol{u}}
\newcommand{\bnu}{\boldsymbol{\nu}}
\newcommand{\bzero}{\boldsymbol{0}}
\usepackage{JASA_manu}

\DeclareMathOperator*{\argmin}{arg\,min}

\newcommand{\bOmega}{\boldsymbol{\Omega}}
\newcommand{\bPsi}{\boldsymbol{\Psi}}
\newcommand{\bbeta}{\boldsymbol{\beta}}
\newcommand{\bgamma}{\boldsymbol{\gamma}}

\def\be{\mbox{Be}}

\def\bQ{\mbox{\bf Q}}
\def\bY{\mbox{\bf Y}}
\def\bE{\mbox{\bf E}}

\def\bS{\mbox{\bf S}}
\def\bU{\mbox{\bf U}}
\def\bzero{\mbox{\bf 0}}
\def\bone{\mbox{\bf 1}}
\def\by{\mbox{\bf y}}
\def\be{\mbox{\bf e}}
\def\bu{\mbox{\bf u}}
\def\bX{\mbox{\bf X}}

\def\bx{\mathbf{\bf x}}

\newcommand{\calM}{\mathcal{M}}

\newcommand{\calT}{\mathcal{T}}

\newcommand{\calS}{\mathcal{S}}
\newcommand{\calB}{\mathcal{B}}

\begin{document}
\title{Bayesian Semiparametric Functional Mixed Models for Serially Correlated Functional Data, with Application to Glaucoma Data}
\author{Wonyul Lee  \\
Department of Biostatistics \\ University of Texas M.D. Anderson Cancer Center, Houston, TX 77230 \\
email: \texttt{freshlwy@gmail.com} \\
\vspace*{-8pt}
{Michelle F. Miranda} \\
Department of Biostatistics\\ University of Texas M.D. Anderson Cancer Center, Houston, TX 77230 \\
email: \texttt{MFMiranda@mdanderson.org} \\
\vspace*{-8pt}
{Philip Rausch} \\
Department of Psychology\\ Institut f\"{u}r Psychologie \\ Humboldt-Universit\"{a}t zu Berlin , Germany \\
email: \texttt{philip.rausch@gmail.com}\\
\vspace*{-8pt}
{Veerabhadran} Baladandayuthapani  \\
Department of Biostatistics\\ University of Texas M.D. Anderson Cancer Center, Houston, TX 77230  \\
email: \texttt{veera@mdanderson.org} \\
\vspace*{-8pt}
Massimo Fazio  \\
Department of Ophthalmology\\ University of Alabama at Birmingham, Birmingham, AL 35294  \\
email: \texttt{massimof@uab.edu} \\
\vspace*{-8pt}
{J. Crawford Downs}  \\
Department of Ophthalmology\\ University of Alabama at Birmingham, Birmingham, AL 35294  \\
email: \texttt{cdowns@uab.edu} \\
\vspace*{-8pt}
Jeffrey S. Morris$^*$  \\
Department of Biostatistics\\ University of Texas M.D. Anderson Cancer Center, Houston, TX 77230  \\
email: \texttt{jefmorris@mdanderson.org} \\
}

\maketitle
\thispagestyle{empty}

\newpage

\mbox{}
\begin{center}
\textbf{Author's Footnote:}
\end{center}
Wonyul Lee is Postdoctoral Researcher, University of Texas M.D. Anderson Cancer Center, Houston, TX 77230 (email: freshlwy@gmail.com); Michelle Miranda is Postdoctoral Researcher, University of Texas M.D. Anderson Cancer Center, Houston, TX 77230 (email: MFMiranda@gmail.com); Philip Rausch is Researcher, Humboldt-Universit\"{a}t zu Berlin (email: philip.rausch@gmail.com); Veera Baladandayuthapani is Associate Professor, University of Texas M.D. Anderson Cancer Center, Houston, TX 77230 (email: veera@mdanderson.org); Massimo Fazio is Assistant Professor, University of Alabama at Birmingham, Birmingham, AL 35294 (email: massimof@uab.edu); and {J. Crawford Downs} is Professor, University of Alabama at Birmingham, Birmingham, AL 35294 (email: cdowns@uab.edu); Jeffrey S. Morris is Del and Dennis McCarthy Distinguished Professor, University of Texas M.D. Anderson Cancer Center, Houston, TX 77230 (email: jefmorris@mdanderson.org). This work is supported by grants from the National Cancer Institute (R01-CA178744, P30-CA016672, {R01-CA160736}), the National Science Foundation (1550088), and {the National Eye Institute (R01-EY18926)}. The authors also thank Richard Herrick and Nan Chen for computational assistance, and two very thorough reviewers and associate editor whose insightful queries led to a greatly improved paper.

\thispagestyle{empty}

\newpage
\begin{center}
\textbf{Abstract}
\end{center}

Glaucoma, a leading cause of blindness, is characterized by optic nerve damage related to {intraocular} pressure (IOP), but its full etiology is unknown. Researchers at UAB have devised a custom device to measure scleral strain continuously around the eye under fixed levels of IOP, which here is used to assess how strain varies around the {posterior pole}, with IOP, and across glaucoma risk factors such as age.  The hypothesis is that scleral strain decreases with age, which could {alter biomechanics of the optic nerve head} and cause damage that could eventually lead to glaucoma.  To evaluate this hypothesis, we adapted Bayesian Functional Mixed Models to model these complex data consisting of correlated functions on spherical scleral surface, with nonparametric age effects allowed to vary in magnitude and smoothness across the scleral surface, multi-level random effect functions to capture within-subject correlation, and functional growth curve terms to capture serial correlation across IOPs that can vary around the scleral surface.  Our method yields fully Bayesian inference on the scleral surface or any aggregation or transformation thereof, and reveals interesting insights into the biomechanical etiology of glaucoma.  The general modeling framework described is very flexible and applicable to many complex, high-dimensional functional data.

\vspace*{.3in}

\thispagestyle{empty}

\newpage
\setcounter{page}{1}

\section{Introduction}
\label{s:intro}

Glaucoma is one of the leading causes of blindness in the world. While its etiology is not fully understood, it is known to be caused by damage to the optic nerve head (ONH) that can be induced by intraocular pressure (IOP).  Researchers hypothesized that {the} biomechanics of {the peripapillary (PP)} scleral region close to the ONH, shown to be an important determinant of ONH biomechanics, may play a major role in glaucoma pathogenesis and progression \citep{Sigal05,Burgoyne08}.  Recently, novel custom instrumentation was developed that can induce a fixed level of IOP and precisely measure the mechanical strain in the posterior human sclera \citep{Fazio12a,Fazio12b}. A commercial laser speckle interferometer (ESPI) is used to measure the IOP-induced scleral displacement continuously around the {posterior} eye.  The scleral displacement is processed to estimate the mechanical strain tensor on a grid of points on the scleral surface, and the largest eigenvalue of this strain tensor computed to yield the maximum principal strain (MPS), a scalar measurement for each location on the scleral surface that summarizes the magnitude of {tensile} strain at that location. Intuitively, scleral regions with higher MPS for a given level of IOP are more pliable, which in principle {could either} relieve IOP and reduce the potential for ONH damage {or focus strain at the ONH and increase glaucoma damage}.  Age is a primary glaucoma risk factor and age-related stiffening occurs in other load-bearing soft tissues.  Thus, the researchers hypothesized that age-related changes in scleral stiffness might contribute to the known age-related increase in glaucoma risk.  

Utilizing this custom instrumentation, \citet{Fazio14} conducted a study to test the hypothesis that scleral stiffness increases with age, with the expectation that future work will follow to elucidate the specific role of scleral stiffness in glaucoma. They obtained twenty pairs of eyes from normal human donors in the Lions Eye Bank of Oregon in Portland, Oregon, and the Alabama Eye Bank in Birmingham, Alabama. 
From each subject, the MPS was measured {in the posterior globes of} both left and right eyes on {a} partial spherical domain with 120 circumferential locations $\phi\in(\SI{ 0}{\degree},\SI{360}{\degree})$ and 120 meridional locations $\theta \in(\SI{ 9}{\degree},\SI{24}{\degree})$, where $\theta=\SI{0}{\degree}$ corresponds to the ONH. For each eye, MPS was measured under nine different IOP levels (7, 10, 15, 20, 25, 30, 35, 40, and {45 mmHg}). {Further scientific and technical details can be found in Section \ref{s:Application}}. Figure \ref{fg1} plots a polar azimuthal projection of MPS functions for one subject under {45 mmHg} IOP level. The center of each panel corresponds to the ONH of each eye.  The hypothesis is that MPS will decrease with age, especially in scleral regions closer to the ONH.

\begin{figure}
\centerline{\includegraphics[scale=0.6]{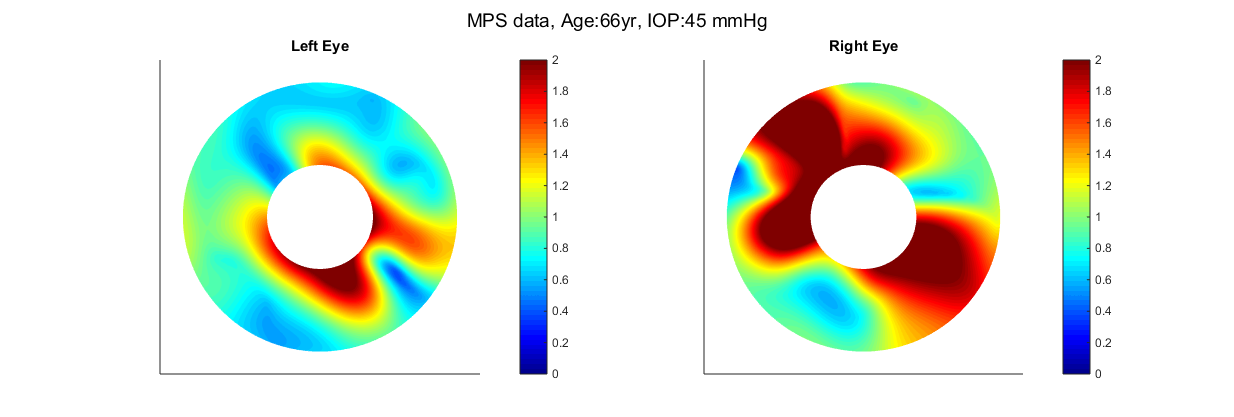}}
\caption{\footnotesize \textbf{Data plot:} Polar azimuthal projection of partial spherical MPS function for both left and right eyes from one subject of age 66yr under {45 mmHg} of intraocular pressure.}\label{fg1}
\end{figure}

The resulting data set is complex and high dimensional with many layers of structure.  First, for each eye at each IOP level, the MPS measurements constitute a function on a two-dimensional partial spherical domain for which one needs to account for within-function (intrafunctional) correlations. Second, there are multiple sources of between-function (interfunctional) correlation.  The measurements from the left and right eye from the same subject are expected to be correlated, and measurements from the same eye across the IOP levels should be serially correlated.  The strength of these nested and serial correlations may potentially vary around the scleral surface. Based on preliminary looks at the data, it appears that the age effect on MPS may not be linear, and also appears to vary around the scleral surface.   This data set is also enormous, with over 4.5 million measurements, which poses practical problems for model building and fitting.

Many researchers, when faced with such complexities, would use one of several strategies to simplify the data so they can analyze them.  Some researchers would get rid of the complexities of the functional data by computing summaries and modeling only those, while discarding the original functional data.  In this application area, some researchers create summaries in the peripapillary (PP) and mid-peripheral (MP) scleral regions by integrating over the region closest to the ONH ($ \SI{9}{\degree} - \SI{17}{\degree}$) and further out ($\SI{17}{\degree} - \SI{24} {\degree}$), respectively, or over several circumferential sections.  This practice would miss any scientific insights not captured by these summaries.  Alternatively, some researchers would only model a subset of the data (e.g. only a single IOP or single eye per subject) to avoid having to deal with potentially complex interfunctional correlation, or ignore this correlation entirely by modeling IOP within eyes or eyes within subject as independent.  Some would ignore intrafunctional correlation by modeling individual pixels on the image independently.    Many researchers would also only consider parametric or linear effects for covariates such as age without considering whether more complex nonparametric covariate effects might be necessary to capture the true relationship.   All of these simplification strategies have potential statistical downsides. 

In our opinion, it would be far preferable to model this complex function data set in its entirety using a statistical model that is flexible enough to capture all of its potentially complex intrafunctional and interfunctional structure.  However, to the best of our knowledge,  no statistical model has been presented in existing literature that simultaneously handles all of this structure.  In this paper, we use the Bayesian functional mixed model (BayesFMM) framework, introduced in \cite{Morris06} and further developed in subsequent papers, to model these data, which has sufficient flexibility to capture nonparametric covariate effects, serial and nested interfunctional correlation, functions on a fine 2d partial spherical domain, and is computationally efficient enough to scale up to this enormous size and produce Bayesian inference for functional parameters as well as any desired summaries.  This requires a careful description of how to utilize the BayesFMM framework to model smooth nonparametric effects and serial interfunctional correlation, which has not been done in any existing literature to date.

While motivated by and applied to the glaucoma scleral strain data, the BayesFMM framework we present here is more generally applicable to many types of complex, high dimensional functional data of modern interest including wearable computing data, genome-wide data, proteomics data, geospatial time series data and neuroimaging data.  Our hope is that besides revealing scientific insights into glaucoma, this paper can serve as a template for performing a state-of-the-art functional response regression analysis for other complex functional data sets.

The rest of the paper is organized as follows. Section 2 contains a brief review of some relevant literature in functional regression, including existing methods for modeling nested or serial interfunctional correlation and nonparametric smooth covariate functional effects.  Section 3 contains methods, overviewing the BayesFMM framework and providing methodological details for how to fit serially correlated functions and nonparametric smooth functional effects in this framework and discussing its properties.  Section 4 describes a model selection heuristic that can be used to assess which of the various potential complex modeling structures are needed for the current data before having to run a full MCMC.    Section 5 presents the details of our analysis of the glaucoma scleral strain data, including details on basis function and modeling components chosen, a summary of results, and various sensitivity analyses.  
Section 6 contains a discussion of general scientific conclusions from our analysis and an assessment of the strengths and weaknesses of the BayesFMM framework for this setting of complex functional regression models.   Supplementary materials provide additional computational details and graphical results of the analysis, as well as code to fit the models contained in the paper.

\section{Literature Review}
There is a rich and rapidly expanding literature on methods to perform functional regression, which includes \textit{functional predictor regression} (scalar-on-function), \textit{functional response regression} (function-on-scalar), and \textit{function-on-function regression}.  For example, see \cite{Morris15} for an extensive review of work in this area, which has exploded in the past decade.  In this paper, our goal is to regress the scleral strain MPS functions on predictors age and IOP, so we are interested in functional response regression.  Here, we will summarize some key existing literature relevant to the structures present in the glaucoma scleral strain data set, including methods that account for nested and serial interfunctional correlation and smooth nonparametric covariate effects.

As summarized in \cite{Morris15}, much of the work on functional response regression assumes independently sampled functions, but a number of published models can handle interfunctional correlation.  Many are focused on nested or crossed sampling designs that induce compound symmetry covariance structures among functions sampled within the same cluster, including \cite{Brumback98}, \cite{Morris2003}, \cite{Berhane08},  \cite{Aston10}, and \cite{Goldsmith16}.  There are relatively few that model serially correlated functions, for which functions are observed at multiple levels of some continuous variable for the same subject.  Most commonly, the functions occur along a grid of time points, which could be called \textit{longitudinally correlated functional data}, but can also occur over other continuous variables such as IOP for our glaucoma data.  There are a number of papers that focus on functional predictor regression \citep{Goldsmith2012, Gertheiss2013, Kundu2016, Islam2016} or estimate multi-level principal components \citep{Greven10, Zipunnikov2011, Chen2012, Li2014, Zipunnikov2014, Park2015, Shou2015, Hasenstab2017} for serially correlated functions, but these papers do not deal directly with functional response regression, i.e. do not regress the functions on covariates while accounting for this serial correlation in the error structure.  
Any methods built for a general functional mixed modeling framework containing multiple levels of general random effect functions, including the BayesFMM framework first introduced by \cite{Morris06} and the functional additive mixed model (FAMM) framework first introduced by \cite{Scheipl15}, can be used for functional response regression while accounting for serial interfunctional correlation if an acceptable parametric form for the serial variable can be found.  However, no specific models or examples in existing papers have presented such a case. In this paper, we will demonstrate in detail how to account for serial correlation within the BayesFMM framework.

Most functional response regression work, while modeling coefficients as nonparametric in the functional domain $t$, has focused on models linear in covariates $x$, e.g. with terms such as $x \beta(t)$.  A number of methods allow functional coefficients that are smooth and nonparametric in covariate $x$ as well, e.g. $f(x, t)$.  The last chapter of \cite{GAM} describes additive mixed models (AMMs) that extend \cite{Laird1982} to include nonparametric fixed effects and parametric random effects.  After specifying a parametric form for $t$ in fixed and random effects, this framework can fit terms $f(x,t)$ that are nonparametric in $x$ but parametric in $t$ using {\tt mgcv} in R.  \cite{Scheipl15} describe a generalization of AMMs to the functional regression settting, yielding \emph{functional additive mixed models (FAMM)} that can fit functional response, functional predictor, or function-on-function regression models, and allowing terms $f(x, t)$ that are nonparametric in both $x$ and $t$, using {\tt mgcv} to fit the underlying models.  Initially designed for splines, this work was extended to allow the use of functional principal components (fPC) for the random effect functions to handle sparse, irregularly sampled outcomes in \cite{Cederbaum2015}, to model generalized outcomes in \cite{Scheipl2016}, and to utilize a new boosting-based fitting procedure for some models that leads to other computational and modeling benefits in \cite{Brockhaus2015}.  This series of work is summarized in a review article \cite{Greven2017}.  These works utilize the fact, going way back to \cite{Wahba1978}, that penalized splines can be represented using a linear mixed model framework.  Using similar representations, the BayesFMM framework first introduced in \cite{Morris06} can also be used to fit nonparametric terms $f(x,t)$ with appropriate specification of the design matrices $X$ and $Z$ in the model, but this has not been done in any existing paper.  In this paper, we will describe in detail how to accommodate smooth nonparametric terms $f(x,t)$ in the BayesFMM framework.

To our knowledge, the only methods in existing literature with the flexibility to both fit smooth nonparametric functional covariate terms $f(x,t)$ and simultaneously account for nested and serial correlation are the FAMM framework \citep{Scheipl15,Greven2017} and the BayesFMM framework \citep{Morris06}.  Both of these frameworks are extremely flexible, based on functional extensions of linear mixed models, but have important differences, which are discussed in detail in \cite{Morris2017} and \cite{Greven2017rejoinder}.  The FAMM framework has several limitations that prevent its application to our glaucoma scleral strain data, including the requirement of functions to be on 1d Euclidean domains, use of spline bases for fixed effects with L2 penalties, and use of a computational approach that may not scale up well to enormous data sets like this.  Thus, to fit these data, we adapt the BayesFMM framework to include smooth nonparametric age effects and serial correlation across IOP, and demonstrate how we can use it to reveal insights into the biomechanical etiology of glaucoma.  


\section{Methods}
\label{s:method}

We first overview the BayesFMM framework in Section 3.1, and then for ease of exposition we build up the necessary components of our model separately before presenting our final general model.  In Section 3.2, we demonstrate how to capture serial interfunctional correlation through functional growth curve models, and then demonstrate how to accommodate smooth nonparametric covariate functional effects in Section 3.3.  Besides presenting the models, we also describe their properties and contrast with existing alternative methods.  In Section 3.4 we present a general alternative form of the BayesFMM that includes smooth nonparametric terms that we use to fit our glaucoma scleral strain data.


\subsection{{Bayesian Functional Mixed Models (BayesFMM)}}
\label{sec:BayesFMM}

Suppose we have a sample of functions $Y_i(\bt), i=1, \ldots, N$ observed on a common fine grid of size $T$ on a domain $\calT$, which is potentially multi-dimensional and/or non-Euclidean \citep{Morris11}.  The FMM introduced by \cite{Morris06} is a functional response regression model given by 
\begin{eqnarray}
Y_i(\bt) &=& \sum_{a=1}^{A} X_{ia} B_a(\bt) + \sum_{h=1}^H \sum_{m=1}^{M_h} Z_{ihm} U_{hm}(\bt) + E_i(\bt). \label{eq:FMM},
\end{eqnarray}
where $B_a(\bt)$ are fixed effect functions that model the effect of covariate $X_{ia}$ on the response $Y$ at position $\bt$ for each covariate $a=1,\ldots,A$, and the $U_{hm}(\bt)$ are random effect functions at level $h=1,\ldots,H$ corresponding to design matrix $Z_{ihm}$, with $m=1,\ldots,M_h$ being the number of random effects at the respective levels.  As in linear mixed models for scalar data, the fixed and random effect predictors can be discrete or continuous, and can involve individual covariates or interactions of multiple covariates.  Although not explicitly portrayed in (\ref{eq:FMM}), this modeling framework also accommodates functional predictors to perform function-on-function regression \citep{Meyer15}.

\textbf{Distributional and Covariance Assumptions:}  Here, for simplicity, we first describe the Gaussian FMM with conditionally independent random effect and residual error functions, and then mention some other alternatives available in this framework.  In this case, the random effect functions $U_{hm}(\bt)$ are iid mean zero Gaussian Processes with intrafunctional covariance cov$\{U_{hm}(\bt_1),U_{hm}(\bt_2)\} = Q_h(\bt_1,\bt_2)$ and the residual error functions $E_i(\bt)$ are iid mean zero Gaussian Processes with intrafunctional covariance cov$\{E_i(\bt_1),E_i(\bt_2)\}=S(\bt_1, \bt_2)$.  Other extensions of this framework allow the option of conditional autoregressive (CAR) \citep{Zhang14} or Matern spatial covariance or AR($p$) temporal interfunctional correlation structures in the residual errors  \citep{Zhu14}.  Although focusing on Gaussian regression here, a robust version of this framework assuming heavier tailed distributions on the random effects or residuals is available \citep{Zhu11} if robustness to outliers is desired, and can also be utilized with any other features or modeling components in the BayesFMM framework.

\textbf{Basis Transform Modeling Approach:} A \textit{basis transform modeling approach} is used to fit model (\ref{eq:FMM}).  This first involves representing the observed functions with a basis expansion with a set of basis functions $\psi_k(\bt), k=1,\ldots, K$:
\begin{eqnarray}
Y_i(\bt) &=& \sum_{k=1}^{K} Y^*_{ik} \psi_k(\bt) \label{eq:basis}
\end{eqnarray}
While initially developed for wavelets \citep{Morris06}, as first discussed in \cite{Morris11}, the modeling approach can be used with any basis.  It is meant to be used with  \textit{lossless} transforms with $Y_i(\bt) \equiv \sum_k Y^*_{ik} \psi_k(\bt)$ for all observed $\bt$, so that the basis coefficients $\{Y^*_{ik}; k=1,\ldots,K\}$ contain all information within the observed functional data $\{Y_i(\bt); \bt = \bt_1, \ldots, \bt_T\}$, or at least \textit{near-lossless} with
\begin{eqnarray}
\left\lVert Y_i(\bt) - \sum_{k=1}^K Y^*_{ik} \psi_k(\bt) \right\lVert < \epsilon \hspace{8pt} \forall i=1,\ldots,N \label{eq:near-lossless}
\end{eqnarray}
for some small value $\epsilon$ and measure $\lVert \bullet \lVert$.  This assures that the chosen basis is sufficiently rich such that for practical purposes it can recapitulate the observed functional data, and visual inspection of the raw functions and basis transformation should reveal virtually no difference.   Any basis functions can be used, including commonly used choices splines, wavelets, Fourier bases, PCs or creatively constructed custom bases, and can be defined on multi-dimensional or non-Euclidean domains $\calT$.

Rather than including the bases in a design matrix and using scalar regression methods to fit the model, our approach is to transform the observed functions into the basis space to obtain the basis coefficients $\bY^*$, a $N \times K$ matrix for which element $(i,k)$ contains the basis coefficient $k$ for observed function $i$, fit a basis-space version of the FMM to these coefficients, and then transform results back to the data space model (\ref{eq:FMM}) for estimation and inference.  With basis representation written in matrix form $\bY=\bY^* \bPsi$ with $\bPsi$ a $K \times T$ matrix of basis functions evaluated on the observational grid with $\bPsi_{kj}=\psi_k(\bt_j)$, the coefficients can be computed by $\bY^*=\bY \bPsi^-$ with $\bPsi^-=\bPsi' (\bPsi \bPsi')^{-1}$ as long as rank$(\bPsi)=K$, or for certain basis functions including wavelets and Fourier bases their special structure enables fast algorithms for computing these coefficients.  

\textbf{Basis Space Model:} Thus,  rather than fitting model (\ref{eq:FMM}) directly, the basis-space version of the model is fit for each basis coefficient $k=1,\ldots,K$:
\begin{eqnarray}
\vspace{-24pt}
Y^*_{ik} = \sum_{a=1}^A X_{ia} B^*_{ak} + \sum_{h=1}^H \sum_{m=1}^{M_h} Z_{ihm} U^*_{hmk} + E^*_{ik}, \label{eq:basis_FMM}
\end{eqnarray}
where $B^*_{ak}$, $U^*_{hmk}$, and $E^*_{ik}$ are basis coefficients for the functional fixed effects $B_a(\bt)=\sum_k B^*_{ak} \psi_k(\bt)$, functional random effects $U_{hm}(\bt)=\sum_k U^*_{hmk} \psi_k(\bt)$, and functional residuals $E_i(\bt)=\sum_k E^*_{ik} \psi_k(\bt)$, respectively.   While in principle correlation across basis coefficients can be accommodated, for complex, high-dimensional functions, it may be beneficial to model these basis coefficients independently.  For the Gaussian FMM with conditionally independent random effect and residual error functions, it is assumed $U^*_{hmk} \sim N(0,q_{hk})$ and $E^*_{ik} \sim N(0,s_k)$ with $q_{hk}$ and $s_k$ scalar variance components.  Although modeling independently in the basis space, this structure induces intrafunctional correlation according to the chosen basis functions.  For example, for the residual error functions, the induced $T \times T$ intrafunctional correlation $\bS$ with $\bS_{ij}=\bS(\bt_i, \bt_j)$ is given by:
\begin{eqnarray}
\bS&=&\bPsi^{'} \bS^* \bPsi, \label{eq:covariance}
\end{eqnarray}
where $\bS^*=\mbox{diag}\{s_k; k=1,\ldots,K\}$, and $\bQ_h$ defined likewise.  For suitably chosen basis functions that effectively capture the characteristic structure of the observed functions $Y_i(\bt)$, this can allow a flexible class of covariance structures indexed by $K$ covariance parameters.  One can assess the suitability of these assumptions by taking the basis-space variance components, computing (\ref{eq:covariance}), and plotting this covariance matrix to see if it appears to capture the salient structure.  Figure \ref{f:combo_plots} panels (e) and (f) plot the intrafunctional correlation structure induced by the chosen tensor wavelet basis for the scleral strain MPS data set for a particular scleral location for the eye-to-eye random intercepts and residual error functions, and the file {\tt intrafunctional\_correlation.mp4} in the supplement  shows a more extensive summary of all random levels across scleral locations.

\textbf{Shrinkage Priors for Regularization of Fixed Effects:} If fitting this model using a frequentist approach, L1 or L2 penalties could be imposed on the basis space fixed effects to induce regularization/smoothing of the fixed effect functions $B_a(\bt)$, as described in \cite{Morris15}.  However, this framework was designed to use a Bayesian modeling approach, in which case regularization of the fixed effect functions $B_a(\bt)$ is accomplished through specification of shrinkage priors for the corresponding basis coefficients:
\begin{eqnarray}
B^*_{ak} \sim g(\bgamma_{aj}) \label{eq:shrinkage_prior}
\end{eqnarray}
for some mean zero distribution $g(\bullet)$ with corresponding regularization parameters $\bgamma_{aj}$ indexed by $j=1,\ldots,J$ that define a partitioning of the basis coefficients $k=1,\ldots,K$ into \textit{regularization sets}, which are subsets of basis coefficients sharing the same regularization parameters.  \cite{Morris06} used the spike-slab prior \citep{George1993} for $g(\bullet)$ and we also use that here, but other alternatives include Gaussian, Laplace \citep{Park2008}, Horseshoe \citep{Carvahlo2010}, Normal-Gamma \citep{Griffin2010}, and Dirichlet-Laplace \citep{Bhattacharya2015}.  The regularization parameters $\bgamma_{aj}$ can be given hyperpriors or be estimated by empirical Bayes, which is described in detail for the spike-slab in \cite{Morris06}.

\textbf{Model Fitting Approach:} A fully Bayesian modeling approach is used to fit a Markov Chain Monte Carlo (MCMC) to the basis space model (\ref{eq:basis_FMM}) for each $k$, and then transforming back to the data space, e.g. using $B_a(\bt)=\sum_k B^*_{ak} \psi_k(\bt)$ to yield posterior samples for each parameter in the data-space FMM (\ref{eq:FMM}).  This requires specification of priors for the variance components; a vague empirical Bayes approach is used that centers the prior on REML starting values of these parameters, minimally informative to be equivalent to two data points of information, as detailed in Section \ref{s:Application}.  BayesFMM fits a marginalized version of model (\ref{eq:basis_FMM}) with all random effects $\bU^*$ integrated out, which integrates over the interfunctional correlation induced by the random effect levels when updating the fixed effects, and speeds convergence of the chain and calculations since the random effect functions themselves need not be sampled.    As detailed in the supplement, the fixed effect updates involve conjugate (spike-slab) Gibbs steps and variance components are updated via Metropolis-Hastings with proposal variances automatically computed based on the corresponding Fisher information.  If desired, posterior samples for random effects can be obtained by sampling from conjugate Gaussians.

\textbf{Bayesian Inference:}  Given these posterior samples, one can compute any desired posterior probabilities and pointwise or joint credible bands that can be used for Bayesian inference.  Joint bands can be constructed using the approach described in \cite{Ruppert03}.  These inferential summaries can be constructed in the data or basis space for any functional of model parameters, including contrasts (e.g. $B_1(\bt) - B_2(\bt)$), nonlinear transformations (e.g. exp$\{B_a(\bt)\}$), derivatives (e.g. $\partial f(x_a,\bt)/\partial x_a)$, or integrals (e.g. $\int_{\bt \in \calT_0} B_a(\bt) d \bt$ for some $\calT_0 \subset \calT$) aggregating information across regions of $\bt$.  We make use of these to produce inference on numerous scientifically interesting summaries for our scleral strain MPS data in Section \ref{s:Application}.

\textbf{Example FMM for Scleral Strain MPS Data:}. To illustrate this framework, suppose that we model MPS functions from both left and right eyes for each subject, but only for a specific IOP level, and suppose we are willing to assume a linear \textit{age} effect.  Let $Y_{i1}(\bt)$ and $Y_{i2}(\bt)$ be MPS functions for the left and right eyes respectively from the $i$-th subject, with $\bt \in \calT$ indexing the scleral domain, with $\bt=(\theta, \phi)$ being spherical coordinates on the scleral surface.   We could represent these data with the following FMM:
\begin{align}
\label{e:FMMd2} Y_{ij}(\bt)=  B_0(\bt)+X_{\mbox{age},i}B_{\mbox{age}}(\bt)+U_{i}(\bt)+E_{ij}(\bt),
\end{align}
with $U_i(\bt) \sim N\{\bzero, Q(\bullet)\}$ and $E_{ij}(\bt) \sim N\{\bzero, S(\bullet)\}$.  The $U_i(\bt)$ induce a generalized compound symmetry covariance structure between the functions from the left and right eyes from the same subject, with cov$\{Y_{ij}(\bt),Y_{ij'}(\bt)\}=Q(\bt,\bt)+S(\bt,\bt)I(j=j')$.  To simultaneously model data for all IOP, we would need to add a random effects level to capture the serial correlation across MPS functions for different IOP for the same eye.  We next describe how this can be done.

\subsection{Accommodating Serial Interfunctional Correlation via Functional Growth Curves}
\label{s:growth}

Recall that in our scleral strain MPS data set, for each eye we have MPS functions from each of a series of IOP levels ranging from {7 mmHg to 45 mmHg}, which induces a serial correlation across scleral strain functions for the same eye according to IOP.  It is important to account for this serial interfunctional correlation in some appropriate fashion in order to obtain efficient estimates and accurate inference for any fixed effect functions in the model.  Here we demonstrate how functional random effects in a functional mixed modeling framework can be used to capture this serial correlation using a type of functional growth curve model, and discuss the properties of this strategy in the context of the BayesFMM framework using a basis transform modeling approach.

\textbf{Basic Growth Curve Model for Scleral Strain MPS Data:}  For illustration, here we only consider the serial effect of IOP and omit any age effect and consider only the left eye for each subject.  Suppose $Y_{ip}(\bt)$ is the MPS function for the $i^{th}$ subject, $i=1,\ldots,N$, after exposure to an IOP level of $p$. {Note that $Y_{ip}(\bt)$ is a function of $\bt$ that varies across serial variable $p$. To capture the serial effect of $p$}, we consider the following \textit{functional growth curve model}:
{\begin{align}
\label{e:gcurve} Y_{ip}(\bt) = m(p, \bt)+u_{i}(p, \bt)+E_{ip}(\bt),
\end{align}}
where {$m(p,\bt)$ is the mean MPS for IOP level $p$ and scleral location $\bt$, $u_{i}(p, \bt)$ is a mean zero random effect for subject $i$ that represents a subject-specific growth curve in $p$ that is allowed to vary across scleral location $\bt$, and $E_{ip}(\bt)$ are residual error functions assumed to be independent and identically distributed mean zero Gaussians with covariance $S(\bullet)$.  If we can find a suitable parametric form for the serial effect of $p$ with basis functions $G_d(p), d=0,1,\ldots,D$, we assume $m(p, \bt)=\sum_{d=0}^{D}B_{d}(\bt)G_d(p)$ and $u_{i}(p, \bt)=\sum_{d=0}^{D}U_{i,d}(\bt)G_d(p)$ with cov$\{U_{i,d}(\bt_1), U_{i,d}(\bt_2)\}=Q_d(\bt_1, \bt_2)$.   In practice, we recommend using basis functions that are orthogonal across $d$, i.e. $\int G_d(p) G_{d'}(p) dp = 0$ for $d \ne d'$  to obviate the need to have cross-covariance terms between $U_{i,d}(\bt)$ and $U_{i,d'}(\bt)$ to avoid additional computational complexity in the model.

The introduction of this eye-level random growth curve induces serial covariance across functions for the same eye at different levels of IOP, with cov$\{Y_{ip}(\bt),Y_{ip'}(\bt)\}=\sum_{d=0}^D{G_d(p)G_d(p')Q_d(\bt,\bt)}.$  Indexed by $\bt$, the strength and shape of this serial covariance can vary across the scleral surface $\bt$.  Figure \ref{f:combo_plots} panel (d) contains the induced serial correlation across IOP for the marked scleral location, and the file {\tt IOP\_corr.mp4} in the supplementary materials is a movie file that demonstrates how this correlation varies over the scleral surface.   Note also that cov$\{Y_{ip}(\bt_1),Y_{ip'}(\bt_2)\}=\sum_{d=0}^D{G_d(p)G_d(p')Q_d(\bt_1,\bt_2)},$ meaning that this structure enables ``borrowing of strength" from nearby $\bt$ in determining the strength and shape of the serial covariance according to the intrafunctional covariance indicated by the off-diagonal elements of $Q_d(\bullet)$.   If model (\ref{e:gcurve}) is marginalized with respect to the random effect functions, the resulting error terms can be seen to contain the induced serial correlation structure, which is subsequently accounted for in any estimation or inference of the fixed effect functions. 


\textbf{Incorporation into BayesFMM framework:}  It can be seen that model (\ref{e:gcurve}) can be written as a functional mixed model with $D+1$ fixed effect predictors and $D+1$ random effect levels, each with $N$ subject-specific random effect functions.  Thus, any FMM framework allowing multiple levels of random effect functions could be used to fit this model.  In the BayesFMM framework, after transforming the observed functions $Y_{ip}(\bt)$ into the basis space through $Y_{ip}(\bt)=\sum_{k=1}^K  Y^*_{ipk} \psi_k(\bt)$ as described in Section \ref{sec:BayesFMM}, the model for coefficient $k$ would be given by:
{\begin{align}
Y^*_{ipk} = \sum_{d=0}^D G_d(p) B^*_{dk}+\sum_{d=0}^D G_d(p) U^*_{idk}+E_{ipk},
\end{align}}
with $U^*_{idk} \sim N(0,q_{dk})$ and $E^*_{ipk} \sim N(0,s_{k})$.  This would induce serial covariance across $p$ in the model for each basis coefficient with cov$(Y^*_{ipk}, Y^*_{ip'k})=\sum_{d=0}^D G_d(p) G_d(p') q_{dk}$, and the marginalized model that is fit with the $U^*_{idk}$ integrated out will contain this serial covariance in the error structure and explicitly account for it when updating the fixed effect coefficients.  This basis space model induces the serial covariance structure described in the data space model (\ref{e:gcurve}), with $Q_d(\bt_1, \bt_2)$ = $\sum_{k=1}^K \psi_k(\bt_1) \psi_k(\bt_2) q_{dk}$.  The heteroscedasticity of the variance components across basis functions $k$ allows the strength and shape of the serial covariance to vary across the scleral surface $\bt$, and effectively borrows strength from nearby $\bt$ through the chosen basis functions as determined by the induced off-diagonal elements of $Q_d(\bt_1, \bt_2)$.   Figure \ref{f:combo_plots} panel (e) plots the intrafunctional correlation structure corresponding the random intercept $Q_0(\bt_1,\bt_2)$ induced by the tensor wavelet basis chosen for the scleral strain MPS data set at the marked scleral location, and file {\tt intrafunctional\_correlation.mp4} in the supplement presents the induced intrafunctional correlation for each of the $Q_d(\bt_1, \bt_2), (d=0,1,2)$ across all scleral locations. 

This strategy can be used with any parametric model indicated by the $G_d(p), d=0,\ldots,D$, preferably orthogonalized.   Section \ref{s:Application} demonstrates that the IOP effect in the scleral strain MPS data is hyperbolic, so we devise an orthogonalized hyperbolic model for these data.  Also, note that while the serial variable IOP is sampled on a common grid across subjects for our data, this strategy can allow each the grid points for the serial variable to vary across subjects.


\subsection{Smooth Nonparametric Covariate Functional Effects}
\label{s:nonpara}

One of the primary scientific goals in the scleral strain MPS data is to study the effect of age on MPS and assess how it varies around the scleral surface.  Preliminary investigations of the data suggest that the age effect might not follow a simple parametric form, and a nonparametric representation might be appropriate.  While the fixed effect functions in the BayesFMM framework are linear in the covariates, using the mixed model representation of penalized splines shown by \cite{Wahba1978}, it is possible to fit a semiparametric functional mixed model with a smooth nonparametric age effect using this framework, as we will demonstrate in this section.  

\textbf{Smooth Nonparametric Age Effect for Scleral Strain MPS data}:  For ease of exposition, in this section we just consider a single smooth nonparametric term with no other covariates or random effects.  Thus, suppose for each subject $i=1,\ldots,N$ we only model a single scleral strain function $Y_i(\bt)$, say for left eye and IOP=45mmHg, with the model:
\begin{align}
\label{e:NPM0} Y_{i}(\bt)=  B_0(\bt) + f(X_{\mbox{age}_i}, \bt)+E_{i}(\bt),
\end{align}
where $B_0(\bt)$ is a functional intercept and $f(X_{\mbox{age}_i}, \bt)$ represents a nonparametric effect of age on MPS at scleral location $\bt$, with $\int f(x,\bt)dx=0 \forall \bt$ and penalizing $\int \{f''(x,\bt)\}^2 dx$ to induce smoothness across $x$ for each $\bt$.  
As we will demonstrate, it is possible to represent this smooth nonparametric term as a sum of a linear fixed effect function for age and spline random effect functions, 
\begin{align}
\label{e:splineFMM} f(X_{\mbox{age},i},\bt)=X_{\mbox{age}_i} B_1(\bt) + \sum_{m=1}^{M+2} Z_{\bsB,m}(X_{\mbox{age}_i}) U_{\calS m}(\bt) 
\end{align}
for some suitably constructed random effect design matrix $\{Z_{\bsB,m}(x), m=1,\ldots,M+2\}$ based on Demmler-Reinsch basis functions \citep{DemmlerReinsch}, with spline random effects $U_{\calS m}(\bt)$ following a mean zero Gaussian with cov$\{U_{\calS m}(\bt_1),U_{\calS m}(\bt_2)\} = Q_\calS(\bt_1,\bt_2)$.  These model components can be incorporated within a FMM framework like BayesFMM, as we now describe.

\textbf{Incorporation into BayesFMM framework:}  To fit model (\ref{e:NPM0}) using the BayesFMM framework, we fit separate penalized splines for each basis coefficient $k$, which induces correlated penalized spline fits for each scleral location $\bt$.  Specifically, after transforming the observed functions $Y_{i}(\bt)$ into the basis space according to the basis representation $Y_{i}(\bt)=\sum_{k=1}^K Y^*_{ik} \psi_k(\bt)$ as described in Section \ref{sec:BayesFMM}, we specify the following model for each basis coefficient $k=1,\ldots,K$:
\begin{align}
\label{e:NPM} Y^*_{ik}=  B^*_{0k} + f^*_k(X_{\mbox{age}_{i}})+E^*_{ik},
\end{align}
with $E^*_{ik} \sim N(0,s_k)$ and $f^*_k(x)$ a smooth nonparametric function of $x$ for basis function $k$.  We pull out the intercept $B^*_{0k}$ and constrain $\int f^*_k(x) dx=0$ to ensure identifiability in additive models that contain multiple smooth nonparametric terms.  We represent $B^*_{0k}+f^*_k(x)$ using B-spline basis functions,  
\begin{align}
\label{e:Bspline} B^*_{0k}+f^*_k(x)=\sum_{m=1}^{M+4}\calB_m(x)\nu^*_{mk},
\end{align}
where $\nu^*_{mk}$ are B-spline coefficients and $\calB_m(x), m=1,\ldots,M+4$ are the cubic B-spline basis functions defined by the knots $\eta_1,\ldots,\eta_{M+8}$ such that
\begin{align}
\nonumber a=\eta_1=\eta_2=\eta_3=\eta_4<\eta_5<\cdots<\eta_{M+4}=\eta_{M+5}=\eta_{M+6}=\eta_{M+7}=\eta_{M+8}=b,
\end{align}
and $a$ and $b$ are two boundary knots \citep{Hastie09}.  We can write model \eqref{e:NPM} in matrix form,
\begin{align}
\label{e:NPM2} \by^*_k=\bsB \bnu^*_k +\be^*_k,
\end{align}
where $\by^*_k=(Y^*_{1k},\ldots,Y^*_{Nk})'$, $\bsB$ is the $N\times(M+4)$ B-spline design matrix with the $(i,m)$-th entry being $\calB_m(X_{\mbox{age},i})$, $\bnu^*_k=(\nu^*_{1k},\ldots,\nu^*_{(M+4)k})'$, and $\be^*_k=(E^*_{1k},\ldots,E_{Nk})'\sim N(\bzero, s_k I_N)$.   Following \cite{Wand08}, we assume the following prior distribution on the 
B-spline coefficients:
\begin{align}
\bnu^*_k \sim MVN(\bzero, q_{\calS k} \bOmega)
\end{align}
where $\bOmega$ is a $(M+4) \times (M+4)$ matrix with $\bOmega_{mm'}=\int_a^b \calB_{m}''(x)\calB_{m'}''(x) dx$.  The resulting posterior mean of the spline random effects is $\hat{\bnu}^*_k=(\bsB'\bsB+\lambda^*_k\Omega)^{-1}\bsB'\by^*_k$, where $\lambda^*_k=s_k/q_{\calS k}$.  It can be shown that $\bsB \hat{\bnu}^*_k$ corresponds to the O'Sullivan penalized spline estimator of $B^*_{0k}+f^*_k(x)$ with penalty term $\lambda^*_k \int_a^b \{f^{*''}_k(x)\}^2 dx$ \citep{Wand08}, and if the knots are placed at each observed $X_{\mbox{age}_i}$, then this corresponds to the cubic smoothing spline estimator.

The spectral decomposition of $\Omega$ allows us to reformulate this prior specification as a mixed model with independent random effects as follows.  It is known that $\mbox{rank}(\Omega)=M+2$. Therefore, the spectral decomposition of $\Omega$ has the form of $\Omega=PDP'$, where $D=\mbox{diag}(0,0,d_1,\ldots,d_{M+2})$ and $P'P=I_{M+4}$. Let $P=(X_{\Omega},Z_{\Omega})$, where $X_{\Omega}$ is a $(M+4)\times 2$ sub-matrix of $P$ corresponding to the first two columns of $P$ and $Z_{\Omega}$ is a $(M+4)\times (M+2)$ sub-matrix of $P$ corresponding to the other columns. Let $\bbeta^*_k$ be a two-dimensional vector, and $\bu^*_{\calS k}=(U^*_{\calS 1k},\ldots,U^*_{\calS (M+2)k})'$ be an $(M+2)$-dimensional random vector.  It can be shown that $\bnu^*_k=X_{\Omega} \bbeta^*_k+Z_{\Omega}\mbox{diag}(d_1^{-1/2},\ldots,d_{M+2}^{-1/2})\bu^*_{\calS k}$ with $\bbeta^*_k$ a fixed effect and $\bu^*_{\calS k} \sim MVN(\bzero, q_{\calS k} I_{M+2})$.

Therefore, we have the following mixed model representation of \eqref{e:NPM},
\begin{align}
\nonumber \by^*_k &=\bsB \bnu^*_k +\be^*_k \\
\nonumber &=\bsB\{X_{\Omega}\bbeta^*_k+Z_{\Omega}\mbox{diag}(d_1^{-1/2},\ldots,d_{M+2}^{-1/2})\bu^*_{\calS k}\}+\be^*_k\\
\label{e:Bspline2} &=X_{\bsB}\bbeta^*_k+Z_{\bsB}\bu^*_{\calS k}+\be^*_k,
\end{align}
where $X_{\bsB}=\bsB X_{\Omega}$ and $Z_{\bsB}=\bsB Z_{\Omega}\mbox{diag}(d_1^{-1/2},\ldots,d_{M+2}^{-1/2})$.  The $Z_{\bsB}$ are called the Demmler-Reinsch spline bases \citep{DemmlerReinsch}.  It can be shown that $X_{\bsB}$ is a basis for the space of the straight line, so model \label{e:Bspline2} can equivalently be rewritten as
\begin{align}
\label{e:Bspline3} \by^*_k= \bone_N B^*_{0k}+\bx_{\mbox{age}} B^*_{1k}+Z_{\bsB}\bu^*_{\calS k}+\be^*_k,
\end{align}
where $\bone_N$ is an $N$-dimensional vector consisting of 1's and $\bx_{\mbox{age}}=(X_{\mbox{age},1},\ldots,X_{\mbox{age},N})'$.   We see that this is the form of a linear mixed model with a level of spline random effects with design matrix $Z_{\bsB}$ and random effects $\bu^*_{\calS k} \sim MVN(\bzero, q_{\calS k} I_{M+2})$.  With the intercept term pulled out as implied by the $\int f^*_k(x) dx=0$ assumption, the term $f^*_k(X_{\mbox{age}_i})$ in (\ref{e:NPM}) is given by $X_{\mbox{age}_i} B^*_{1k} + Z_{\bsB}(X_{\mbox{age}_i}) \bu^*_{\calS k}$, and thus in the BayesFMM framework we can incorporate a nonparametric term $f^*_k(x)$ by simply including a linear fixed effect $x B^*_{1k}$  plus a level of random effects with the corresponding Demmler-Reinsch design matrix $\sum_{m=1}^{M+2} Z_{\bsB m}(x) U^*_{\calS mk}$.  These penalized splines for each basis $k$, when projected back to the function space, induce a smooth nonparametric functional effect $f(X_{\mbox{age}_i},\bt)$ given by (\ref{e:splineFMM}), with $Q_\calS(\bt_1,\bt_2)=\sum_k \psi_k(\bt_1) \psi_k(\bt_2) q_{\calS k}$.  Based on these derivations, we can add any additional smooth nonparametric term $f(z,\bt)$ to the FMM framework by simply adding a linear fixed effect function $z B_2(\bt)$ and an additional level of spline random effects $\sum_{m=1}^{M_z+2} Z_{\bsB(z)} U_{\calS_z m}(\bt)$ with $U_{\calS_z m}(\bt)$ a mean zero Gaussian with covariance cov$\{U_{\calS_z m}(\bt_1),U_{\calS_z m}(\bt_2)\}=Q_{\calS_z}(\bt_1, \bt_2).$

A similar procedure could be followed to utilize other spline modeling approaches within this framework, e.g. P-splines \citep{PSplines} with differencing penalties or truncated polynomial splines \citep{Ruppert03}, but we prefer the O'Sullivan splines \citep{Wand08} given their natural second derivative penalty and formal connection to smoothing splines.

\textbf{Intrafunctional Correlation of $f(x,\bt)$ Across $\bt$:}  This framework allows the nonparametric smooth effect $f(x,\bt)$ of $x$ to vary over $\bt$, but is not the same as modeling independent splines for each $\bt$.  Because the splines are fit in the basis space, the nonparametric fits are correlated intrafunctionally by:
\begin{align}
\label{e:SplineCorr} \mbox{cov}\{f(x,\bt_1), f(x,\bt_2)|B_1(\bt_1),B_1(\bt_2),q_{\calS m}\} = \sum_{k=1}^K \sum_{m=1}^{M+2} \psi_k(\bt_1) \psi_k(\bt_2) \{Z_{\calB_m(x)}\}^2 q_{\calS m}.
\end{align}
This means that the spline fit for $\bt_1$ borrows strength from other functional locations $\bt_2$ according to the effective intrafunctional covariance structure $Q_{\calS}(\bt_1, \bt_2)$ that induces smoothing across $\bt$ in the spline fits of $f(x,\bt)$.

\textbf{Smoothing Parameter of $x$, $\lambda(\bt)$, is Nonstationary and Smooth Across $\bt$}: In our BayesFMM implementation of the smooth nonparametric term $f(x,\bt)$, we allow the penalized spline for each basis function $k$ to have its own smoothing parameter $\lambda_k=s_{k}/q_{\calS k}$.  The basis space model induces a residual error covariance matrix cov$\{E_i(\bt_1), E_i(\bt_2)\} = S_{\bt_1, \bt_2}$ back in the data space, with diagonal elements $s(\bt)$, and a spline random effect covariance matrix cov$\{U_{\calS m}(\bt_1),U_{\calS m}(\bt_2)\}=Q_{\calS}(\bt_1, \bt_2)$ back in the data space, with diagonal elements $q_{\calS}(\bt)$.  Thus, the effective smoothing parameter for the induced spline fit $f(x,\bt)$ at location $\bt$ is given by
\begin{align}
\label{e:lambdat} \lambda(\bt) = s(\bt)/q_{\calS}(\bt),
\end{align}
meaning that the smoothness in $x$ is allowed to vary across $\bt$, enabling some parts of the function to be linear with large $\lambda(\bt)$ and others to be nonlinear with small $\lambda(\bt)$.  Also, this smoothing parameter is not estimated independently for each $\bt$, but the off-diagonal elements of $\bS$ and $\bQ_{\calS}$ imply a dependency across $\bt$ in $\lambda(\bt)$, meaning that the model ``borrows strength'' across $\bt$ leading to smoothness in $\lambda(\bt)$ across $\bt$. 

We believe this to be the first presentation of a model with such flexibility in the literature, i.e. with $f(x,\bt)$ varying smoothly across $\bt$ with the smoothing parameter in $x$, $\lambda(\bt)$, also varying smoothly across $\bt$.  The FAMM models of \cite{Scheipl15} and \cite{Greven2017} estimate terms like $f(x,\bt)$ that are smooth across both $x$ and $\bt$, but utilize an additive penalty term involving marginal smoothing parameters in the $x$ and $\bt$ directions, $\lambda_x$ and $\lambda_{\bt}$.  This structure does not allow the type of nonstationarities enabled here, which in Section 5 of the supplement we demonstrate are necessary to accurately model the scleral strain MPS data.  It may be possible in the FAMM framework to accommodate this type of flexibility by putting a spline on $\lambda_x$ that varies smoothly across $\bt$, but this has not been done in any published paper to date, and it is not clear whether such an approach would be computationally feasible for large functional data sets.

\textbf{Degrees of Freedom Function $DF(\bt)$:}   In the penalized spline literature with penalized spline estimator given by $\hat{f}(x)=\calB(\calB' \calB + \lambda \Omega)^{-1} \calB' \by=\bX(\lambda) \by$, a standard summary of the nonlinearity of the fit is given by the dimensionality of the projection space given by $DF=\mbox{trace}\{\bX(\lambda)\}$, called the \textit{degrees of freedom} of the fit.  A $DF=2$ indicates a linear model and $DF \gg 2$ indicates significant nonlinearity.  To assess how the degree of nonlinearity of the spline fit $f(x,\bt)$ varies over $\bt$, we can compute the \textit{degrees of freedom function} $DF(\bt)$ marginally across $\bt$ by
\begin{align}
\label{e:dft} DF(\bt) = \mbox{trace}[X\{\lambda({\bt})\}] = \mbox{trace}[\calB\{\calB'\calB + \lambda({\bt}) \Omega\}^{-1} \calB'],
\end{align}
with $\lambda({\bt})$ defined as in (\ref{e:lambdat}).  In general semiparametric functional mixed models with other levels of random effects to account for interfunctional covariance according to model (\ref{e:SPFMM}) below, as necessary for modeling our scleral strain MPS data, the derivation for $DF(\bt)$ is more complex, and outlined in Section 2 of the supplementary materials. Panel (c) of Figure \ref{f:combo_plots} presents $DF(\bt)$ for the MPS data.

\subsection{General Bayesian Semiparametric Functional Mixed Model}

In order to model the highly structured scleral strain MPS data set, we need include all of the modeling structures described in the preceding sections, including random effects to capture nested and serial interfunctional correlation and smooth nonparametric smooth covariate effect functions, together in a common BayesFMM model.   To highlight its ability to model smooth nonparametric structures as described in Section \ref{s:nonpara}, it is useful to adapt the notation of the core BayesFMM  model to explicitly include these terms.  We term this version of the FMM a \textit{semiparametric functional mixed model} since it includes both linear and smooth covariate effects.  

Given a sample of functions $Y_{i}(\bt); i=1,\ldots,N; \bt \in \mathscr{T}$, with covariates for fixed linear effects $X_{ia_l}, a_l=1,\ldots,A_l$, smooth nonparametric effects  $X_{ia_n}, a_n=1,\ldots,A_n$, and $H$ levels of random effect covariates $Z_{ihm}, h=1,\ldots, H; m=1,\ldots, M_h$, we have the following semiparametric FMM:
\begin{align}
Y_{i}(\bt)=&\sum_{a_l=1}^{A_l} X_{ia_l} B_{a_l}(\bt) + \sum_{a_n=1}^{A_n} f(X_{ia_n},\bt) + \sum_{h=1}^{H} \sum_{m=1}^{M_h} Z_{ihm} U_{hm}(\bt) + E_{i}(\bt),  \label{e:SPFMM}
\end{align}
with $U_{hm}(\bt) \sim GP(\mathbf{0}, Q_h)$ and $E_{i}(\bt) \sim GP(\mathbf{0}, S)$ being mean zero Gaussian processes with covariance surfaces $Q_h, h=1,\ldots,H$ and $S$ defined on $\mathscr{T} \times \mathscr{T}$.

Using the structures defined in Section \ref{s:nonpara}, this model can be directly fit by the BayesFMM software of \cite{Morris06} using the following FMM:
\begin{align}
\label{e:SPFMM2} Y_{i}(\bt)=&\sum_{a_l=1}^{A_l} X_{ia_l} B_{a_l}(\bt) + \sum_{a_n=1}^{A_n} X_{ia_n} B_{a_n}(\bt) + \\
\nonumber &\sum_{a_n=1}^{A_n} \sum_{m_{a_n}=1}^{M_{a_n}+2} Z_{\calB m_{a_n}}(X_{i a_n}) U_{\calS a_n m_{a_n}}(\bt) +  \sum_{h=1}^{H} \sum_{m=1}^{M_h} Z_{ihm} U_{hm}(\bt) + E_{i}(\bt),  
\end{align}
with $M_{a_n}$ being the number of interior knots for the spline for $X_{i a_n}$, $Z_{\calB m_{a_n}}(X_{ia_n})$ the corresponding Demmler-Reinsch design matrix, $U_{\calS a_n m_{a_n}} (\bt) \sim GP(\mathbf{0}, Q_{\calS a_n})$ the corresponding spline random effect functions, and $U_{hm}(\bt) \sim GP(\mathbf{0}, Q_h)$ and $E_{i}(\bt) \sim GP(\mathbf{0}, S)$ modeling the interfunction covariance structure.   As described above, the model would be fit in the transformed basis space.  We first fit the basis space model with all random effects integrated out, and then sample the spline random effects from their complete conditional distribution while integrating out the other $H$ levels of random effects that capture any interfunctional covariance, and then project back to the function space in order to construct posterior samples of $f(x_{a_n}, \bt)$ on any desired grid of $\bt$.

While omitted from (\ref{e:SPFMM}) for ease of presentation, this model can also be easily made to include any desired parametric-nonparametric interaction terms, with interaction of parametrically modeled covariate $X_{ia_l}$ and nonparametrically modeled covariate $X_{ia_n}$ being represented by the term $X_{ia_l} f_{a_l}(X_{ia_n},\bt)$.  For $X_{ia_l}$ that are categorical dummy variables, this allows separate nonparametric fits of $(X_{ia_n},\bt)$ for different levels of the dummy variable.    For continuous $X_{ia_l}$, this allows the corresponding slope to vary smoothly and nonparametrically with both $X_{ia_n}$ and $\bt$.  For example, in our scleral strain data, one may wish to include an interaction term to allow the nonparametric age effect to vary across IOP levels.  If dummy variables were specified for each IOP level, this would allow separate independent nonparametric age effects for each IOP level.  If IOP is modeled continuously via a parametric model like the hyperbolic model described in Section \ref{s:Application}, this would allow the hyperbolic coefficients to vary smoothly by age and scleral position, which would be equivalent to nonparametric age effects that vary across IOP but borrow strength from nearby IOP according to the structure induced by the hyperbolic model.  In either case, the fixed effect and random spline design matrices corresponding to the $X_{ia_l} f_{a_l}(X_{ia_n},\bt)$ would be given by $X_{ia_l} X_{ia_n}$ and $X_{ia_l} Z_{\calS m}(X_{i a_n})$, respectively, which are straightforwardly included in the FMM.  As described in Section \ref{s:Application}, we considered these interaction structures, but found they did not appear necessary for representing the scleral strain MPS so were not included in the final model.

\section{Model Selection Heuristic for Semiparametric BayesFMM}
\label{s:model_sel}

Given the extensive flexibility of the semiparametric BayesFMM framework, there are a large number of modeling decisions to make.  For example, in our sclera strain MPS data set, should the age effect be linear or nonparametric?  If nonparametric, should the smoothing parameter in age be allowed to vary across the scleral surface $\bt$, or is a common smoothing parameter across all scleral locations sufficient?  Should the fixed IOP effect be linear, hyperbolic, or nonparametric?  Should there be an interaction of age and IOP?  Should there be a fixed left vs. right eye effect?  For the random effect levels, is the subject-specific random effect necessary to account for correlation between right and left eyes for the same subject?  Is the correlation across functions from multiple IOP for the same eye sufficiently handled by a compound symmetry structure assuming equal correlations, or is a structure allowing serial correlation necessary?  Should this serial correlation be based on a linear, parabolic, or hyperbolic model?  These decisions are challenging to make in a simple generalized additive mixed model framework with scalar responses, and become even more challenging in the current setting with complex, high-dimensional functional responses.

There are a some papers in the frequentist literature for performing variable selection in functional regression contexts \citep{Scheipl2013, Gertheiss2013b, Brockhaus2015}.  However, there is a lack of functional regression model selection methods for MCMC-based fully Bayesian models such as the semiparametric BayesFMM, which present special challenges.  One could split the data into training and validation data sets, fit separate MCMC for each prospective model in the training data, and then compute ratios of predicted marginal densities, integrating over MCMC posterior samples, for the validation data, as done in \cite{Zhu14}, for example.  These predictive Bayes Factors would provide a rigorous model selection measure, or alternatively parallel MCMC could be run for each prospective model, and a multinomial random variable with Dirichlet prior used to select and perform Bayesian model averaging across models as the chains progress.  These strategies might work fine for simple, low dimensional data sets or settings with only a few prospective models, but for the current setting with complex, high-dimensional data, they are impractical.

In this section, we present a model selection heuristic that we have developed that can explore a number of potential model structures to find which seem to be most appropriate for the given data without running any MCMC, and also provides ML and REML estimates that can be used as starting values for the parameters in the BayesFMM.  This heuristic is admittedly \textit{ad hoc}, but is based on standard methods and appears to perform well in simulations, and so we believe can be a useful tool for modelers to assess which structures to included in their semiparametric FMM.

Our overall approach is to fit linear mixed models (LMM) to each basis coefficient $k$ using the {\tt lme} function in R \citep{nlme} for each prospective model, and then use a weighted voting scheme based on importance weights for each basis and an adapted Bayesian Information Criterion (aBIC) to obtain probability scores for each prospective model.  
Here we outline the steps in detail.

\begin{enumerate}
\item{\textbf{Basis transform and importance weights:}} Transform the raw functions $Y_i(\bt), i=1,\ldots, N$ to the basis space $Y^*_{ik}, k=1,\ldots,K$, and compute a series of weights $w_k$ that measure the relative importance of each basis for representing the data set.  These weights can be computed by $w_k=\sum_{i}{Y_{ik}^*}^2/\sum_{i}\sum_{k}{Y_{ik}^*}^2$, with $\sum_k w_k=1$.  For orthogonal $\psi_k(\bt)$, the $w_k$ represent the relative percent energy captured by basis coefficient $k$.
\item{\textbf{Fit basis-specific LMM  and compute aBIC scores:}} For each prospective model $\calM_c, c=1,\ldots,C$, use {\tt lme} in R \citep{nlme} to fit the corresponding LMM to the data for each basis coefficient $k=1,\ldots,K$, and compute an adapted version of the BIC ($aBIC_{ck}$), which we define as:
\begin{align}
aBIC_{ck} = -2 \mbox{log-likelihood}_{ck} + n_{\mbox{par},c}\mbox{log}(N) \nonumber,
\end{align}
where the log-likelihood is the marginal likelihood of the fixed effect and variance components of the model with the non-spline random effects integrated out conditional on the data for basis $k$, $N$ is the total number of observations in the dataset for basis $k$, and $n_{\mbox{par},c}$ is the total number of parameters of model $c$.  As discussed by \cite{Vaida2005} and \cite{DIC}, selection of the effective number of parameters for LMM or Bayesian hierarchical models is tricky and context dependent.  If inference is desired on the random effects themselves, then counting only fixed effects and variance components as parameters is not appropriate.  In our setting, we are not interested in the random effects at levels capturing interfunctional correlation as we work with the marginalized model, but for nonparametric terms $f(x,\bt)$ we are clearly interested in the ``random effects" corresponding to the spline coefficients.  Thus, we count the number of parameters to be the sum of the number of fixed effects, the number of variance components and the estimated degrees of freedom for each nonparametric term.  This last term adjusts appropriately for the extra parameters of the spline fits even thought they are captured as random effects in the LMM.
\item{\textbf{Use weighted voting scheme to rank models:}}  We compute a probability weight $P_c$ for each model $\calM_c; c=1,\ldots,C$:
\begin{align}
\nonumber P_c=\sum_{k=1}^K w_k I\{c=\argmin_{c'}aBIC_{c'k}\}
\end{align}
\end{enumerate}
This procedure is applied in two steps: first assessing different fixed effect models (including parametric and/or nonparametric effects), and second assessing various random effect structures for capturing interfunctional variability while conditioning on the best fixed effect model.

In principle, $I\{c=\argmin_{c'}aBIC_{c'k}\}$ indicates whether the model $\mathcal{M}_c$ is the best in terms of $aBIC$ for the data set on the $k^{th}$ basis. Therefore, the $P_c$  is computed via a weighted voting scheme, an aggregated measure of proportion of times $\mathcal{M}_c$ is the best model across the all basis coefficients, with basis coefficients weighted by $w_k$.  
In this way, the model fit for basis coefficients that account for a larger proportion of the total variability in the data count more towards the overall model selection.  Empirically, we have found this weighted voting scheme seems to work well, as it is robust in the sense of not allowing any one basis function, especially one explaining a relatively low proportion of total energy for the given data set, to dominate the model selection because of an extreme $aBIC$ score.  This can also be applied using alternative measures (e.g. $aAIC$).  We acknowledge that this strategy is \textit{ad hoc} and more rigorous model selection methods for settings like this are needed, but we believe it can provide useful guidance for modelers and performs well in simulations, as described below.

\textbf{A word of caution:}. This heuristic is meant for selecting among various different modeling structures for specified covariates as done for our case study, or perhaps could be used to select among a few covariates, but \textit{it is not intended for high-dimensional variable selection across many potential predictors}.  In such settings, consideration of a large number of models and only fitting the ``best'' one can dramatically inflate type I error rates, and post-selection inference as described in \cite{Berk2013} would need to be considered. 

\textbf{Simulation Study on Model Selection:}  We conducted a simple simulation study to investigate the performance of this model selection heuristic.  We considered four different models:
\begin{itemize}
	\item Model 1 (null model):  $Y_{ij}(\bt) = B_0(\bt)+E_{ij}(\bt)$,
	 \item Model 2 (linear age effect):  $Y_{ij}(\bt) = B_0(\bt)+X_{\mbox{age},i}B_{\mbox{age}}(\bt)+E_{ij}(\bt)$,
	 \item Model 3 (nonparametric age effect): $Y_{i}(\bt) = f(X_{\mbox{age},i}, \bt)+E_{ij}(\bt)$, and
	 \item Model 4 (linear age effect, random effect): $Y_{ij}(\bt) = B_0(\bt)+X_{\mbox{age},i}B_{\mbox{age}}(\bt)+U_{j}(\bt)+E_{ij}(\bt)$.
\end{itemize}
We fit each of these models to the scleral strain data and used the fitted model as the truth, and simulated 100 replicate data sets for each model.  For each simulated data set, we fit each of these four models and performed the model selection procedure using $P_c$ to select the best model.  In all four scenarios, this procedure selected the correct model 100/100 times. {Average values of $P_c$ for each model can be found in Section 4 of the supplementary materials, and Section 6 of the supplement investigates issues that can arise in variable selection of GAMMs when considering nonparametric smooth terms of subject-specific covariates in models including subject-level random effects.}


\section{Glaucoma Scleral Strain MPS Case Study} \label{s:Application}

\subsection{Overview of Glaucoma Scleral Strain MPS Data}  

As described in Section \ref{s:intro}, glaucoma is characterized by ONH damage related to IOP but its etiology is not fully known.  Researchers have hypothesized that biomechanics of the scleral region close to the ONH may modulate the effect of IOP on the ONH, and thus may play an important role in glaucoma.  In particular, the scleral surface is elastic so deforms under pressure, which can partially relieve IOP-induced forces on the eye, including the ONH.  Thus, studies of these properties could reveal insights into the etiology of glaucoma.

Recently, novel custom instrumentation was developed that can precisely measure the mechanical strain in the posterior human sclera at a fixed level of IOP \citep{Fazio12a,Fazio12b}.  Briefly, the posterior 1/3 of the eye is clamped, sealed, and pressurized.  Next, the eye is {preconditioned}, and then pressurized from {7 mmHg to 45 mmHg} using an automated system with computer feedback control, while scleral surface displacements are measured by a laser speckle interferometer.  This device measures a light interference distribution that is used to reconstruct the surface displacement field in three dimensions {with nanometer-scale precision}.  These displacements were processed as described in \citet{Fazio12a} to compute the 3D strain tensor, a $3 \times 3$ matrix summarizing the displacement in the meridional, circumferential, and radial directions, continuously around the outer scleral surface.  The leading eigenvalue of the strain tensor, called the maximum principal strain (MPS), was computed on a grid of scleral locations for 120 circumferential locations $\phi\in(\SI{ 0}{\degree},\SI{360}{\degree})$ and 120 meridional locations $\theta \in(\SI{ 9}{\degree},\SI{24}{\degree})$, where $\theta=\SI{0}{\degree}$ corresponds to the ONH.  This yields MPS functions defined on a grid of 14,400 points on the scleral surface that comprises a partial spherical domain.

Using this custom instrumentation, \citet{Fazio14} conducted a study to investigate age-related changes in the scleral surface strain.  They obtained twenty pairs of eyes from normal human donors in the Lions Eye Bank of Oregon in Portland, OR and the Alabama Eye Bank in Birmingham, AL.  For each subject, the MPS measurements were obtained as described above at nine different levels of IOP ({7, 10, 15, 20, 25, 30, 35, 40, and 45 mmHg}) for both left and right eyes.  The data for both eyes from one subject failed a quality control check, so was excluded from analysis, as did one of the eyes from four other subjects.  Thus, the data we analyzed consisted of 34 eyes from 19 subjects.  With 14,400 measurements for each of 9 IOP levels $\times$ 34 eyes, this data set contained over 4.5 million measurements.  Let $Y_{ijp}(\bt)$ be the MPS for eye $j$ for subject $i$ under IOP level $p$ at scleral location indexed by $\bt=(\theta, \phi)$, which on the sampling grid can be written as a vector $\by_{ijp}$ of length $14,400$.  The primary goals are to study MPS, assessing how it varies around the scleral surface, across IOP, and with age.  The hypothesis is that MPS is greater near the ONH, which could confer a protective effect, and that MPS tends to decrease with age, which could contribute to increased stress on the ONH thus conferring increased glaucoma risk.

\subsection{Model Specification}

\textbf{Basis Transform:}
Various criteria can be considered when choosing which basis to use within the BayesFMM framework, including sparse representation, fast calculation, richness for representing the functional parameters at the various levels of the models, ability to capture the key visual features of the observed functions, and flexibility for representing the intrafunctional correlation in the data.  Multiresolution bases like wavelets have advantages for many of these considerations, so we constructed a custom rectangular wavelet basis defined on the cylindrical spherical projection of the partial scleral space $\bt=(\theta, \phi)$, which is a tensor transform computed by successively applying 1D wavelet transforms to the meridional and circumferential directions.

\textbf{Tensor Wavelets for Scleral Space:}. Specifically, we constructed $\psi_k(\bt)=\psi_k(\theta,\phi)$ as a tensor wavelet, $\psi_k(\bt) = \psi^{\theta}_{k1}(\theta) \otimes \psi^{\phi}_{k2}(\phi)$, with meriodonal wavelet $\psi^{\theta}_{k_1}(\theta)$ being a db3 wavelet basis with three vanishing moments, reflection boundary condition, 5 levels of decomposition and circumferential wavelet $\psi^{\phi}_{k_2}(\phi)$ being a db3 wavelet with three vanishing moments, 5 levels of decomposition and periodic boundary conditions since its domain is circular, covering the entire circumferential space.  This transform yielded a basis  $\{\psi_k(\bt); k=1, \ldots, K=17,185\}$.  While single-indexed here for simplicity of presentation, these basis coefficients can be written as multi-indexed by circumferential scale $j_1=0, \dots, 5$, meriodonal scale $j_2=0, \ldots, 5$, circumferential locations $k_1=1,\ldots,K_{1j_1}$ and meriodonal locations $k_2=1,\ldots,K_{2j_2}$ with $K=\sum_{j_1}\sum_{j_2} K_{1j_1}*K_{2j_2}$.  The levels $j_1=0$ and $j_2=0$ correspond to the father wavelet coefficients at the lowest level of decomposition, and the other $j_1$ and $j_2$ index the corresponding mother wavelets at increasing levels of scale.  With $\by_{ijp}=Y_{ijp}(\bt)$ being the observed function for subject $i$, eye $j$, and IOP $p$ on the scleral surface sampling grid of size $T=14,400$ written in vector form, this basis representation can be written as $\by_{ijp}=\by^*_{ijp} \bPsi$, where $\bPsi$ is a $K \times T$ basis matrix with elements $\psi_k(\bt)$ and $\by^*_{ijp}$ is a vector of $K$ corresponding basis coefficients.  Because of the structure of the tensor transform, if we unstack $\by_{ijp}$ into a $(T_1=120) \times (T_2=120)$ matrix $\bY_{ijp}$ with rows indexed by equally spaced meriodonal locations $\theta_1=\SI{9}{\degree}, \ldots, \theta_{120}=\SI{24}{\degree}$ and columns by equally spaced circumferential locations $\phi_1=\SI{ 0}{\degree}, \ldots, \phi_{120}=\SI{360}{\degree}$, we could write $\by^*_{ijp}=\mbox{vec}(\bPsi_\theta \bY_{ijp} \bPsi'_\phi)$, where $\mbox{vec}(\bullet)$ is the column-stacking vectorizing operator, $\bPsi_\theta$ is the $K_1 \times (T_1=120)$ basis matrix corresponding to the meriodonal wavelet $\psi^\theta_{k_1}(\theta)$ and $\bPsi_\phi$ the $K_2 \times (T_2=120)$ basis matrix corresponding to the circumferential wavelet $\psi^\phi_{k_2}(\phi).$   In principle, spherical wavelets could be used as the transforming basis, but currently available software does not handle transforms for part of the sphere, and since we only model $\theta$ over a limited range of $\SI{ 9}{\degree}$ - $\SI{24}{\degree}$, the distortion from using the basis on the projection and not the true spherical geodesic is not great.

Some eyes had technical processing artifacts that resulted in a spike of extremely high MPS at some local set of scleral locations, typically close to the boundary.  Given the multiresolution nature of the wavelet transform, these artifacts were captured by wavelet coefficients at extremely high frequency scales.  Given the relatively smooth nature of most of the MPS functions, these wavelet coefficients were essentially zero for all eyes except for those with the artifact which yielded very large coefficients.  Thus, we removed these artifacts by filtering out any  wavelet coefficients with \textit{extremely} skewed distributions for which the mean across all samples was more than {$100 \times$} the median.  As seen in the supplementary file \textit{RawMPScurves.zip}, with illustration in Supplemental Figure 26, this strategy effectively removed the outlying spikes without substantively affecting MPS values for other scleral locations.   We also applied the joint wavelet compression strategy described in \citet{Morris11} to obtain a reduced dimension near-lossless basis function to use, and found a subset of 269 wavelet coefficients that jointly preserved $>99.5\%$ of the total signal energy for each eye, and an average of $>99.9\%$, leading to $>50:1$ compression.  As shown in Supplemental Figure 26 and \textit{RawMPScurves.zip}, the data projected into the basis is essentially identical to the raw data, demonstrating its near-lossless nature.  We considered this basis for our model.

We modeled these basis coefficients using the basis transform modeling approach described in Section \ref{sec:BayesFMM}.  Besides providing a relatively sparse representation and enabling the adaptive removal of spiky artifacts, this transform being a location-scale decomposition allowed nonstationary intra-scleral correlations and adaptive borrowing of strength across scleral locations.  File {\tt intrafunctional\_correlation.mp4} in the supplement contains a movie file demonstrating the form of the intrafunctional correlation structure induced by this choice,  computed by constructing the basis and basis transform matrices $\bPsi$ and $\bPsi^-$, respectively, and applying (\ref{eq:covariance}) to the basis space covariances at the various random effect and residual error levels of the model.  For illustration, panels (e) and (f) of Figure \ref{f:combo_plots} contains plots of this surface at a particular scleral location for two of the random levels.

We also considered using principal components computed on the wavelet-transformed (and compressed) data, similar to the strategy used in \cite{Meyer15}, which implies applying a singular value decomposition to the wavelet-space data matrix, and then using the resulting eigenvectors to construct the empirical basis functions $\psi_k(\bt)$ that are used for the BayesFMM modeling.   In this case, we kept $K=29$ basis functions that explained $>99.5\%$ of the total variability in the data set according to the scree plot, which as estimated by four-fold cross validation retained a minimum of $96.7\%$ of the total energy for each eye, so is somewhat near-lossless.  We used the wavelets for our primary analysis given that it yielded a richer basis set for representing the various functional parameters at various levels of the models, but for sensitivity we also presented results using the BayesFMM using these wavelet-regularized principal components in Section 7 of the supplement, as well as other summaries including the induced intrafunctional correlation structures. 


\textbf{Model Selection:}  We applied the model selection heuristic described in Section \ref{s:model_sel} to help select the structures in the semiparametric FMM to include in the FMM. We first determined which fixed effect covariates to include and what their functional forms should be. Here we summarize the results, for which more details are provided in Section 3 of the supplementary materials.  Three different fixed effects were considered: age, IOP, and eye (left vs. right). For the form of the age effect, we considered two possibilities: linear or nonparametric. For the form of the IOP effect, we considered three different possibilities: linear, hyperbola, or nonparametric. Models without the eye effect were also compared. As a result, we compared 12 different models for the fixed effect selection. It turned out that the model with the nonparametric age effect, the hyperbolic IOP effect, and no eye effect showed the highest $P_c$ when using $aBIC$, and the model with nonparametric age effect, hyperbolic IOP effect, and a left vs. right eye effect had the highest $P_c$ when using $aAIC$.  For our primary analysis, we considered the model with no left vs. right eye effect, since there is no strong scientific rationale for such an effect, and present the other model as a sensitivity analysis in Section 7 of the supplementary materials.   We also assessed whether the smoothing parameter for the nonparametric age effect should be constant or vary around the sclera, and found that the sclerally varying smoothing parameter was clearly necessary for good fit, as detailed in Section 5 of the supplement. Once we selected the main fixed effects, we assessed whether the interaction term between age and IOP was needed, and our model selection heuristic suggested the interaction was not necessary. Finally, with the selected fixed terms, we compared several different random effect distributions to capture the interfunctional covariance structure. Two different levels of random effects were considered: the subject-level random effect and the serial eye-level random effect. For the form of the eye-level random effect in terms of IOP as illustrated in Section \ref{s:growth}, we considered three different forms: constant (compound symmetry), linear, or hyperbola. Our model selection heuristic selected the eye-level random effect with the hyperbolic IOP effect, but not the subject-level random effect.

\textbf{Model:} Thus, the final fitted semiparametric FMM was:
\begin{align}
\nonumber Y_{ijp}(\bt) =& B_0(\bt) +B_{1}(\bt)G_1(p)+B_{2}(\bt)G_2(p)+f(X_{\mbox{age},i}, \bt)+\\
\label{e:finalmodel4} &U_{ij}(\bt)+U_{ij1}(\bt)G_1(p)+U_{ij2}(\bt)G_2(p)+E_{ijp}(\bt),
\end{align}
with $X_{\mbox{age},i }$ is the age for subject $i$, and $G_1(p)$ and $G_2(p)$ are the values of the orthogonalized hyperbolic basis corresponding to $\mbox{IOP}=p$ as described below.  This is equivalent to the FMM:
\begin{align}
\label{e:finalFMM} Y_{ijp}(\bt) =& B_0(\bt) + B_{1}(\bt)G_1(p)+B_{2}(\bt)G_2(p)+B_3(\bt) X_{\mbox{age},i} + \sum_{m=1}^{M+2} Z_{\calB m}(X_{\mbox{age},i}) U_{\calS m}(\bt) + \\
\nonumber &U_{ij}(\bt)+U_{ij1}(\bt)G_1(p)+U_{ij2}(\bt)G_2(p)+E_{ijp}(\bt),
\end{align}
with $Z_{\calB,m}(x)$ the Demmler-Reinsch basis functions corresponding to $M$ interior knots on $x$, $U_{\calS m}(\bt) \sim GP\{\mathbf{0}, Q_{\calS}\}, U_{ij}(\bt) \sim GP\{\bzero, Q_0\}, U_{ij1}(\bt) \sim GP\{\bzero,Q_1\}, U_{ij2}(\bt) \sim GP\{\bzero, Q_2\}$, and $E_{i}(\bt) \sim GP(\mathbf{0}, S)$, and with $Q_{\calS}, Q_0, Q_1, Q_2$, and $S$ being covariance surfaces defined on $\calT \times \calT$.  Following the guidelines suggested by \cite{Ruppert03}, we chose $M=5$ equally spaced knots over $X_{\mbox{age}}$.

\textbf{Parameterization of IOP effect:}. From a preliminary investigation in which we fit separate models to each scleral location $\bt$, we found that the serial IOP effects were well modeled by a hyperbola of special form $Y=b_0+b_1 p + b_2 p^{-1}$, with an average $R^2$ of $0.98$ across all eyes and scleral locations, and this form was also chosen by the model selection heuristic.    To accommodate model fitting without having to include a covariance between $b_1$ and $b_2$, we utilized orthogonalized versions of these predictors:  $G_1(p)=\sqrt{2}/2 X_{1,p} -\sqrt{2}/2 X_{2,p}$ and $G_2(p)=\sqrt{2}/2 X_{1,p} +\sqrt{2}/2 X_{2,p}$ where $X_{1,p}$ is a standardized version of $p$ and $X_{2,p}$ is a standardized version of $p^{-1}$.   

\textbf{Basis Space Model:}  We used the fast rectangular 2D wavelet transform to compute the basis coefficients from the raw functions, equivalent to the matrix multiplication $\by^*_{ijp}=\by_{ijp} \bPsi^{-}$ with $\bPsi^-=\bPsi'(\bPsi \bPsi')^{-1}$ with $\by^*_{ijp}$ a vector of length $K=269$ with elements $Y^*_{ijpk}, k=1,\ldots,K$.  We then fit the basis-space version of model (\ref{e:finalFMM}):
\begin{align}
\label{e:finalFMMk} Y^*_{ijpk} =& B^*_{0k} + B^*_{1k} G_1(p)+B^*_{2k}G_2(p)+B^*_{3k} X_{\mbox{age},i} + \sum_{m=1}^{M+2} Z_{\calB m}(X_{\mbox{age},i}) U^*_{\calS mk} \\
\nonumber &+U^*_{ijk}+U^*_{ij1k}G_1(p)+U^*_{ij2k}G_2(p)+E^*_{ijpk}, \mbox{  with}
\end{align}
$U^*_{\calS mk} \sim N(0,q_{\calS k}), U^*_{ijk} \sim N(0,q_{0k}), U^*_{ij1k} \sim N(0,q_{1k}), U^*_{ij2k} \sim N(0,q_{2k})$, and $E^*_{ijpk} \sim N(0,s_k)$.

\textbf{Prior Specification:} We specified vague conjugate inverse Gamma priors for each basis space variance component $\{q_{\calS k}, q_{0k}, q_{1k}, q_{2k}, s_k\}$  in the model, with prior mode being the REML starting values and with effective sample size of 2, e.g. $s_k \sim \mbox{InverseGamma}(a_s, b_s)$ with $a_s=2$ and $b_s=3*\hat{s}_k$ where $\hat{s}_k$ is the REML starting values for $s_k$.   We used spike-slab priors for the basis-space fixed effects $\{B^*_{ak}, a=0, \ldots, 3\}$, with regularization parameters $\{\pi_{aj}, \tau_{aj}\}$ varying over predictor $a=0, \ldots, 3$, with \textit{regularization sets} $j=1, \ldots, J=36$ determined by the tensor wavelet scale levels, with $j=0$ for $j_1=j_2=0$, $j=1$ for $j_1=0, j_2=1$, \ldots, $j=J=36$ for $j_1=j_2=5$.   We estimated the regularization parameters using the empirical Bayes algorithm specified in \cite{Morris06}.  To assess sensitivity of results to these choices of regularization parameters, we also ran the model doing no additional regularization (beyond wavelet compression) by setting $\pi_{aj}\equiv 1$ and $\tau_{aj}\equiv 10^6$.  Results are provided in Section 7 of the supplementary materials.

\textbf{Model Fitting:}  We ran an MCMC to obtain posterior samples of the parameters of model (\ref{e:finalFMMk}) $\{B^*_{\bullet k}, q_{\bullet k}, s_k\}$ from the marginalized version of this model with $U^*_{\bullet k}$ all integrated out. 
We fit a total of 10,000 posterior samples after a burn-in of 5000, thinning by keeping every 10.  We then sampled the spline random effects $U^*_{\calS m k}$ from their complete conditional distributions with the other random effects still integrated out, which are conjugate multivariate normal Gibbs steps as detailed in Section 1 of the supplementary materials, from which posterior samples of $f^*_k(x)=xB^*_{3k} + \sum_m Z_{\calB m}(x) U^*_{\calS mk}$ were subsequently constructed for a grid of ages $x$ of size $71$ corresponding to ages 20-90.  Let $\mathbf{F}^*_{g}$ be a $71 \times (K=269)$ matrix representing posterior sample $g$ of the basis space nonparametric age effect, $g=1,\ldots, 1000$.  We then transformed this back to the data space via $\mathbf{F}_g=\mathbf{F}^*_{g} \Psi$ to obtain the $71 \times (T=14,400)$ matrix of posterior samples of the nonparametric age effect $f(X_{\mbox{age}},\bt)$ in model (\ref{e:finalmodel4}), and similarly transforming the other fixed effects back to the data space to get posterior samples of $\{B_0(\bt), B_1(\bt), B_2(\bt)\}$ on the sampling grid of $\bt$ and used for posterior inference.  

On a laptop computer, the entire analysis took 7hr39min on a single core,  with the basis transform taking 1m37s, model selection 22m48s, each MCMC iteration 0.77s with 15,000 iterations taking 3hr15m, and the postprocessing including inverse basis transform of posterior samples and key inferential summary calculations 4hr.  Many other summaries and plots were computed for the purposes of this paper for sensitivity and illustration of the deep properties of the modeling framework at more computational expense, but these additional analyses are not necessary for analysis of the data.  The Metropolis-Hastings acceptance probabilities ($\approx 0.85-0.95$) were reasonable, and Geweke convergence statistics (median 0.01, $Q_{.025}=-1.98, Q_{0.975}=1.94$) across the many parameters in the model showed that most were within roughly 2 standard deviations of zero, and only $4.4\%$ of the corresponding p-values were less than 0.05, so suggest reasonable MCMC convergence (see Section 9 of the supplementary materials for more details).  
We simulated virtual MPS functions for hypothetical subjects with specified age and IOP from the posterior predictive distribution of the data (see Section 10 of the supplementary materials), and found that the simulated MPS data are visually similar to the MPS data from real eyes, suggesting the BayesFMM model with the tensor wavelet bases was sufficiently flexible to capture the salient features of these data, and lending support to its use for inference.  We share these pseudo data on github (\url{https://github.com/MorrisStatLab/SemiparametricFMM}), as well as the real data and scripts to perform all analyses.


\begin{figure} 
\centerline{\includegraphics[scale=0.7]{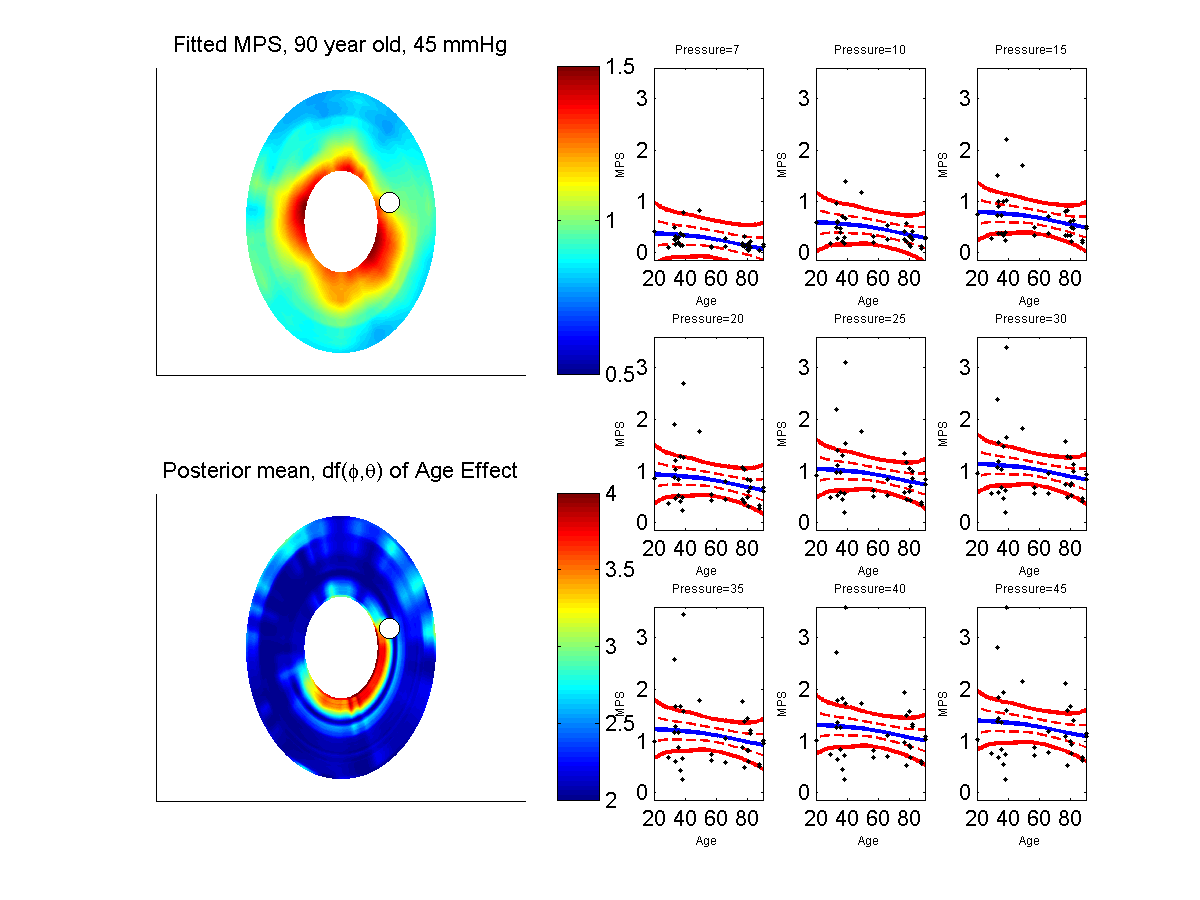}}
\caption{\footnotesize (a) Polar azimuthal projection of fitted MPS function for left eye from one subject of age 90yr under {45 mmHG} of IOP. (b) posterior mean of degrees of freedom of nonparametric MPS fit of age. {The right nine panels} depict estimated nonparametric MPS fit of age for all nine IOP levels at the scleral location indicated by {the white dot} in (a) and (b), along with the raw data indicated by the black dots.}\label{f:MPS_age_pressure}
\end{figure}

\subsection{Scientific Results} \label{sec:Results}

{\bf Nonparametric MPS function of age:} First, we computed the posterior mean MPS curve as a function of age over the entire meridional and circumferential domain at each level of IOP.   As a thorough summary of the model fit, we generated a plot of these fits for each scleral location, and joined these together to make a digital movie file {\tt MPSvAGE-wave.mp4} contained in the supplemental materials.  
Figure \ref{f:MPS_age_pressure} shows a snapshot of the movie at the scleral location indicated by the white dot. Panel (a) depicts a polar azimuthal projection of the fitted MPS function for a left eye from a subject of age 90yr under {45 mmHg} of IOP.  Note that MPS is higher near the ONH, as expected as a protective effect.  Panel (b) shows the posterior mean degrees of freedom (DF) of the nonparametric age effect. We see strong nonlinear age effects in scleral regions close to the ONH and towards the inferior and nasal regions of the sclera, while many other regions show linear or almost linear age effects. The right nine panels contain the fitted nonparametric age effect at nine different IOP levels at the scleral position indicated by the white dot. In each panel, the black dots are the raw data, the solid blue line is the estimated nonparametric mean MPS function of age, and the solid and dashed red lines correspond to joint and point-wise 95\% credible bands of the mean MPS curve, with joint bands computed as described in \citet{Meyer15}.  We see from this plot how the MPS increases with IOP, and that the hyperbolic model seems to capture the rate of increase very well.   From this plot and the movie in the supplement showing results stepping across the scleral locations, we see this model fits the data for all IOP and scleral locations remarkably well in spite of the fact that independent splines were not fit to the data for each IOP and scleral location separately, but rather is the result of the complex joint unified model \eqref{e:finalmodel4} that borrows strength from other IOP according to the modeled hyperbolic serial effect and from other scleral locations according to the basis functions. The model also borrows strength from other nearby scleral locations according to the basis functions in estimating the nonlinearity of the age effect, as seen in the local smoothness of the $DF(\bt)$ plot.  

\begin{figure}
\centerline{\includegraphics[scale=1]{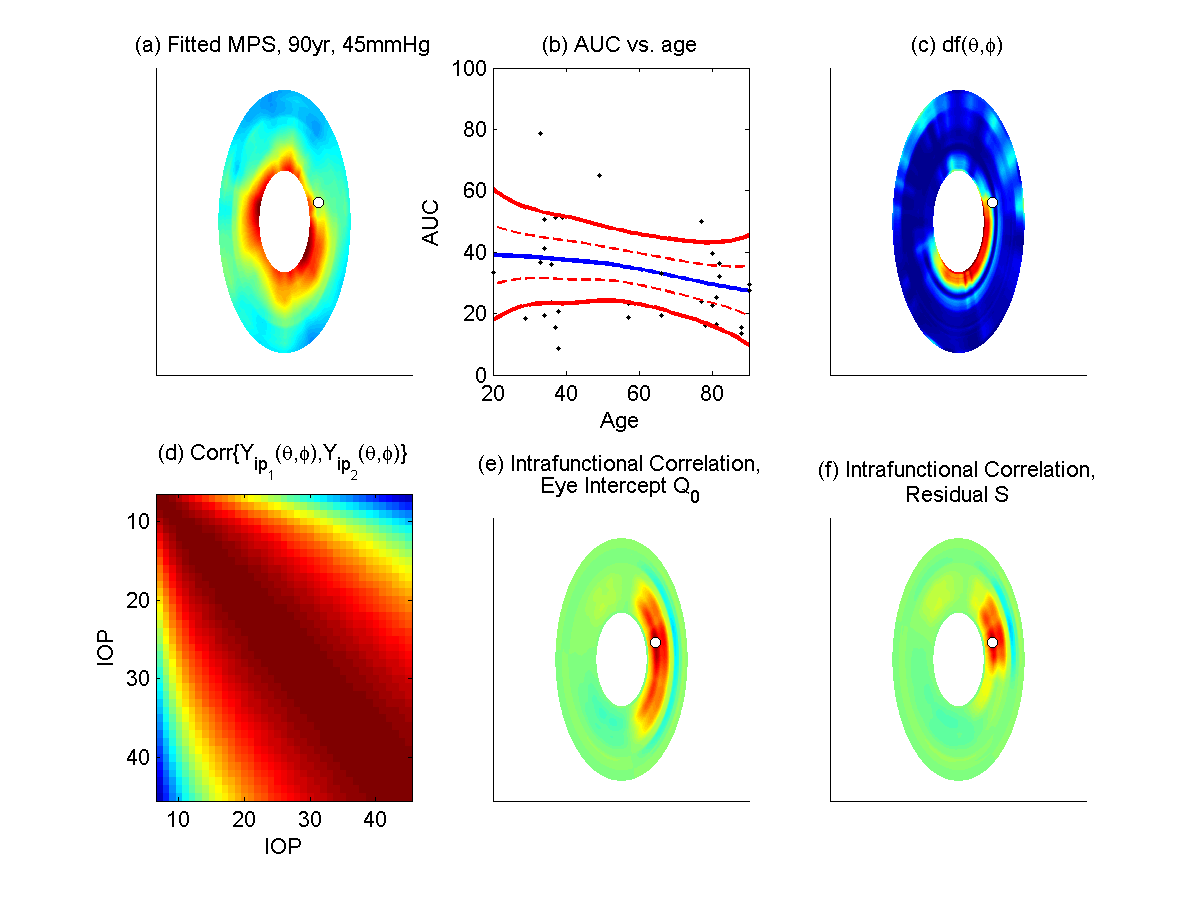}} 
\caption{ \footnotesize \textbf{Key Summaries of Fitted Model}.  Shows key summaries at the scleral position marked by the open circle, including (a) the fitted MPS for a 90yr old with IOP=45mmHg, (b) the nonparametric MPS vs. age curve using the AUC to integrate over IOP, with blue line being posterior mean, dotted and solid red lines pointwise and joint credible bands, and the raw data (computing AUC for this scleral location for each eye) indicated by dots, (c) the degrees of freedom of the nonparametric age fit as a function of the scleral location, (d) the serial correlation across IOP induced by the model at this scleral position, and (e) and (f) being the intrafunctional correlation surface induced by our model and choice of tensor basis for the eye-to-eye random intercept and residual error levels, respectively, at this scleral position.  The file {\tt combo\_plot.mp4} in the supplement is a movie file showing how these summaries vary across scleral locations.}\label{f:combo_plots}
\end{figure}

{\bf Induced Intrafunctional and Interfunctional Covariance Structures:}  To assess the intrafunctional covariance structures induced by our tensor wavelet bases, we used equation (\ref{eq:covariance}) to estimate the scleral space intrafunctional covariance matrices $Q_d(\bt_1,\bt_2), d=0,1,2$ and $S(\bt_1,\bt_2)$.  Supplementary Figure 22 plots the diagonals of these matrices representing the variances at the various hierarchical levels as a function of scleral location $\bt$.  Note that the  variance for the eye intercept $Q_0(\bt,\bt)$ is an order of magnitude greater than that of the eye-level IOP coefficients $Q_1(\bt,\bt)$ and $Q_2(\bt,\bt)$, which are in turn an order of magnitude greater than the residual error variance $S(\bt,\bt)$.  Also, note that these covariances vary around the scleral surface, with locations near the ONH having greater levels of variability.  The supplement also contains a movie file {\tt interfunctional\_cor.mp4} that represents the corresponding intrafunctional correlation surfaces induced by our model by stepping around scleral locations and for each plotting a heatmap representing the correlation of the indicated location with all other scleral locations.  For illustration, panels (e) and (f) of Figure \ref{f:combo_plots} show the correlation of a specific scleral location $\bt^*$ (indicated by the white dot) with all other scleral locations $\bt$ at the eye intercept $Q_0(\bt^*, \bt)$ and residual error $S(\bt^*,\bt)$ levels, and Supplementary Figure 18 contains these plus those for the eye hyperbolic random effects $Q_1(\bt^*,\bt)$ and $Q_2(\bt^*,\bt)$ for this location, and {\tt Intrafunctional\_correlations.mp4} contains a movie file showing all scleral locations.  Note how our model captures local intrafunctional correlation, and the strength and tails of this correlation are allowed to vary by scleral location and hierarchical level.  Supplementary Figure 19 includes equivalent plots for the wavelet-regularized principal component basis functions, and {\tt Intrafunctional\_correlations-pc.mp4} contains a movie showing all scleral locations.  Note that the PC basis functions, although global, induce intrafunctional correlation surfaces that are dominated by the local correlation among nearby scleral locations that is also captured by the tensor wavelet bases.

We also computed the induced interfunctional serial correlation across MPS curves for different IOP for the same eye using the formulas contained in Section \ref{s:growth}.  The supplement contains a movie file {\tt Intra\_IOP\_corr.mp4} that plots the variance, var$\{Y_{ijp}(\bt)|IOP=p,\bt\}$, as a function of IOP and scleral location and the serial correlation across IOP, corr$\{Y_{ijp}(\bt),Y_{ijp'}(\bt)\}$, as a function of IOP as it varies around the scleral surface $\bt$.  Note the form of the serial correlation induced by the hyperbolic model, and how it is able to vary yet borrow strength across scleral locations.  Panel (d) of Figure \ref{f:combo_plots} portrays this serial correlation at a single scleral location.

{\bf Inference on functionals of the parameters:}  While Figure \ref{f:MPS_age_pressure} and the accompanying movie file provide a thorough summary of the age effect on MPS estimated by our model, for interpretability it may be useful to aggregate results over IOP and/or scleral locations.  One major advantage of our fully Bayesian approach is that we are able to compute posterior samples and obtain estimates and inference for any functional of the model parameters.

First, to aggregate information across all IOP, we considered the area under the MPS vs. IOP curve, defined as follows:
\begin{align}
    \nonumber\mbox{AUC}(X_{\mbox{age}},\bt)=f(X_{\mbox{age}},\bt)+\int_{7}^{45} \{B_1(\bt)G_1(p) +B_2(\bt)G_2(p)\} dp.
\end{align}
This integral summarizes the total MPS behavior over the range of IOP in the study, which covers the practical range of IOP values in this context. The integral is estimated numerically for each MCMC sample to yield posterior inference on AUC.  We plotted the posterior mean AUC vs. age curve and corresponding posterior pointwise and joint credible bands for each scleral location, and assembled into a digital movie file {\tt AUCvAGE-wave.mp4} in the supplementary materials, and for illustration panel (b) of Figure \ref{f:combo_plots} contains this plot for a single scleral location.  Although our model was not fit to the AUC data, but rather the raw data for each IOP, note how the model provides a nonparametric smooth fit of AUC vs. age for each scleral location, and again these fits borrow strength across scleral locations.   From these results, we see that that aggregating over IOP, the MPS tends to decrease with age at most scleral locations, especially near the ONH.  With MPS decreasing with age, the eye seemingly becomes less elastic and less able to absorb IOP, potentially exposing the ONH to IOP-induced damage over time.

Given the nonlinearity of the age fit, it may be instructive to {directly} look at the rate of decline of MPS over age.  We can do this by computing inference on the derivative of the AUC curve with respect to $X_{\mbox{age}}$.  Given the lack of an IOP $\times$ age interaction in our final fitted model, it follows that $\partial\mbox{AUC}(X_{\mbox{age}},\bt)/\partial{X_{\mbox{age}}}=\partial f(X_{\mbox{age}},\bt)/\partial{X_{\mbox{age}}}$. Using the fact that $f(X_{\mbox{age}},\bt)=\beta_{0}(\bt)+X_{\mbox{age}}\beta_{3}(\bt)+Z_{\bsB}(X_{\mbox{age}})\bu_{\calS}(\bt)$ as described in Section \ref{s:nonpara}, with $Z_{\bsB}(X_{\mbox{age}}) = \{\mathcal{B}_1(X_{\mbox{age}}),\ldots$
$,\mathcal{B}_{M+4}(X_{\mbox{age}})\}'Z_{\Omega}\mbox{diag}(d_1^{-1/2},\ldots,d_{M+2}^{-1/2})$ , it can be easily seen that
\begin{align}
\frac{\partial \mbox{AUC}(X_{\mbox{age}},\bt)}{\partial X_{\mbox{age}}}=\beta_{3}(\bt)+\frac{\partial{Z_{\bsB}(X_{\mbox{age}})}}{\partial{X_{\mbox{age}}}}\bu_{\calS}(\bt). \label{eq:deriv}
\end{align}
Our R scripts contain calculations of $\frac{\partial{Z_{\bsB}(X_{\mbox{age}})}}{\partial{X_{\mbox{age}}}}$, which come from \citet{Wand08}.  By applying (\ref{eq:deriv}) to the posterior samples of $\beta_3(\bt)$ and $\bu_{\calS}(\bt)$, we could obtain posterior samples of this derviative for each scleral location, although we do not present those results here.  Since the nonparametric fits were done placing a smoothness penalty on the age effects themselves, the derivatives may appear slightly undersmoothed.  If one had primary interest in estimating these derivatives smoothly, they could do so by simply choosing higher order penalties in the spline fits.  A third order penalty would produce smoothness in the first derivative.

{\bf Aggregated summaries over functional regions:}  Although our model fits the entire data set over all scleral locations, for ease of interpretation {researchers} at times would like to look at estimates and inference for aggregated scleral regions.  Given the hypothesis that scleral strain is most important near the ONH, we averaged results over all circumferential regions to obtain results as a continuous function of meridional distance from the optic nerve head.  Figure \ref{fg4}  depicts the posterior mean AUC as a function of age and distance from the ONH, aggregating over circumferential regions.  We can see how MPS is higher near the ONH and decreases moving away from the ONH, potentially providing a protective effect to the ONH.  Younger individuals have high MPS levels at scleral locations extending well out from the ONH, while for middle age individuals the regions of high MPS does not extend out far from the ONH, and for older individuals the MPS is quite low even close to the ONH.  This coincides with the increased glaucoma risk in older individuals.

\begin{figure}
\centerline{\includegraphics[scale=0.5]{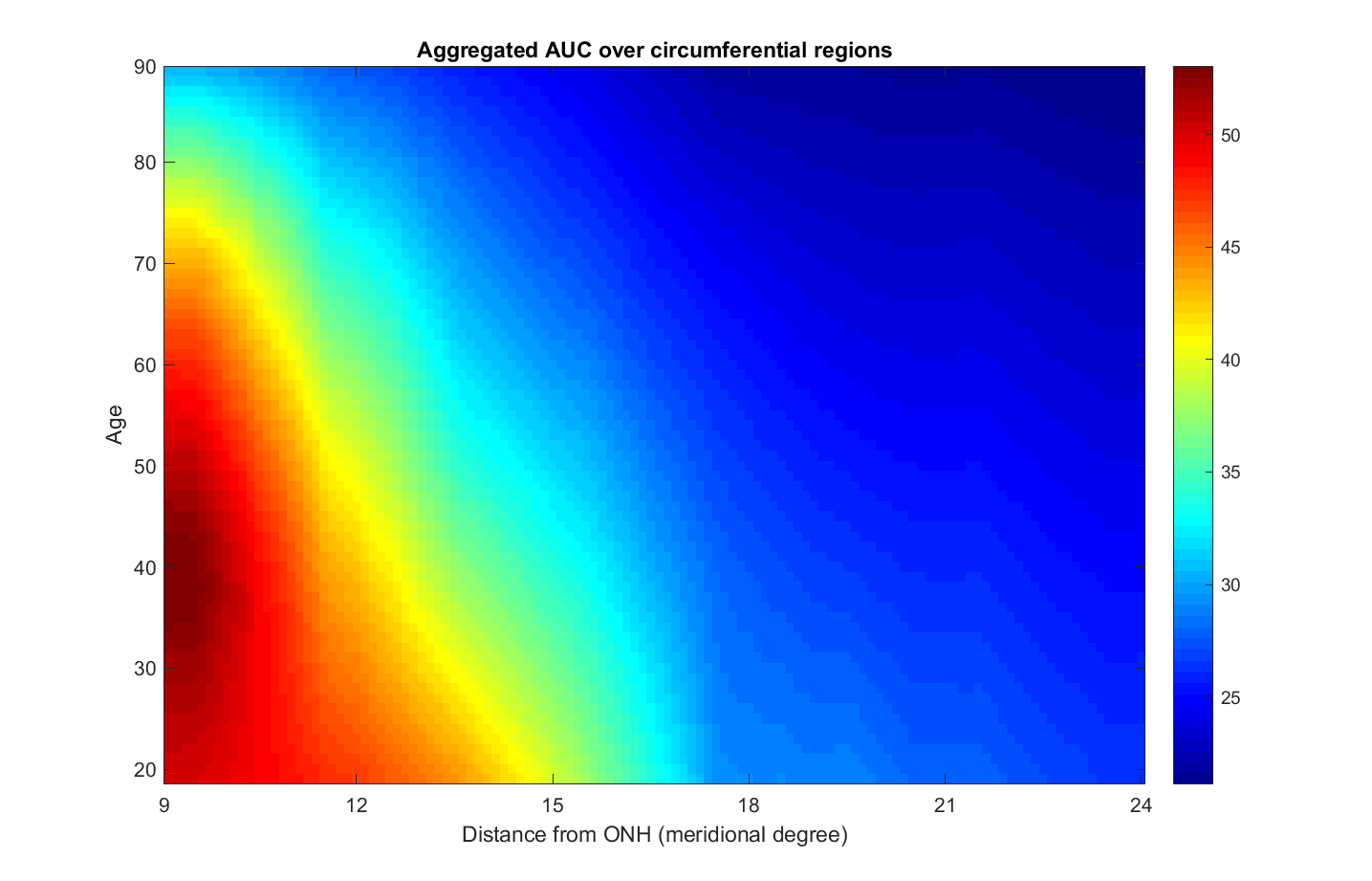}}
\caption{\label{fig:combo_plots} \footnotesize \textbf{Posterior mean AUC (of MPS) as a function of age and distance from ONH.}  These results are obtained by aggregating posterior samples over all circumferential regions.}\label{fg4}
\end{figure}

We also summarized the results aggregating over the innermost region closest to the optic nerve head, which is called peripapillary(PP) region and also within the adjacent region, called the mid-peripheral(MP) region. Here we present aggregated AUC results in both PP and MP regions. Figure \ref{fg5} depicts the age effects aggregating AUC over the PP and MP regions, with the top two plots containing the posterior mean fits and the bottom containing the derivatives with respect to age.  The blue line contains the posterior mean fit. The dotted and solid red lines indicating 95\% pointwise and joint credible intervals, respectively.  From this we can clearly see that the MPS is systematically higher in the PP region closest to the ONH than in the MP region further away, potentially conferring a protective effect.  The MPS decreases with age in both regions, but the decrease is substantially steeper for the all-important PP region close to the ONH.  This effect is nonlinear, with the rate of decrease accelerating throughout middle age (40-60 years old), an age at which glaucoma risk increases substantially.

\begin{figure}
\centerline{\includegraphics[scale=0.6]{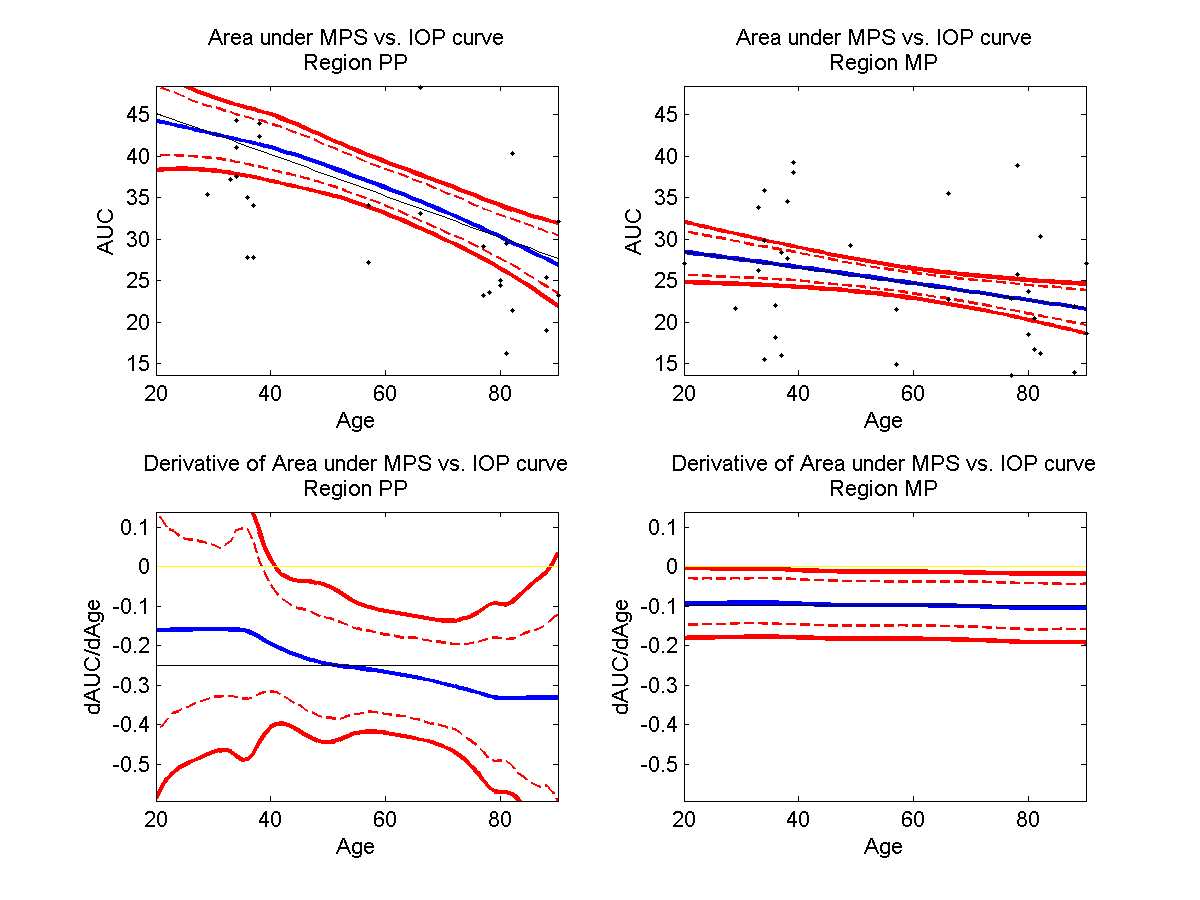}}
\caption{\footnotesize \textbf{Aggregated AUC summaries:} AUC aggregated over peripapillary (PP) region {adjacent to the ONH} and {the} mid-peripheral(MP) region just beyond PP. (a) and (b) show AUC summaries over the two regions. (c) and (d) show derivatives of aggregated AUC {summaries} presented in (a) and (b).}\label{fg5}
\end{figure}

These analyses confirm the hypothesis that MPS is higher in scleral regions closer to the ONH, decreases with age, and this decrease with age is more pronounced near the ONH.  This agrees with the notion that biomechanical changes in the sclera may contribute to increased glaucoma risk.

\textbf{Sensitivity Analyses:} Section 7 of the Supplement and Supplementary Figures 2-25 present extensive results for the alternative model with the left vs. eye effect, with the alternative values for the prior shrinkage hyperparameters $\{\tau_{aj},\pi_{aj}\}$, and using the wavelet-regularized principal components as the projected basis.  Substantive results do not change, so we see our conclusions are not driven by these choices.

\section{Discussion}

In this paper, we demonstrated how to adapt the BayesFMM modeling framework to account for serial interfunctional correlation and smooth nonparametric covariate functional effects and applied it to an innovative glaucoma study investigating MPS of scleral strain tensors. We found that MPS is maximized near {the} ONH. We also found that MPS decreases with age, especially in regions closest to the ONH, {and the decrease in MPS accelerates throughout middle-age. This could contribute to the increased glaucoma risk seen in elderly.} The age effect on MPS tends to be non-linear near the ONH, especially towards the inferior and nasal sides, while the other scleral regions show the age effect close to linear. {Focal glaucoma damage is most often observed in the inferior quadrant of the ONH.}

While motivated by the glaucoma data application, the BayesFMM framework presented here is extremely general, with the ability to be used with any near-lossless basis tranform and applicable {to} many types of complex, high dimensional functional data of modern interest including wearable computing data, genome-wide data, {proteomics} data, geospatial time series data, neuroimaging data, and many others. Using this framework, one can not only accommodate nonparametric functional fixed effects, but also model serially correlated functions through functional growth curve effects.  Our model allows the nonparametric fits, smoothness, and interfunctional correlations to potentially vary over the functional domain, which is necessary for good fit to these data and likely many other complex functional data sets.   In addition, we introduced a model selection heuristic that can be used to select among fixed and random effects and decide whether they should be linear, parametric, or nonparametric, {and whether the fits or smoothness should be constant or vary around the functional domain} before running the MCMC.

On our github page (\url{https://github.com/MorrisStatLab/SemiparametricFMM}), we share the Matlab files required to fit these models, including scripts to apply the model selection heuristic, links to automated software to perform the MCMC, and scripts to compute posterior samples and posterior inference for the nonparametric functional effects and all summary plots included in this paper.  The model is set up using $lmer$ style model statements \citep{lmer}.  The software can be used to fit models with any fixed or random effect function covariates, with different basis functions, and for functions on domains of any dimension or measure given suitable basis functions for the corresponding space.  The method is efficient enough to feasibly apply to very large data sets like our glaucoma data set here with over 4.5 million observations.  While it took a relatively long time to run the MCMC using a single core, we do not think this is inordinate considering the cost and time to collect these data and the extensive inferential summaries provided by the model fits.   This run time was orders of magnitude less than a conceptually simpler approach of applying {\tt lme} to each scleral location, applying a 2d smooth, and using a bootstrap for inference, which by our calculations would take 10 weeks to compute using 1000 bootstrap samples on the same computer we used for the BayesFMM model. If further speed is desired for our method, the model fitting is highly parallelizable and the MCMC code has cluster computing capabilities so can be sped up using GPU or cluster computing resources.  While to code we share only utilizes a single core, future updates of the software will enable distributed computing for faster calculations for big data sets.
We are in the process of extending the package to fit these models and including many other features of the BayesFMM framework in other publications but not included in this paper.  We anticipate this package, which will be available in Matlab and R, will greatly enhance the usability of the method, and expect this package to be completed and freely available in the near future.

One significant benefit of our fully Bayesian approach to fitting these semiparametric functional mixed models is that we can produce inference on any parameters in the model, or any functional or aggregation of these parameters, and this inference integrates over the various sources of interfunctional variability in the model and over the uncertainty in estimating the covariance parameters and level of nonlinearity of the spline fits.   This allows us to perform a flexible, thorough analysis in the entire functional domain, yet produce inference and results for aggregated summaries that may be more interpretable to investigators.  This raises the obvious question, ``What is the advantage of fitting the entire functional model?  Why not just compute the aggregated summaries and model those using standard tools?"  The answer to this question is multi-faceted.

First, if one only looked at specific aggregated summaries, they may miss insights that could have been gleaned from their data but were not captured by these summaries.  For example, if only looking at the PP and MP regions, one may miss out on different MPS behavior in the inferior and nasal regions of the sclera, which may be important.  By modeling the scleral function in its entirety, we are able to examine the entire domain to ensure we are not missing out on any insights, and then we can still produce inference for any desired aggregated summaries for ease of interpretation, as well.  This concept is also relevant in other application areas where summaries are used because of the complexity and high dimensionality of the raw functional data.  This approach of flexible modeling followed by inferenct on summary measures can capture the best of both worlds, analyzing the entire function yet producing inference for summary measures interpretable for the scientific subject area.  

Second, any aggregation draws arbitrary boundaries in the function space.  MPS for scleral locations at the boundary of the PP and MP regions are highly correlated with each other, {and yet} PP and MP extracted summaries arbitrarily separate them from each other.  The flexible functional modeling framework introduced in this paper models the entire functional space, yet captures and accounts for intrafunctional correlation through the chosen basis functions.  This allows a smooth borrowing of strength from nearby locations, and if location-scale bases like wavelets are used, then this borrowing of strength can be adaptive, able to accommodate spatially heterogenous functions for which some functional regions are more correlated than others.  In principle, accounting for this intrafunctional correlation leads to greater efficiency, as has been shown in various contexts.

While flexible, the BayesFMM framework presented here has some limitations and drawbacks, including the need to choose a common basis to use for transformation at all levels of the model, the independence in the basis space assumption that can limit certain types of intrafunctional covariance structure depending on the choice of basis, and the computational intensity from having to run a full MCMC to obtain estimates and inference for model quantities.  It is designed for representing functional data sampled on a common fine grid, so is not suitable for sparsely sampled functional data or functional data for which the individual functions are sampled on wildly different sampling grids, a setting in which the \cite{Scheipl15} and \cite{Greven2017} framework is well-suited.  The model selection heuristic is \textit{ad hoc}, and should not be used to select over a large number of variables.  

In spite of these limitations and drawbacks, the BayesFMM modeling framework is very general, and this paper can serve as a template for how to utilize this modeling framework to model complex functions with various types of interfunctional correlation structures.  It has the potential to impact many areas of science yielding complex functional data, and the analysis presented here illustrates a rigorous, thorough workflow to analyze the entire data set, extract many types of information from them, yet provide interpretable graphical summaries and inferential results desired by investigators.

\newpage
\begin{center}
{\large\bf SUPPLEMENTARY MATERIALS}
\end{center}

\begin{description}

\item[Supplement.pdf:] This document is organized as follows. Section 1 describes details of MCMC update steps. Section 2 provides derivation of $\mbox{DF}(\bt)$ for Model (\ref{e:finalmodel4}).  Sections 3-4 include details of model selection results. In Section 5, we assess whether the smoothing parameter for the nonparametric age effect should be constant or vary around the scleral surface. In Section 6, we provide description of overall procedure to fit our model and obtain inferential results with a simulation dataset. Section 7 contains sensitivity analyses to various modeling assumptions, including results for other basis (wavelet-regularized PC), other model (model that also includes left vs. right eye effect), and other regularization hyperparameters (choice of values with no additional shrinkage provided by prior).  Section 8 describes some supplementary files presenting additional results discussed in the paper, and Section 9 describes the simulated pseudo-data and demonstrates that it appears to capture the features of the real MPS scleral strain data reasonably well.

\item[RawMPScurves.zip:] It includes plots of raw MPS curves, results after tensor wavelet compression, and results after robust filtering to remove spiky artifacts.

\item[movies.zip:] Includes various {\tt .mp4} movie files illustrating various detailed results from the paper, including the following:

\item[MPSvsAge-wave.mp4:] Movie of MPS vs. age for each IOP based on the tensor wavelet basis function.

\item[Combo\_plots.mp4:] Movie of key summary results based on the tensor wavelet basis functions.

\item[Intrafunctional\_correlations.mp4] Movie showing intrafunctional correlations induced by tensor wavelet basis functions.

\item[Intra\_IOP\_corr.mp4] Movie showing interfunctional variance and serial correlation across IOP from same eye based on the tensor wavelet basis functions.
\ 
\item[MPSvsAge-pc.mp4:] Movie of MPS vs. age for each IOP based on the principal component basis functions.

\item[AUCvsAge-pc.mp4:] Movie of AUC vs. age based on the principal component basis functions.

\item[Intrafunctional\_correlations-pc.mp4] Movie showing intrafunctional correlations induced by the principal component basis functions.


\item[MPSvsAge-eye.mp4:] Movie of MPS vs. age for each IOP based on tensor wavelet basis functions and model including left vs. right eye effect.

\item[AUCvsAge-eye.mp4:] Movie of AUC vs. age based on tensor wavelet basis functions and model including left vs. right eye effect.

\item[Intrafunctional\_correlations-eye.mp4] Movie showing intrafunctional correlations induced by the tensor wavelet basis functions and model including left vs. right eye effect


\item[MPSvsAge-nosmooth.mp4:] Movie of MPS vs. age for each IOP based on tensor wavelet basis functions and model with no smoothing ($\pi_{\cdot}=1, \tau_{\cdot}=10^6$).

\item[AUCvsAge-nosmooth.mp4:] Movie of AUC vs. age based on tensor wavelet basis functions and model with no smoothing ($\pi_{\cdot}=1, \tau_{\cdot}=10^6$).

\item[Intrafunctional\_correlations-nosmooth.mp4] Movie showing intrafunctional correlations induced by the tensor wavelet basis functions and model with no smoothing ($\pi_{\cdot}=1, \tau_{\cdot}=10^6$).


\item[EYE\_toolbox.zip] Contains all of the files necessary to run the methods presented in the paper, including the raw glaucoma data, pseudo data and full scripts to run the analyses and produce the plots contained in the paper.  This includes {\tt wfmm\_install.pdf} that contains step-by-step instructions on how to install the R package \textit{wfmm} and associcated executable, and {\tt Analysis\_of\_Pseudo\_Data.pdf} that contains detailed step-by-step instructions for running a complete analysis on pseudo data generated to mimic the real data in the application in Matlab, including the basis transform, model selection heuristic, MCMC in basis space, projection of posterior samples back to data space, MCMC convergence diagnostics, and producing all inferential summaries and plots contained in this paper that present results and illustrate properties of the model, with run time estimates for each step.  We also include {\tt Producing Plots for Main Data Analysis in Paper.pdf} that gives instructions for running scripts to reproduce the figures for the real data analysis for the main model used for the MPS scleral strain data analysis contained in the paper.  This toolbox and data are available on our github (\url{https://github.com/MorrisStatLab/SemiparametricFMM}).


\end{description}





\section*{Supplementary Materials}

\section*{Details of the MCMC Algorithm}
We utilize a Markov chain Monte Carlo algorithm to draw posterior samples for the parameters in our model (12). First, we sample the fixed effects and variance components alternatively using the conditional posterior distribution which is marginalized over the random effects. Then we later sample the random effect $u_{m,k}^*$ which is needed for estimating nonparametric age effect. It is sampled using the full conditional distribution which is still marginalized over the other random effects. This sampling algorithm improves the mixing properties of the MCMC chains and speeds up the MCMCM algorithm as illustrated in \citet{Morris06}. The following are the details of the MCMC.

Note that there are a total of 306 (9 IOP levels $\times$ 34 eyes) observed MPS functions. For easy presentation of the MCMC, we rewrite our model (12) in a matrix formula including all observations:
\begin{align}
    \nonumber\bY^*_k = \bX\bb^*_k + \bZ_{\bsB}\bu^*_k +\sum_{h=0}^2\bZ_h\bu^*_{h,k}+\bE^*_k,
\end{align}
where $\bY^*_k$ is the $306\times 1$ matrix of $Y^*_{ijp,k}$, $\bX$ is the $306\times 4$ design matrix of fixed effects, $\bb^*_k=(\beta^*_{0,k},\beta^*_{1,k},B^*_{1,k},B^*_{2,k})'$, $\bZ_{\bsB}$ is the $306\times (M+2)$ design matrix for $\bu^*_k=(u_{1,k}^*,\ldots,u_{(M+2),k}^*)'$, $\bZ_0$ is the $306\times 38$ design matrix for $\bu^*_{0,k}=(U^*_{1,1,k},\ldots,U^*_{19,1,k},U^*_{1,2,k},\ldots,U^*_{19,2,k})'$, $\bZ_1$ is the $306\times 38$ design matrix for the random effect for $\mbox{IOP}_1$, $\bu^*_{1,k}=(U^*_{1,1,1,k},\ldots,U^*_{19,1,1,k},U^*_{1,2,1,k},\ldots,U^*_{19,2,1,k})'$, $\bZ_2$ is the $306\times 38$ design matrix for the random effect for $\mbox{IOP}_2$, $\bu^*_{2,k}=(U^*_{1,1,2,k},\ldots,U^*_{19,1,2,k},U^*_{1,2,2,k},\ldots,U^*_{19,2,2,k})'$, and $\bE^*_k$ is the $306\times 1$ matrix of $
E^*_{ijp,k}$. For simplicity, let $\bb^*_k=(b^*_{1,k},\ldots,b^*_{4,k})'$. Recall that sparsity priors are placed on the fixed effects:
\begin{equation}
\nonumber b^*_{a,k} = \gamma^*_{a,k}N(0,\tau_{a,k})+(1-\gamma^*_{a,k})I_0,\quad \gamma^*_{a,k}=\mbox{Bernoulli}(\pi_{a,k}), \quad a=1,\ldots,4,
\end{equation}
where $I_0$ is a point mass at zero. Assumptions made on the random terms are that $\bu^*_{k}\sim N(\bzero,q^*_kI_{M+2})$, $\bu^*_{h,k}\sim N(\bzero,q^*_{h,k}I_{38})$ $(h=0,1,2)$, and $\bE_k^*\sim N(\bzero,s^*_{k}I_{306})$ where $I_d$ is a $d\times d$ identity matrix. Let $\Omega_k^*=(q^*_k,q^*_{0,k},q^*_{1,k},q^*_{2,k},s^*_k)'$

\begin{enumerate}
\item[Step 1.] For each $a$, draw a sample of $b^*_{a,k}$ from $f(b^*_{a,k}|\bY^*_k,\bb^*_{-a,k},\Omega_k^*)$, where $\bb^*_{-a,k}$ is the set of all fixed effects except $b^*_{a,k}$. This distribution is a mixture of a point mass at zero and a Gaussian distribution with the Gaussian proportion $\alpha_{a,k}$:
    \begin{align}
    \nonumber &\gamma_{a,k} \sim \mbox{Bernoulli}(\alpha_{a,k}),\\
    \nonumber &b^*_{a,k} = \gamma_{a,k}N(\mu_{a,k},v_{a,k})+(1-\gamma_{a,k})I_0,
    \end{align}
    where
    \begin{align}
    \nonumber &\mu_{a,k}=\hat{b}_{a,k,MLE}^*(1+V_{a,k}/\tau_{a,k})^{-1},\\
    \nonumber &v_{a,k} = V_{a,k}(1+V_{a,k}/\tau_{a,k})^{-1},\\
    \nonumber &\alpha_{a,k} = \frac{\pi_{a,k}}{1-\pi_{a,k}}*(1+V_{a,k}/\tau_{a,k})^{-1/2}\exp\{\frac{1}{2}\zeta_{a,k}^2(1+V_{ak}/\tau_{a,k})^{-1})\}, \mbox{and} \\
    \nonumber &\zeta_{a,k} = \hat{b}_{a,k,MLE}^*/\sqrt{V_{a,k}}.
    \end{align}
    Here $\hat{b}_{a,k,MLE}^*$ is the maximum likelihood estimate of $b^*_{a,k}$ which is
    \begin{align}
    \nonumber &\hat{b}_{a,k,MLE}^*=(\bX_a'\Sigma_k^{-1}\bX_a)^{-1}\bX_a'\Sigma_k^{-1}(\bY^*_k-\bX_{-a}\bb^*_{-a,k}), \\
    \nonumber &V_{a,k} = (\bX_a'\Sigma_k^{-1}\bX_a)^{-1}, \mbox{and}\\
    \nonumber &\Sigma_k = q^*_k\bZ_{\bsB}\bZ_{\bsB}' + \sum_{h=0}^2 q^*_{h,k}\bZ_{h}\bZ_{h}' + s^*_kI_{306}
    \end{align}
    where $\bX_a$ is the $a$-th column of $\bX$ and $\bX_{-a}$ is the $\bX$ with the $a$-th column removed.
\item[Step 2.] Draw a sample of $\Omega_k^*$ by using a random-walk Metropolis-Hastings step from the full conditional distribution
    \begin{align}
    \nonumber f(\Omega_k^*|\bY^*_k,\bb^*_{k})\propto |\Sigma_{k}|^{-1/2}\exp\{\frac{1}{2}(\bY^*_k-\bX\bb^*_k)'\Sigma_k^{-1}(\bY^*_k-\bX\bb^*_k)\}f(\Omega_k).
    \end{align}
    We use an independent zero-truncated Gaussian distribution as the proposal for each parameter. The proposal variances are automatically estimated from the data by using the maximum likelihood estimates \citep{Wolfinger94}.
\item[Step 3.] Sample the random effect $\bu_{k}^*$ related to the nonparametric age effect from its fully conditional distribution marginalized over the other random effects by integrating them out. The distribution can be easily seen to be multivariate Gaussian
    \begin{align}
    \nonumber f(\bu_{k}^*|\bY^*_k,\bb^*_{k},\Omega_{k}^*) \sim N(\boldsymbol{m}_k,V_{k}),
    \end{align}
    where
    \begin{align}
    \nonumber &\boldsymbol{m}_k=V_k\bZ_{\bsB}'(\sum_{h=0}^2q^*_{h,k}\bZ_h\bZ_h'+s^*_kI_{306})^{-1}(\bY_k-\bX\bb^*_k),\\
    \nonumber &V_{k}=[\Psi^{-1}_k+(q^*_kI_{306})^{-1}]^{-1} , \mbox{and}\\
    \nonumber &\Psi^{-1}_k = \bZ_{\bsB}'(\sum_{h=0}^2q^*_{h,k}\bZ_h\bZ_h'+s^*_kI_{306})^{-1}\bZ_{\bsB}.
    \end{align}
\end{enumerate}
Note that the MCMC algorithm can be performed for each $k$ separately. Thus, the MCMC fitting can be done using parallel processing using multiple cores or clusters.

\section*{Derivation of $\mbox{DF}(\bt)$ for Model (9)}
Here we derive the form of $\mbox{DF}(\bt)$ for Model (9) that is introduced in Section 2.4 of the main paper. Recall that Model (9) is given as
\begin{align}
\nonumber Y_{ijp}(\bt) =& f(X_{\mbox{age},i}, \bt)+B_{1}(\bt)\mbox{IOP}_1+B_{2}(\bt)\mbox{IOP}_2+\\
\label{e:finalmodel} &U_{ij}(\bt)+U_{ij1}(\bt)\mbox{IOP}_1+U_{ij2}(\bt)\mbox{IOP}_2+E_{ijp}(\bt),
\end{align}
where $i=1,\ldots,n$; $j=1, 2$; $p=7,10,15,\ldots,45$; and $\mbox{IOP}_1$ and $\mbox{IOP}_2$ together represent the orthogonalized hyperbola terms. Under the assumption of (5) in Section 2.3 of the main paper, the model \eqref{e:finalmodel} for a given $\bt$ can be rewritten in a matrix form,
\begin{align}
\label{e:NPM2} \boldsymbol{y}({\bt})&=\bsB\bnu(\bt) +\bX_{\mbox{IOP}}\bB(\bt)+\bZ_{\mbox{IOP}}\bU(\bt)+\boldsymbol{\epsilon}{(\bt)},
\end{align}
where $\bsB$ is the B-spline design matrix,  $\bX_{\mbox{IOP}}$ is the design matrix for the fixed hyperbolic IOP effect, and $\bZ_{\mbox{IOP}}$ is the design matrix for all random effects. Let $\tilde{\boldsymbol{y}}({\bt})=\boldsymbol{y}({\bt})-\bX_{\mbox{IOP}}\bB(\bt)$ and $\tilde{\boldsymbol{\epsilon}}{(\bt)}=\bZ_{\mbox{IOP}}\bU(\bt)+\boldsymbol{\epsilon}{(\bt)}$. Then, the model \eqref{e:NPM2} becomes
\begin{align}
\label{e:NPM3}  \tilde{\boldsymbol{y}}({\bt}) = \bsB\bnu(\bt) + \tilde{\boldsymbol{\epsilon}}{(\bt)}.
\end{align}
Let $\bW$ be the inverse of covariance of $\tilde{\boldsymbol{\epsilon}}{(\bt)}$, $\bW=\mbox{Cov}( \tilde{\boldsymbol{\epsilon}}{(\bt)})^{-1}$. By multiplying $\bW^{1/2}$ to both sides of \eqref{e:NPM3}, one can have
\begin{align}
\label{e:NPM4}  \bW^{1/2}\tilde{\boldsymbol{y}}({\bt}) = \bW^{1/2}\bsB\bnu(\bt) + \bW^{1/2}\tilde{\boldsymbol{\epsilon}}{(\bt)}.
\end{align}
A smoothing spline for this model can be obtained by the following optimization problem,
\begin{align}
\nonumber \mbox{min} \{{\parallel \bW^{1/2}(\tilde{\boldsymbol{y}}({\bt})-\bsB\bnu(\bt)) \parallel}^2 +\lambda_{\bt}\bnu(\bt)'\Omega\bnu(\bt)\}.
\end{align} One can show that the resulting spline estimator is $\hat{\bnu}(\bt)=(\bsB'\bW\bsB+\lambda_{\bt}\Omega)^{-1}\bsB'\bW\tilde{\boldsymbol{y}}({\bt})$. Following the same arguments in Section 2.3, one can easily see that this penalized spline estimator is equivalent to the posterior mean of $\bnu(\bt)$ given $\tilde{\boldsymbol{y}}({\bt})$ with prior specification $g(\bnu(\bt))\propto \exp(-\frac{1}{2\sigma_{\bt}^2}\bnu(\bt)'\Omega\bnu(\bt))$ where $\lambda_{\bt}={1}/{\sigma_\bt^2}$. Again, following the same arguments in Section 2.3, one can show that the model \eqref{e:NPM4} can be formulated as the mixed model (10) in Section 2.3 of the main paper. As the penalized spline fit is given as $\bsB\hat{\bnu}(\bt)=\bsB(\bsB'\bW\bsB+\lambda_{\bt}\Omega)^{-1}\bsB'\bW\tilde{\boldsymbol{y}}({\bt})$, the $\mbox{DF}(\bt)$ for Model (9) is given as
\begin{align}
\nonumber \mbox{DF}(\bt)= \mbox{trace}\{\bsB(\bsB'\bW\bsB+\lambda_\bt\Omega)^{-1}\bsB'\bW\}.
\end{align}

\newpage
\section*{Model Selection Results for Glaucoma Data}
We performed the model selection strategy described in Section 4 to find the final model to be fitted via MCMC. We first selected fixed effects to be included in the final model and their forms. Three different fixed effects were considered: age, IOP, and eye (left vs. right). For the form of the age effect, we considered two possibilities: linear or nonparametric. For the form of the IOP effect, we considered three different possibilities: linear, hyperbola, or nonparametric. Models without the eye effect were also compared. As a result, we compared 12 different models for the fixed effect selection. The $P^h$ scores are given in Table \ref{tb:fixed}. The model with the nonparametric age effect, the hyperbolic IOP effect, and no eye effect showed the highest $P^h$.

\begin{table}[h]
\begin{center}
\begin{tabular}{|l|l|c|c|c|c| }\hline
\multicolumn{2}{|c|}{Form of fixed effect} & \multicolumn{2}{c|}{$P^h, aBIC$} & \multicolumn{2}{c|}{$P^h, aAIC$}\\ \cline{1-2}\cline{3-4}\cline{5-6}
Age & IOP & without Eye effect & with Eye effect & without Eye effect & with Eye effect\\\hline
Nonparametric & Nonparametric & 0.005 & 0.020 & 0.003 & 0.013 \\
Nonparametric & Hyperbolic       & 0.543 & 0.377 & 0.022 & 0.958\\
Nonparametric & Linear              & 0.000 & 0.000 & 0.000 & 0.000\\
Linear            & Nonparametric & 0.013 & 0.043 & 0.000 & 0.003\\
Linear            & Hyperbolic       & 0.000 & 0.000 & 0,000 & 0,000\\
Linear            & Linear              & 0.000 & 0.000 & 0.000 & 0.000\\ \hline
\end{tabular}
\caption{ \footnotesize Comparison among fixed effects. }\label{tb:fixed}
\end{center}
\end{table}

Once we selected the fixed effects, we assessed whether the interaction term between age and IOP is needed. As shown in Table \ref{tb:interaction}, our model selection criteria made it clear the interaction was not necessary.

\begin{table}[h]
\begin{center}
\begin{tabular}{|l|c|}\hline
Interaction term in model & $P^h$ \\ \hline
No interaction & 1 \\
$f(age)*\mbox{IOP}_1$ & 0 \\
$f(age)*\mbox{IOP}_2$ & 0 \\
$f(age)*\mbox{IOP}_1$+$f(age)*\mbox{IOP}_2$ & 0 \\ \hline
\end{tabular}
\caption{ \footnotesize Comparison among interaction terms, results same for $aBIC$ and $aAIC$. }\label{tb:interaction}
\end{center}
\end{table}

Finally, with the selected fixed terms, we compared several models by varying random effects. Two different levels of random effects were considered: the subject-level random effect and the longitudinal eye-level random effect. For the form of the eye-level random effect in terms of IOP as illustrated in Section 2.2, we considered three different forms: constant, linear, or hyperbola. The results are given in Table \ref{tb:random}. Our model selection strategy selected the eye-level random effect with the form of hyperbola in terms of IOP.

\begin{table}[h]
\begin{center}
\begin{tabular}{|l|c|c|c|c|}\hline
\multirow{2}{*}{Form of IOP random effect} & \multicolumn{2}{|c|}{$P^h aBIC$} & \multicolumn{2}{c|}{$P^h aAIC$} \\ \cline{2-3} \cline{4-5}
 & without subject RE & with subject RE & without subject RE & with subject RE\\\hline
No IOP random effect               & 0.000 & 0.000 & 0.000 & 0.000\\
constant 			                & 0.000 & 0.000 & 0.000 & 0.000\\
Linear                                        & 0.005 & 0.000 & 0.002 & 0.000\\
Hyperbolic without intercept      & 0.000 & 0.000 & 0.000 & 0.000\\
Hyperbolic with intercept           & 0.991 & 0.003 & 0.798 & 0.200\\ \hline
\end{tabular}
\caption{ \footnotesize Comparison among random effects. }\label{tb:random}
\end{center}
\end{table}

\section*{Model Selection Results for Simulation Study}

As described in Section 4, we conducted a simulation study to investigate the performance of the model selection approach. Table \ref{tb:sim} provides average $P^h$ over 100 replications for each model in each scenario. For each simulation setting, the $P^h$ was maximized by the correct model in all 100/100 replications.

\begin{table}[h]
\begin{center}
\begin{tabular}{|l|cccc|}\hline
\multirow{2}{*}{True Model} & \multicolumn{4}{c}{$P^h aBIC$} \\ \cline{2-5}
 & Model 1 & Model 2 & Model 3 & Model 4\\\hline
Model 1 & 0.984 & 0.015 & 0.000 &  0.000\\
Model 2 & 0.273 & 0.723 & 0.003 &  0.002\\
Model 3 & 0.093 & 0.292 & 0.602 &  0.013\\
Model 4 & 0.000 & 0.000 & 0.000 &  1.000\\ \hline
\end{tabular}
\caption{ \footnotesize Average $P^h$ in Simulation Study. }\label{tb:sim}
\end{center}
\end{table}

\section*{Assessing varying smoothness of age effect around sclerum}

As described in Section 2.3, the smoothness penalty parameter $\lambda_\bt$ is equivalent to ${\sigma_{\epsilon,\bt}^2}/{\sigma_\bt^2}$ where ${\sigma_{\epsilon,\bt}^2}$ and $\sigma_\bt^2$ are variances for the random effect and residual error respectively. A key strength of our framework is that it allows the smoothness penalty parameter to vary across functional domains as ${\sigma_{\epsilon,\bt}^2}/{\sigma_\bt^2}$ depends on $\bt$. If one wants to use the smoothness penalty parameter which is common across $\bt$, it is basically equivalent to assume that ${\sigma_{\epsilon,\bt}^2}=\lambda\sigma_\bt^2$ in the estimation procedure. Our model selection strategy can be also used to assess whether the data favors this varying smoothness penalty parameter against a common smoothness parameter. In particular, consider the model (8) introduced in Section 2.3:
\begin{align}
\label{e:Bspline3} \boldsymbol{y}(\bt)= 1_n\beta_{0}(\bt)+X_{\mbox{age}}\beta_{1}(\bt)+Z_{\bsB}\boldsymbol{u}(\bt)+\boldsymbol{\epsilon}(\bt),
\end{align}
where $\boldsymbol{u}(\bt) \sim N(0,\sigma_\bt^2I)$ and $\boldsymbol{\epsilon}(\bt) \sim N(0,\sigma_{\epsilon,\bt}^2I)$. To compute the BIC of the model with a common smoothness penalty $\lambda$, we can rewrite the model \eqref{e:Bspline3} as
\begin{align}
\label{e:Bspline4} \boldsymbol{y}(\bt)= 1_n\beta_{0}(\bt)+X_{\mbox{age}}\beta_{1}(\bt)+\tilde{\boldsymbol{\epsilon}}(\bt),
\end{align}
where $\tilde{\boldsymbol{\epsilon}}(\bt)\sim N(0,\tilde{V})$ and $\tilde{V}=\sigma_\bt^2(Z_{\bsB}Z_{\bsB}'+\lambda I)$ by using the fact that ${\sigma_{\epsilon,\bt}^2}=\lambda\sigma_\bt^2$. This model can be easily fitted after multiplying by $(Z_{\bsB}Z_{\bsB}'+\lambda I)^{-1/2}$ to both sides of \eqref{e:Bspline4}. The resulting BIC can be compared with the BIC of the model with varying smoothness penalty via our model selection procedure.

We conducted simple simulation study to investigate how well the $P^h$ works in the comparison between common and varying smoothness. We considered two different models:
\begin{itemize}
	\item Model 1. $\lambda_\bt={\sigma_{\epsilon,\bt}^2}/{\sigma_\bt^2}$ (varying smoothness),
	\item Model 2. $\lambda={\sigma_{\epsilon,\bt}^2}/{\sigma_\bt^2}$ (common smoothness).
\end{itemize}
First, we fitted the glaucoma data using \eqref{e:Bspline3}. When generating simulation dataset, we used \eqref{e:Bspline3} for Model 1 and \eqref{e:Bspline4} for Model 2. We generated 100 simulated data sets from each true model.  For each simulated data set, we performed the model selection procedure using $P^h$ as described in Section 4 to see how well the procedure choose the true model. It turned out that the model selection procedure always chose the correct model.

We applied this comparison as well to the glaucoma data. Figure \ref{fg1} shows comparison of computed $P^h$ with various values of common smoothness penalty $\lambda$. With all values considered here, our model selection procedure favors the fit with varying smoothness against the fit with common smoothness.

\begin{figure}[h]
\centerline{\includegraphics[scale=0.7]{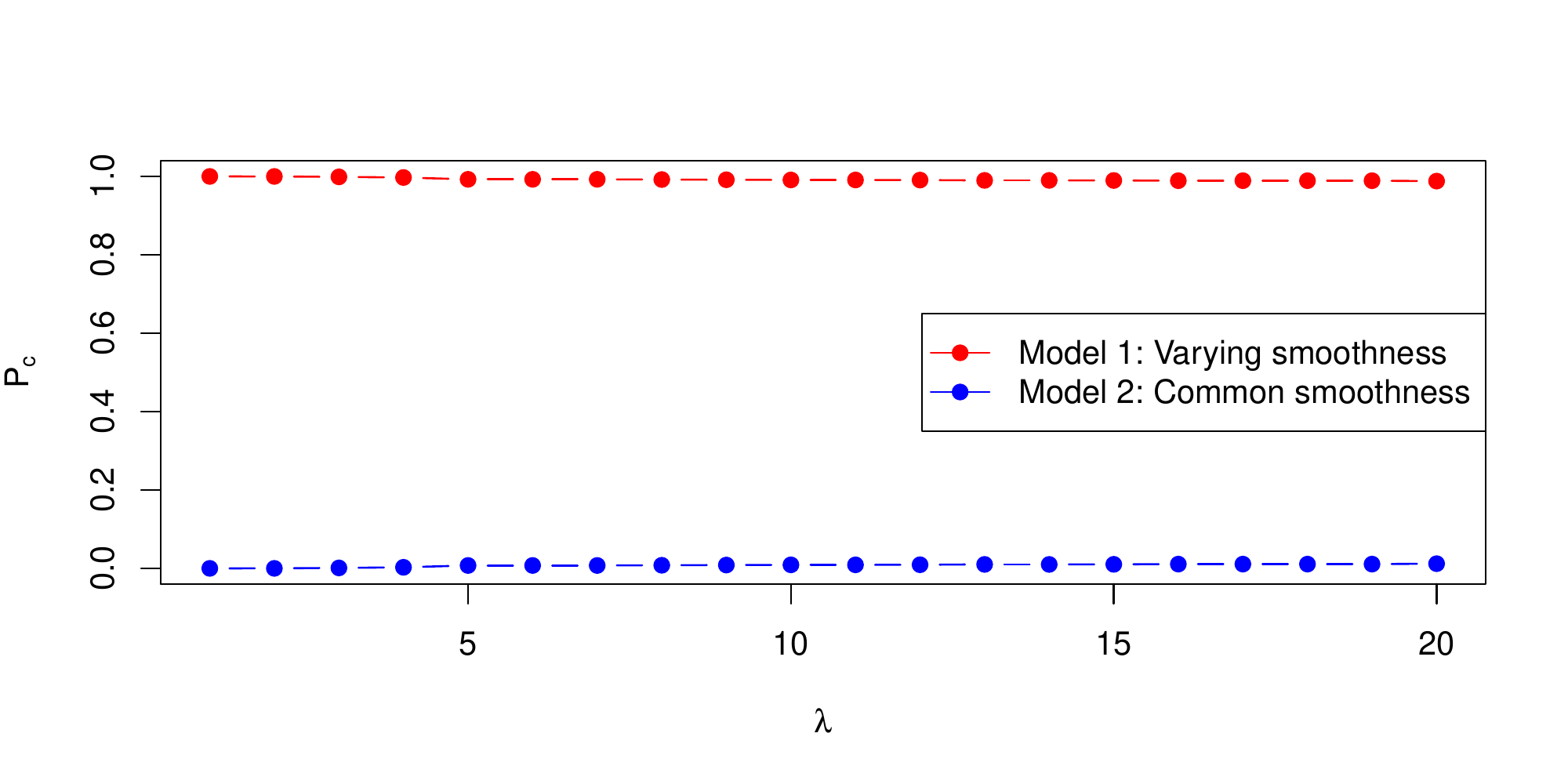}}
\vspace{-0.2in}
\caption{ \footnotesize Comparison between common and varying smoothness}\label{fg1}
\end{figure}

\section*{Simulation Study to Examine Potential Identifiability Issue in Generalized Additive Mixed Model (GAMM)}

As pointed out in Section 4,  one may have difficulty in detecting a nonparametric age effect when subject-specific random effects are included in the model at the same time. However, this is a general issue of GAMMs with a smooth nonparametric effect for a subject-specific covariate that includes repeated measures per subject and a subject-specific random effect. To investigate this issue, we conducted an additional simulation study as follows. We considered the following three different models:

\begin{itemize}
	 \item Model 1 (subject-specific random effect):  $Y_{ijp} = \mu+U_{i}+E_{ijp}$,
	 \item Model 2 (nonparametric age effect): $Y_{ijp} = f(X_{\mbox{age},i})+E_{ijp}$, and
	 \item Model 3 (nonparametric age effect+subject-specific random effect): $Y_{ijp} = f(X_{\mbox{age},i})+U_{i}+E_{ijp}$.
\end{itemize}

Here, $Y_{ijp}$ was the first basis coefficient of the scleral strain data in the wavelet space. Each of Models 1-3 was fit to the real scleral strain data and we generated 100 simulation data sets from the fitted model. For each simulation data set, we performed model selection using (1) the marginal BIC (BIC$_1$) or (2) the modified BIC (BIC$_2$) where the the number of parameters in the penalty was counted using the degree of freedom of the nonparametric fit. Table \ref{tb:sim2} shows the number of times that each model was selected out of 100 replications if the random and fixed effects were evaluated together. We can clearly see that the identifiability issue occured when the true model is Model 3. In this case, neither BIC$_1$ nor BIC$_2$ was able to correctly find the true model (Model 3). However, this issue could be rectified when we performed the fixed effect selection first followed by the random effect selection. Using such a two-step selection procedure, we were able to find the correct model for most times (99/100 for BIC$_1$ and 95/100 for BIC$_2$).  This is the strategy we used for our analysis, and appears to be a fine fix for this problem, but clearly this problem should be further investigated in the setting of GAMMs.

\begin{table}[h]
\begin{center}
\begin{tabular}{|l|l|cccc|}\hline
\multirow{2}{*}{True Model } & \multirow{2}{*}{Criterion} & \multicolumn{3}{c}{Selected Model} \\ \cline{3-5}
 & & Model 1 & Model 2 & Model 3 \\\hline
\multirow{2}{*}{Model 1} & BIC$_1$ &100 & 0 & 0 \\
										 & BIC$_2$ & 100 & 0 & 0 \\ \hline
\multirow{2}{*}{Model 2} & BIC$_1$ & 0 & 100 & 0 \\
										 & BIC$_2$ & 53 & 47 & 0 \\ \hline
\multirow{2}{*}{Model 3} & BIC$_1$ & 98 & 0 & 2 \\
										 & BIC$_2$ & 99 & 0 & 1 \\ \hline
\end{tabular}
\caption{ \footnotesize The number of times that each model was selected }\label{tb:sim2}
\end{center}
\end{table}

\newpage
\section*{Sensitivity Analyses}

\subsection*{Model Choice: Including Left vs. Eye Fixed Effect Function}

Recall that based on the model selectin heuristic, when using $aBIC$ the second-best model also included an ``eye'' effect, which consisted of a fixed offset for left vs. right eye.  This model was the best model when using $aAIC$ for the model selection heuristic.  Thus, for this model and the same tensor wavelet basis used for the primary analysis, we repeated the analysis of the data.  The \textit{MPSvsAge-eye.mp4} file shows a movie of the IOP-specific MPS fit based on the model with left vs. eye effect. The \textit{AUCvsAge-eye.mp4} file shows a movie of the AUC summaries based on the model with left vs. eye effect. The \textit{Intrafunctional\_correlations-eye.mp4} file shows a movie of the induced intrafunctional correlation from this model.  The fits based on this model are similar to those based on the primary model used in the paper, so substantive results are not sensitivity to inclusion of this left vs. right eye effect.

\subsection*{Regularization Hyperparameters: Assessing Results with No Shrinkage}

One concern voiced by reviewers was that results could be sensitive to our choice of prior for the fixed effects.  We use a spike-slab sparsity prior, and as pointed out by one reviewer, results of variable selection can be strongly sensitivty to the choice of the regularization hyperparameters $\pi_{aj}$ and $\tau_{aj}$ for this prior.  As described in the main paper, we used an empirical Bayes approach to estimate these parameters from the data.  However, there is some concern that results may be driven by this informative prior.  To assess, we performed another analysis using relatively uninformative hyperpriors with $\pi_{aj} \approx 1$ and $\tau_{aj}=10^6$ for all $a,j$.   This is an extreme case of the spike slab for which the probability of the zero slab is negligible and also the linear shrinkage induced by the Gaussian variance is also mostly negligible.  We have shown in previous publications this prior, which we call the ``no smoothing prior'', gives point estimates the same as if no smoothing is done, and is thus unbiased.  We applied this prior to our data set using the same tensor wavelet basis used for the primary analysis, and repeated all analyses of the data.  The \textit{MPSvsAge-nosmooth.mp4} file shows a movie of the IOP-specific MPS fit based on the model with no shrinkage. The \textit{AUCvsAge-nosmooth.mp4} file shows a movie of the AUC summaries based on the model with no shrinkage. The \textit{Intrafunctional\_correlations-nosmooth.mp4} file shows a movie of the induced intrafunctional correlation from this model for this choice of prior.  The fits based on this model are similar to those based on the primary model used in the paper, so we can see that there is no significant bias induced by our informative priors.  The difference we see is that the fits over MPS are slightly less smooth in MPS.

\subsection*{Choice of Basis: Wavelet-Regularized Principal Components}

Recall that a total of  269 wavelet coefficients after the outlier filtering were used as basis coefficients in our application. As described in Section 5.2, we also considered using principal components (PC) scores computed on these wavelet coefficients as basis coefficients. In particular, we applied a singular value decomposition to $\mathbf{W}$ where $\mathbf{W}$ is the $306\times 269$ matrix of wavelet coefficients. Let $\mathbf{V}$ be the matrix of right singular vectors. The wavelet-space PC scores were computed by $\mathbf{Y^*}=\mathbf{W}\mathbf{V}$ where $\mathbf{Y^*}$ are the wavelet-space PC scores. We kept only the leading 27 columns of $\mathbf{Y^*}$ that account for most of the variability according to the scree plot ($>99.5\%$).The figure below contains the first 9 PCs, and the file \textit{wPCs.pdf} contains plots of all 27 PC bases. The \textit{MPSvsAge-pc.mp4} file shows a movie of the IOP-specific MPS fit based on the PC scores. The \textit{AUCvsAge-pc.mp4} file shows a movie of the AUC summaries based on the PC scores. The \textit{Intrafunctional\_correlations-pc.mp4} file shows a movie of the induced intrafunctional correlation from this choice of basis.  Note that although allowing more global correlations, the strongest correlations include the local correlations at nearby scleral locations that also dominate these correlation surfaces for the tensor wavelet basis.  The fits based on the PC scores are similar to those based on the wavelet coefficients.  The levels of variability are greater for the wPC bases, and the age effects in the regions more distant from the ONH are more nonlinear than for the wavelet bases, possibly from the extra induced correlation of these locations and scleral positions close to the ONH.  Although the PC basis keeps $99.5\%$ of the variability in the data according to the scree plots, the minimum correlation of the raw functions and the PC projections is $<0.97$, which results in some loss of information.  Also, given such a small number of basis functions, $27$, it is not clear whether this is sufficiently flexible to capture all of the structure at the various levels.  These are some of the reasons we prefer the wavelet bases for this application.

\begin{figure}
\minipage{0.45\textwidth}
	\centerline{\includegraphics[scale=0.4]{age_by_pressure_with_df83.png}}
	\caption{ \footnotesize Tensor Wavelet, Main Model, Empirical Bayes Shrinkage}
	\label{fig:sfig1}
\endminipage \hfill
\minipage{0.45\textwidth}
	\centerline{\includegraphics[scale=0.4]{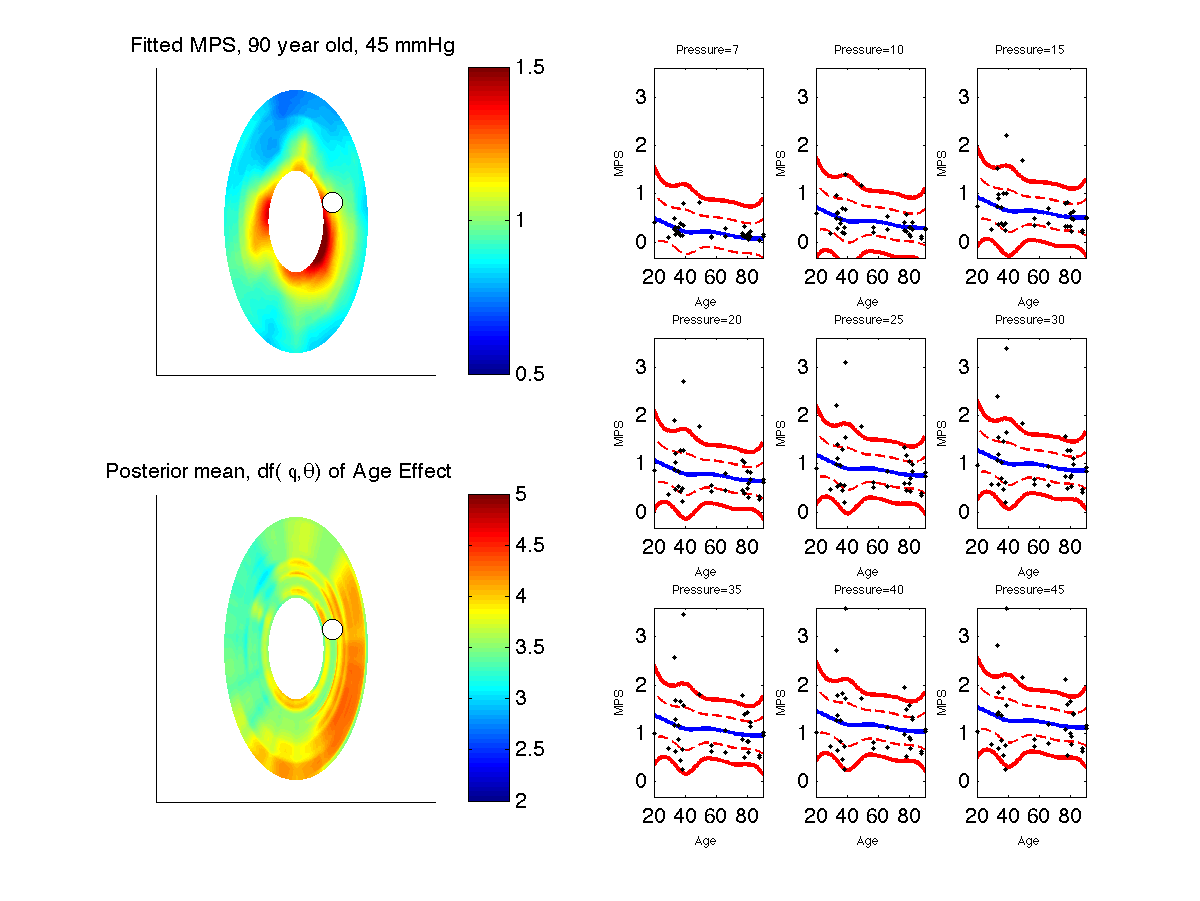}}
	\label{fig:sfig1}
	\caption{ \footnotesize Wavelet-Regularized PC Basis, Main Model}
\endminipage \\
\minipage{0.45\textwidth}
	\centerline{\includegraphics[scale=0.4]{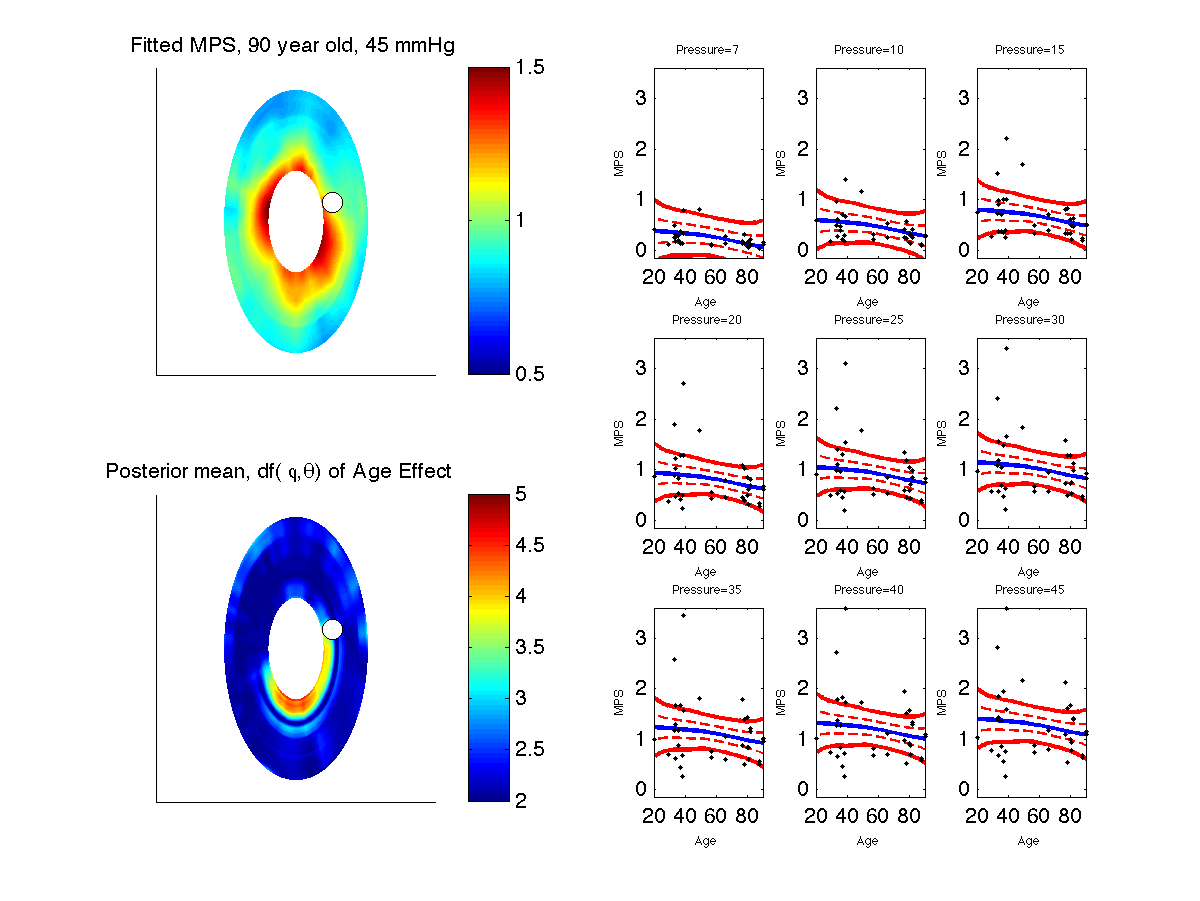}}
	\caption{ \footnotesize Tensor Wavelet, Model with Left vs. Eye Effect, Empirical Bayes Shrinkage}
	\label{fig:sfig1}
\endminipage \hfill
\minipage{0.45\textwidth}
	\centerline{\includegraphics[scale=0.4]{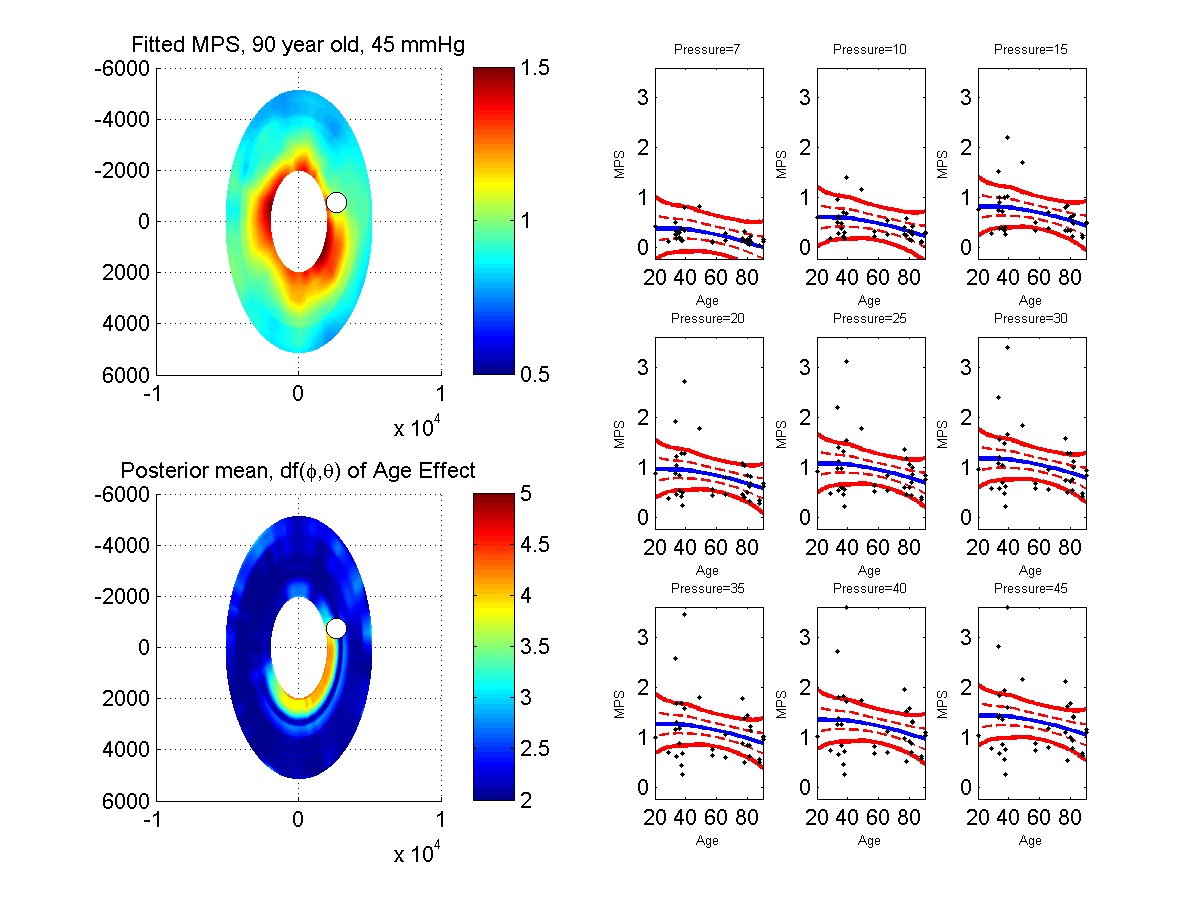}}
	\label{fig:sfig1}
	\caption{ \footnotesize Tensor Wavelet, Main Model, No Shrinkage Prior}
\endminipage
\end{figure}

\begin{figure}
\minipage{0.45\textwidth}
	\centerline{\includegraphics[scale=0.4]{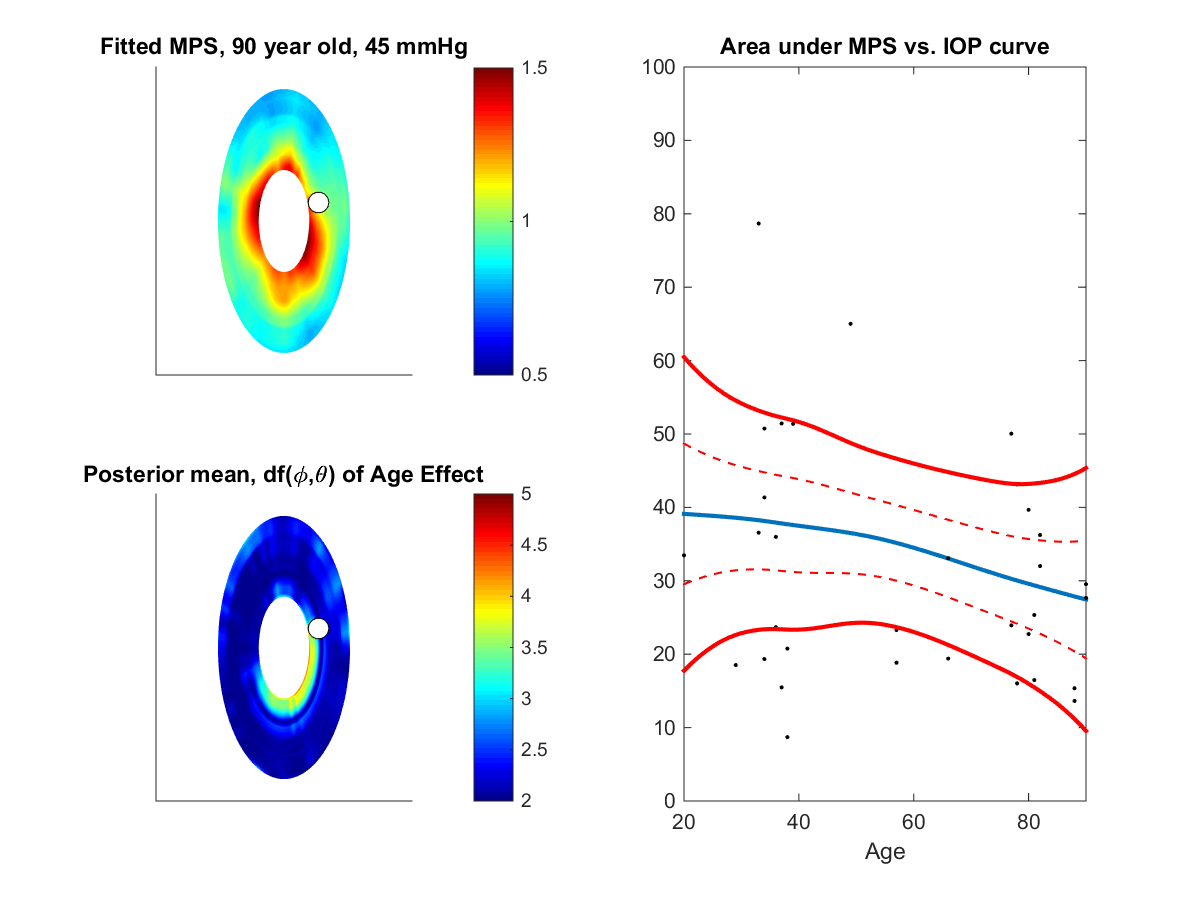}}
	\caption{ \footnotesize Tensor Wavelet, Main Model, Empirical Bayes Shrinkage}
	\label{fig:sfig1}
\endminipage \hfill
\minipage{0.45\textwidth}
	\centerline{\includegraphics[scale=0.4]{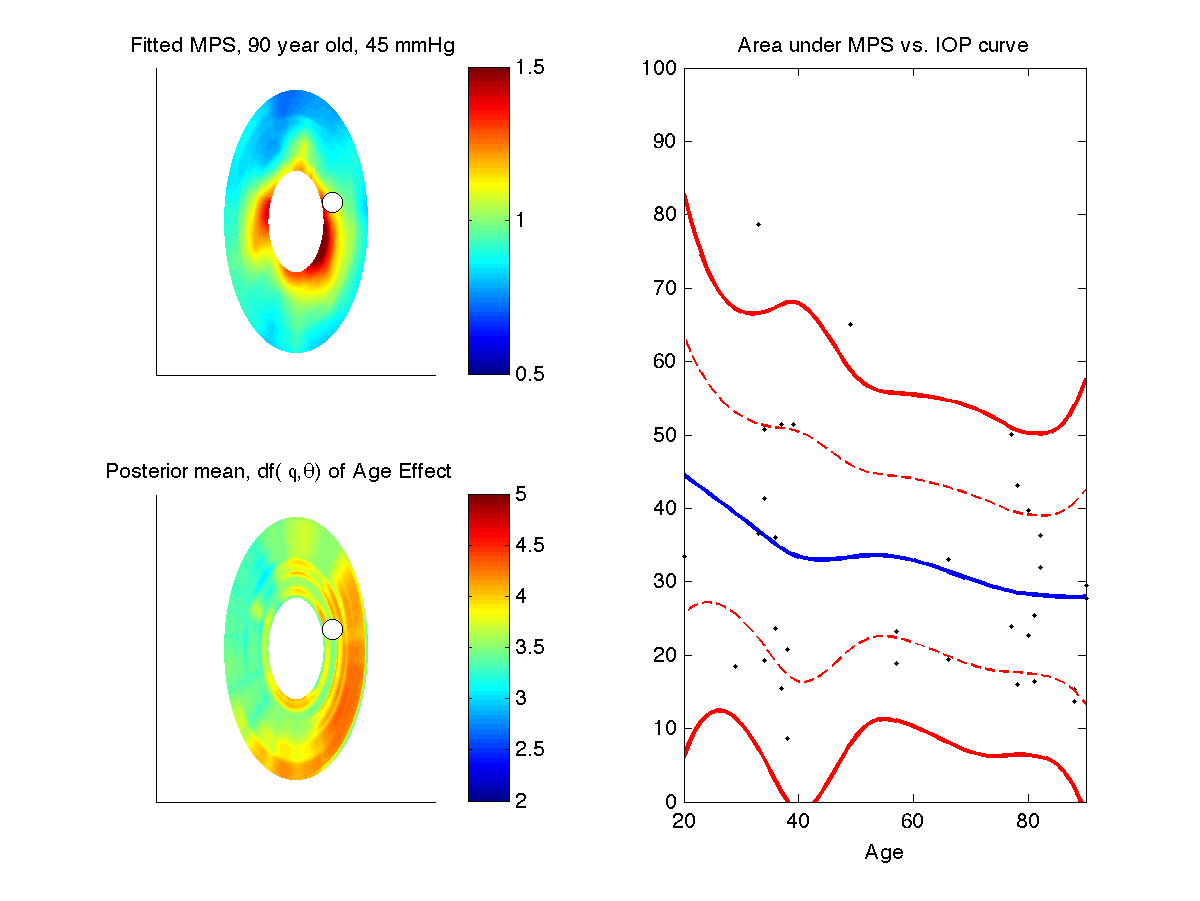}}
	\label{fig:sfig1}
	\caption{ \footnotesize Wavelet-Regularized PC Basis, Main Model}
\endminipage \\
\minipage{0.45\textwidth}
	\centerline{\includegraphics[scale=0.4]{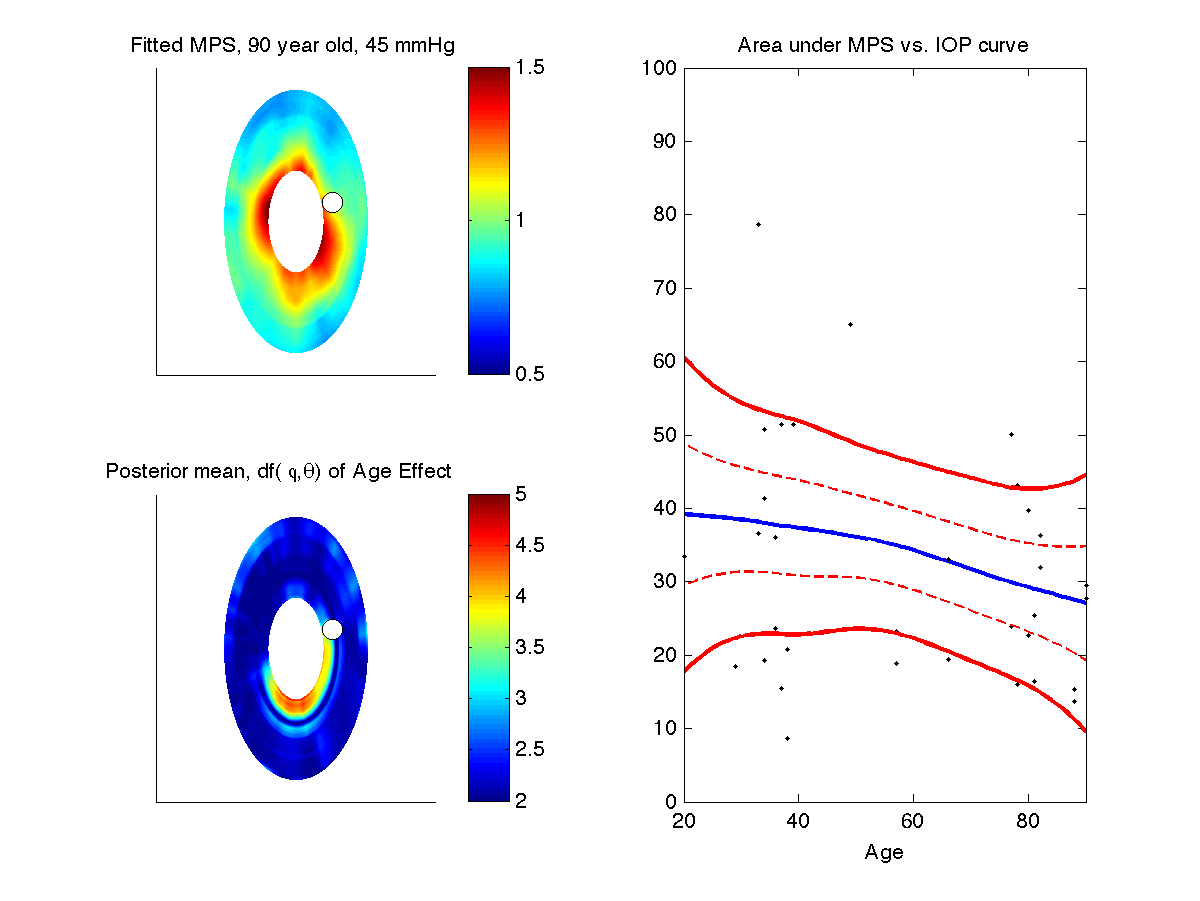}}
	\caption{ \footnotesize Tensor Wavelet, Model with Left vs. Eye Effect, Empirical Bayes Shrinkage}
	\label{fig:sfig1}
\endminipage \hfill
\minipage{0.45\textwidth}
	\centerline{\includegraphics[scale=0.4]{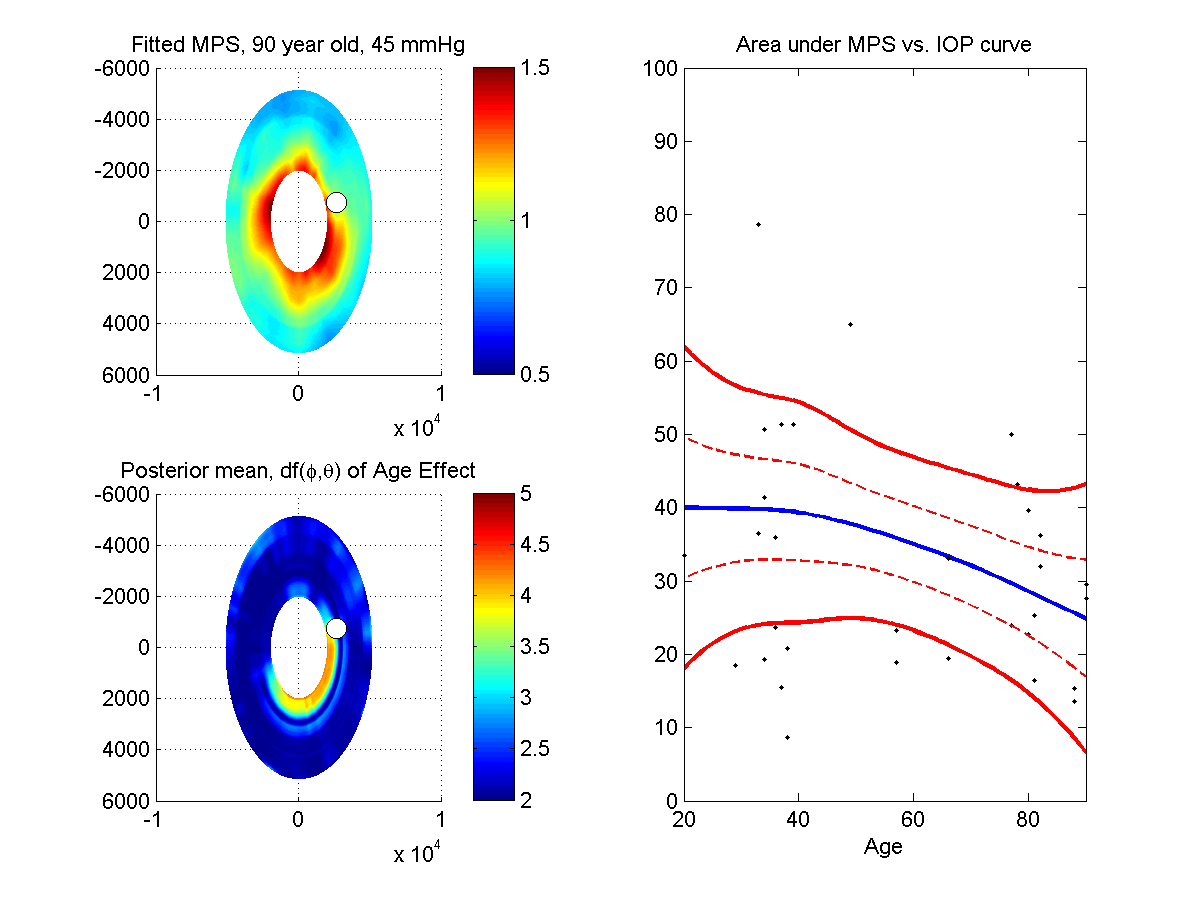}}
	\label{fig:sfig1}
	\caption{ \footnotesize Tensor Wavelet, Main Model, No Shrinkage Prior}
\endminipage
\end{figure}

\begin{figure}
\minipage{0.45\textwidth}
	\centerline{\includegraphics[scale=0.4]{auc_circum.png}}
	\caption{ \footnotesize Tensor Wavelet, Main Model, Empirical Bayes Shrinkage}
	\label{fig:sfig1}
\endminipage \hfill
\minipage{0.45\textwidth}
	\centerline{\includegraphics[scale=0.4]{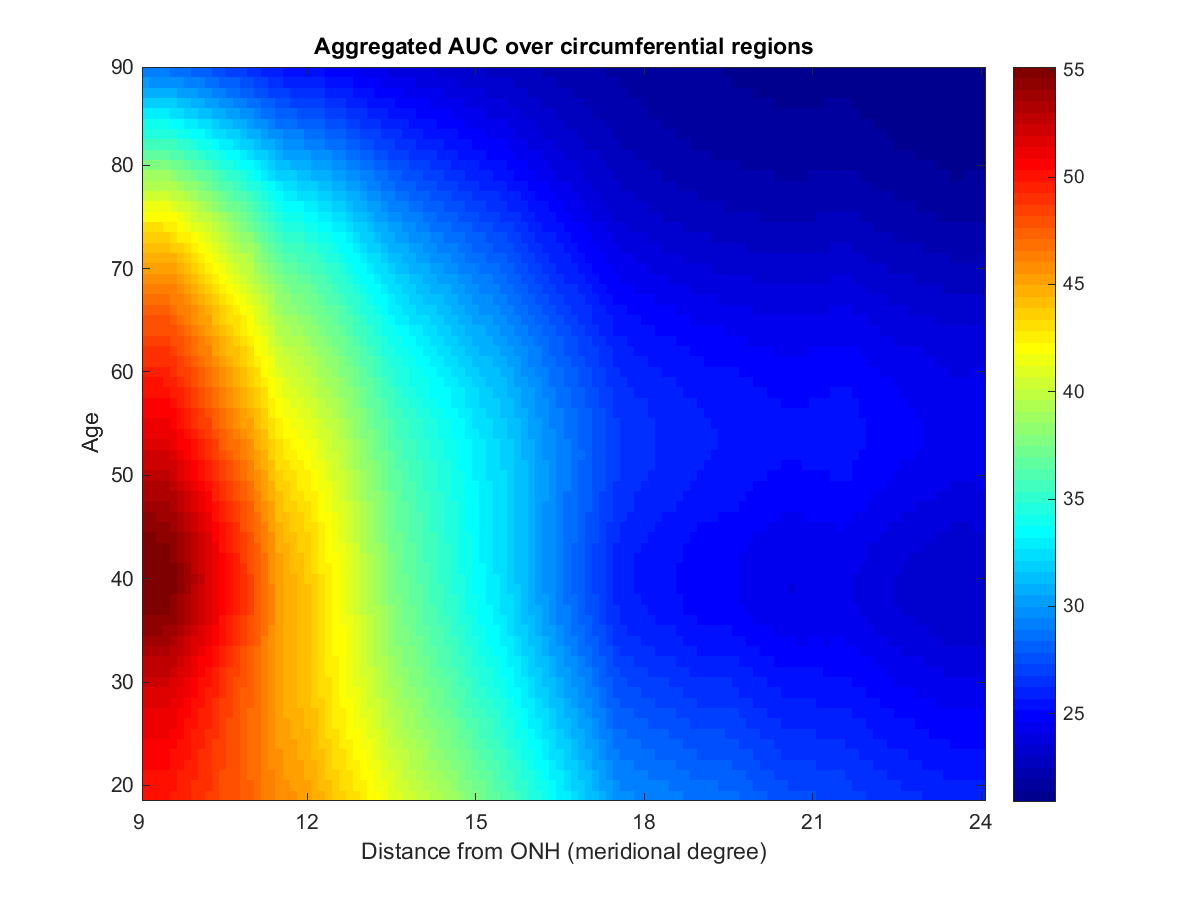}}
	\label{fig:sfig1}
	\caption{ \footnotesize Wavelet-Regularized PC Basis, Main Model}
\endminipage \\
\minipage{0.45\textwidth}
	\centerline{\includegraphics[scale=0.4]{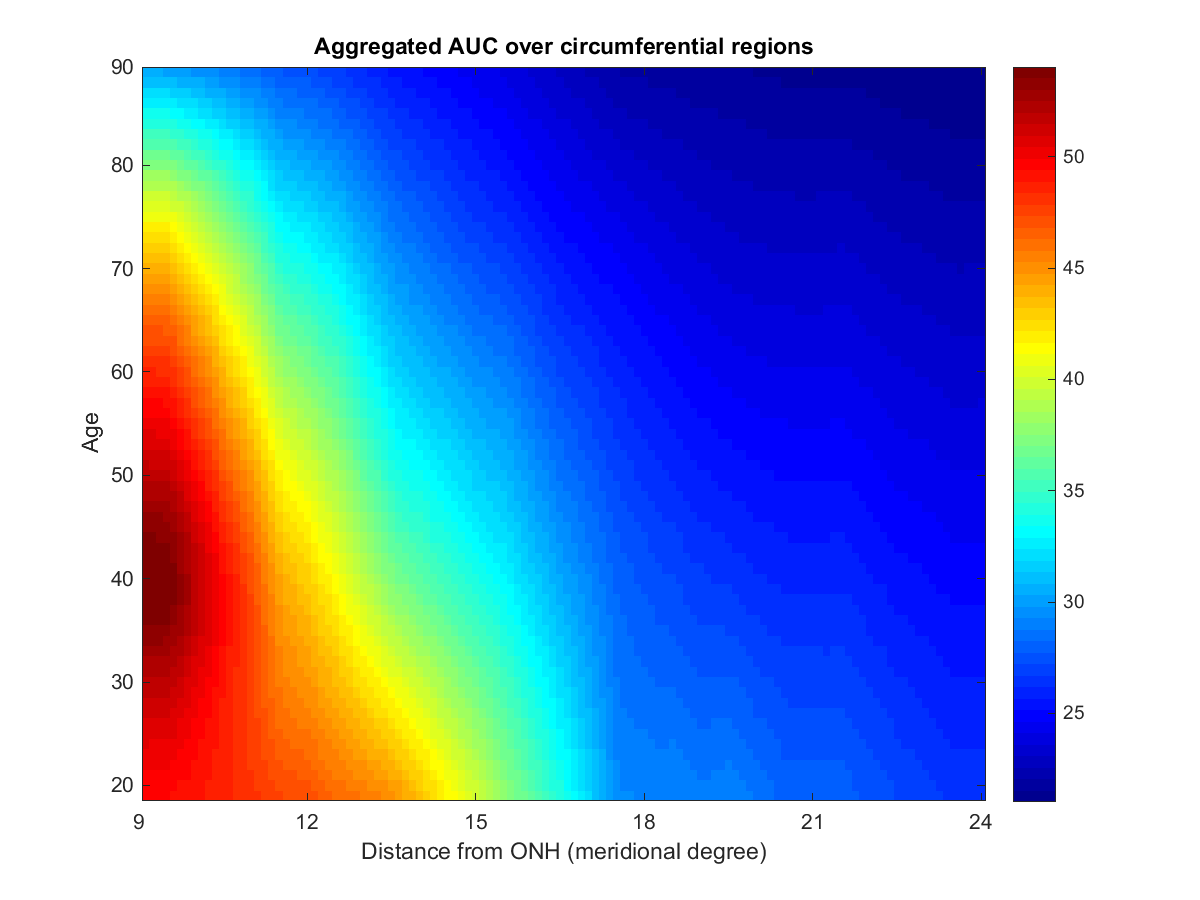}}
	\caption{ \footnotesize Tensor Wavelet, Model with Left vs. Eye Effect, Empirical Bayes Shrinkage}
	\label{fig:sfig1}
\endminipage \hfill
\minipage{0.45\textwidth}
	\centerline{\includegraphics[scale=0.4]{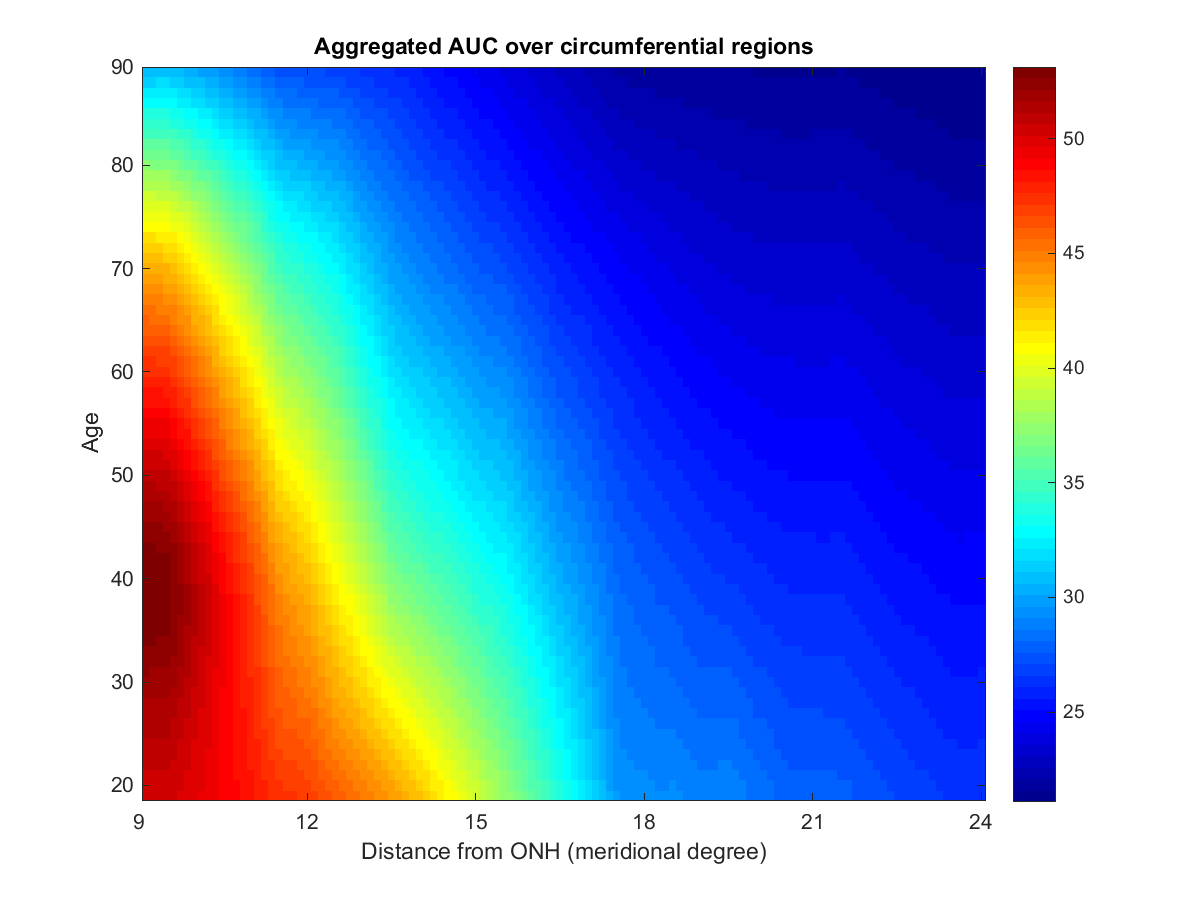}}
	\label{fig:sfig1}
	\caption{ \footnotesize Tensor Wavelet, Main Model, No Shrinkage Prior}
\endminipage
\end{figure}

\begin{figure}
\minipage{0.45\textwidth}
	\centerline{\includegraphics[scale=0.4]{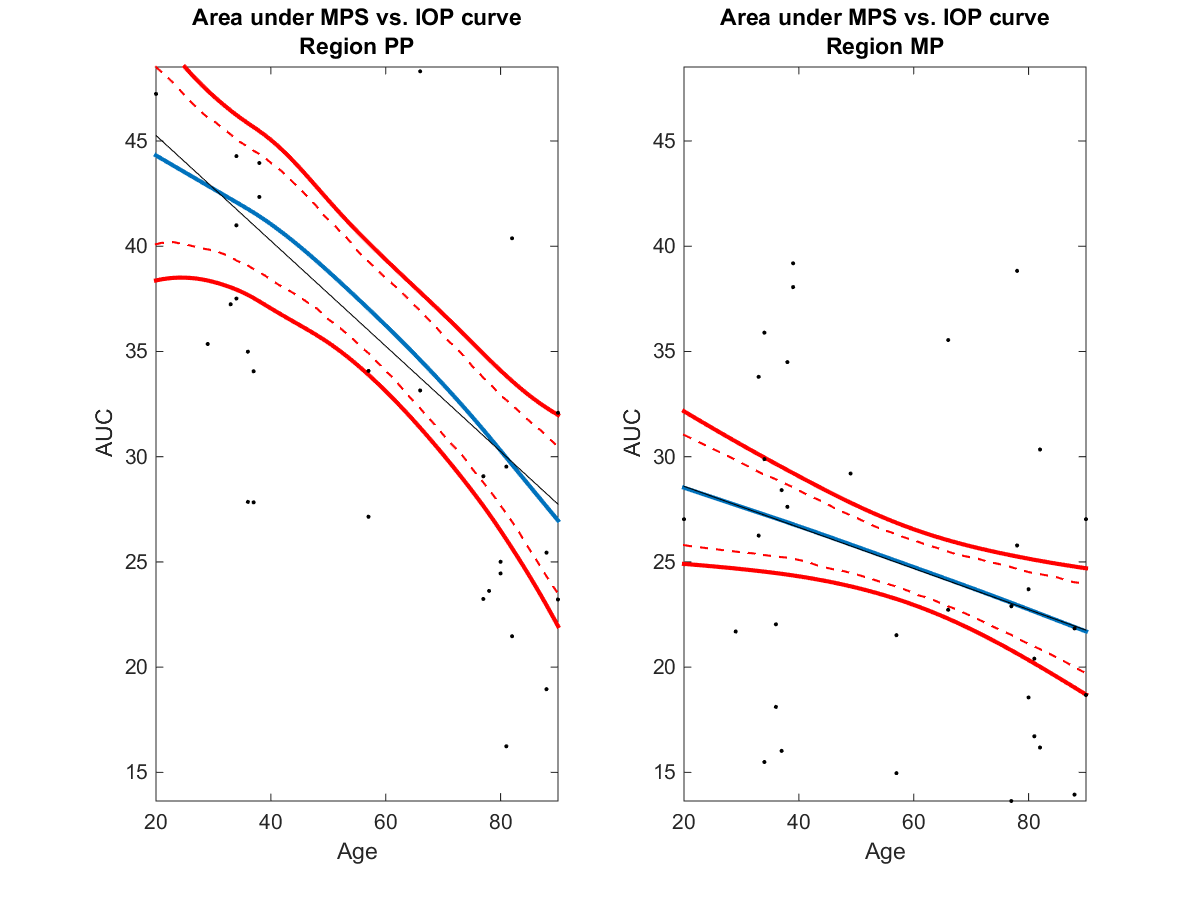}}
	\caption{ \footnotesize Tensor Wavelet, Main Model, Empirical Bayes Shrinkage}
	\label{fig:sfig1}
\endminipage \hfill
\minipage{0.45\textwidth}
	\centerline{\includegraphics[scale=0.4]{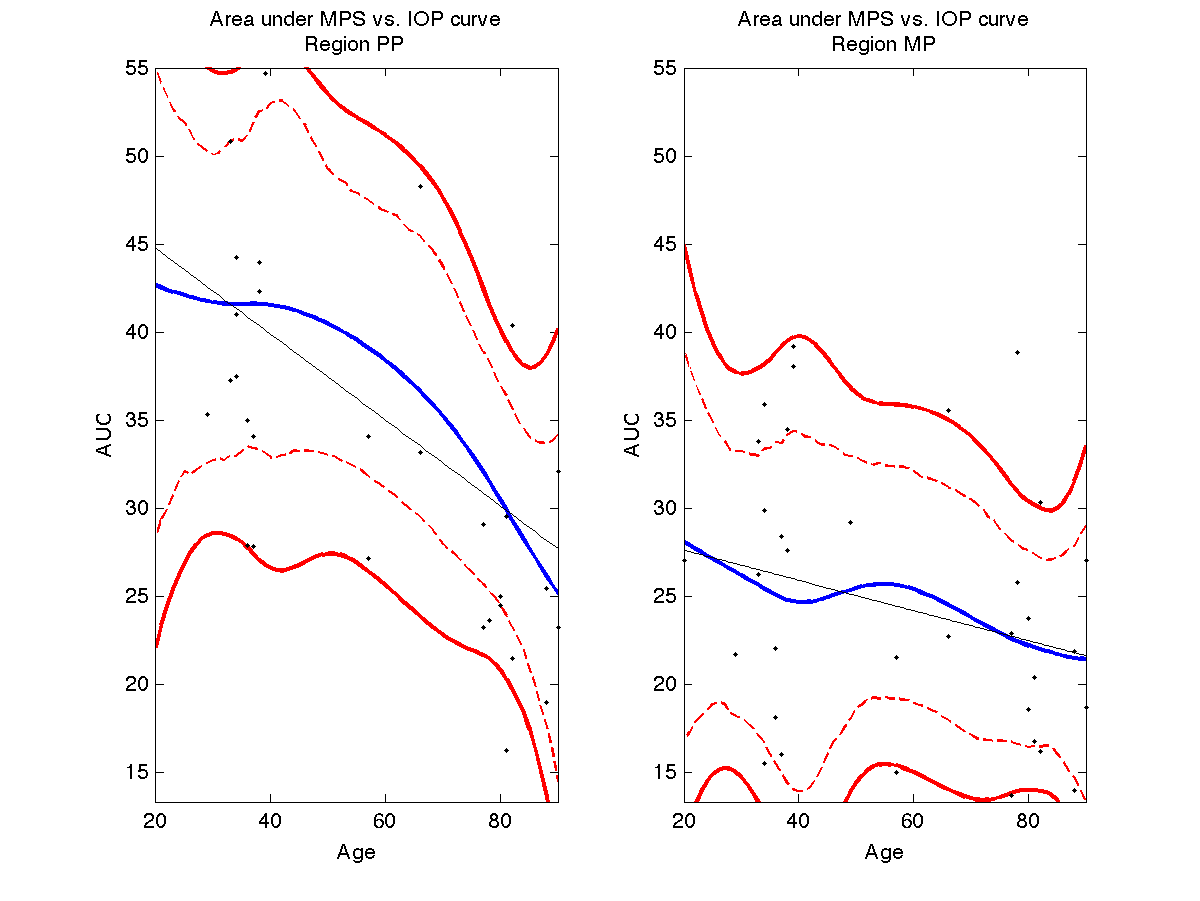}}
	\label{fig:sfig1}
	\caption{ \footnotesize Wavelet-Regularized PC Basis, Main Model}
\endminipage \\
\minipage{0.45\textwidth}
	\centerline{\includegraphics[scale=0.4]{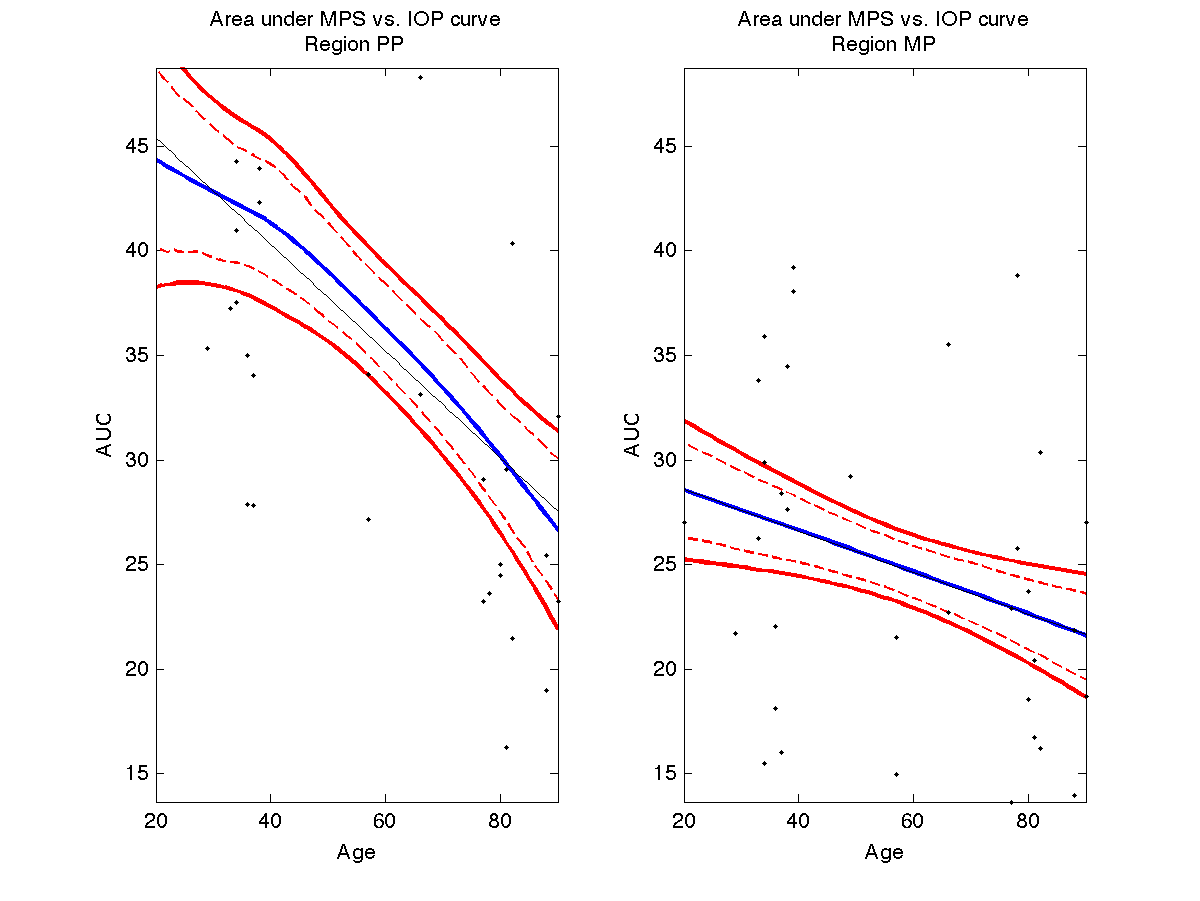}}
	\caption{ \footnotesize Tensor Wavelet, Model with Left vs. Eye Effect, Empirical Bayes Shrinkage}
	\label{fig:sfig1}
\endminipage \hfill
\minipage{0.45\textwidth}
	\centerline{\includegraphics[scale=0.4]{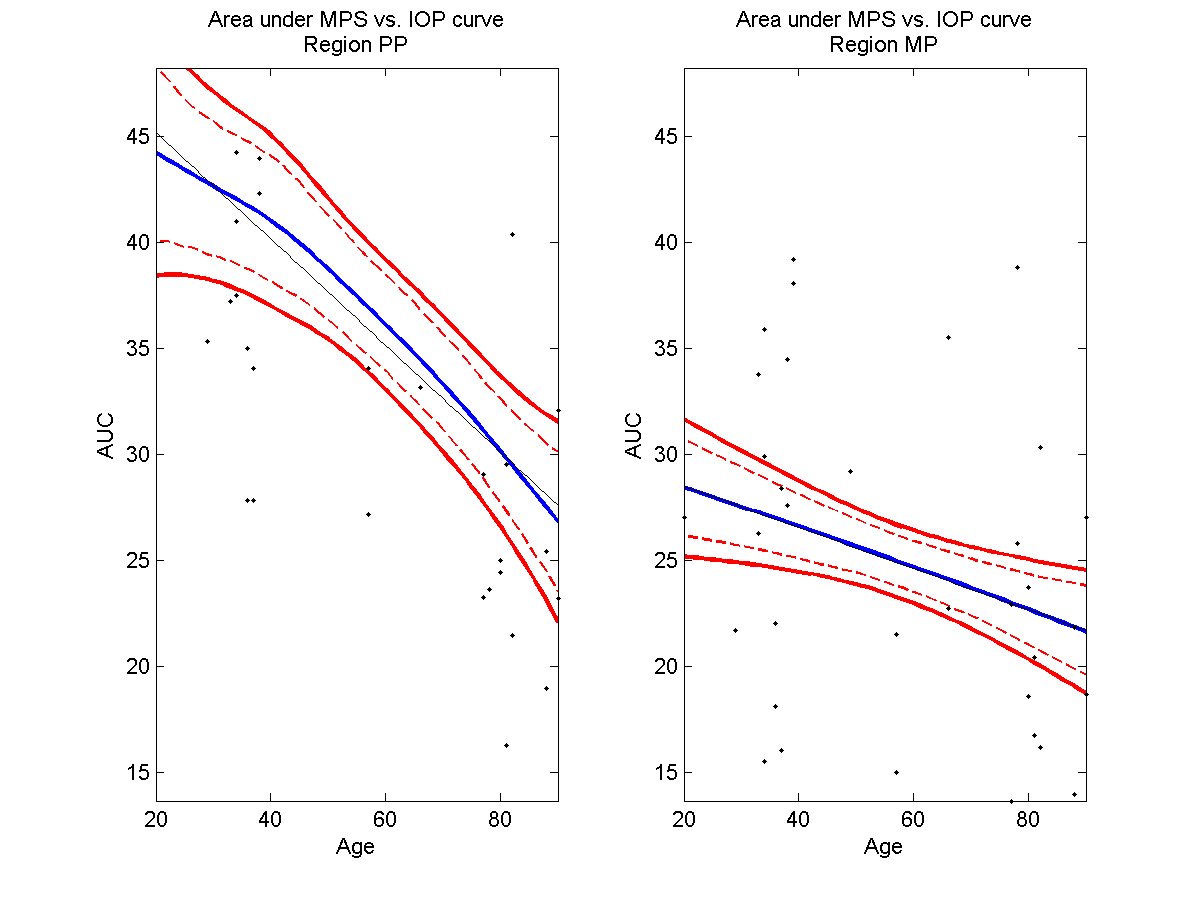}}
	\label{fig:sfig1}
	\caption{ \footnotesize Tensor Wavelet, Main Model, No Shrinkage Prior}
\endminipage
\end{figure}

\begin{figure}
\minipage{0.45\textwidth}
	\centerline{\includegraphics[scale=0.4]{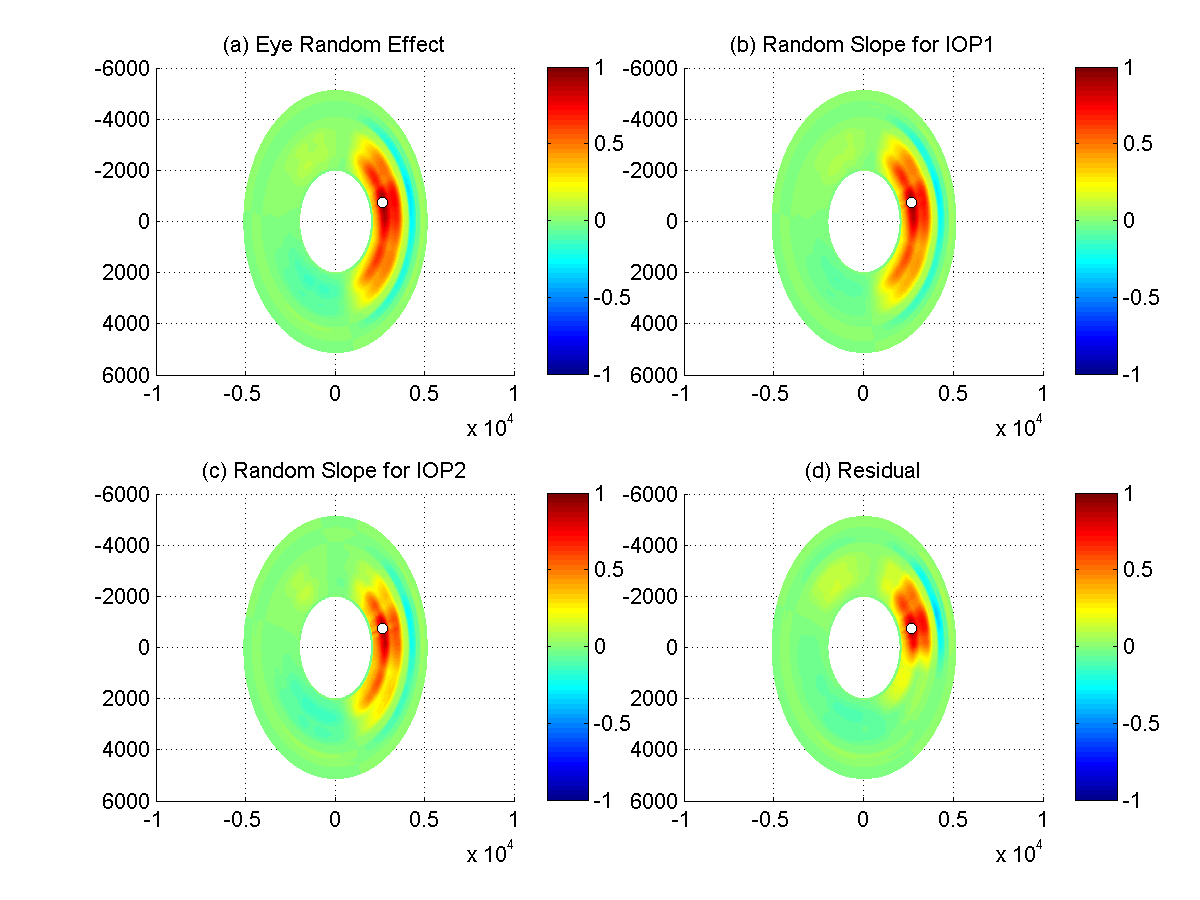}}
	\caption{ \footnotesize Tensor Wavelet, Main Model, Empirical Bayes Shrinkage}
	\label{fig:sfig1}
\endminipage \hfill
\minipage{0.45\textwidth}
	\centerline{\includegraphics[scale=0.4]{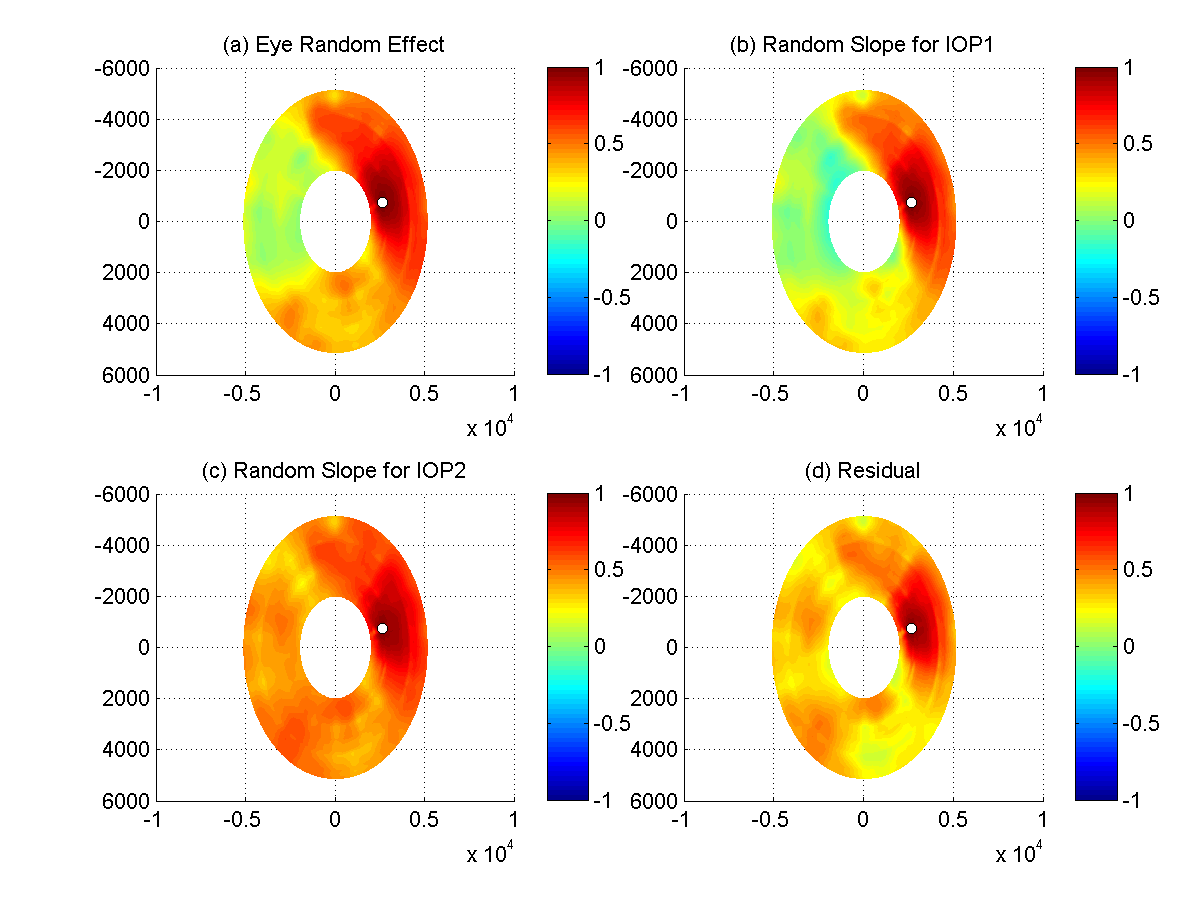}}
	\label{fig:sfig1}
	\caption{ \footnotesize Wavelet-Regularized PC Basis, Main Model}
\endminipage \\
\minipage{0.45\textwidth}
	\centerline{\includegraphics[scale=0.4]{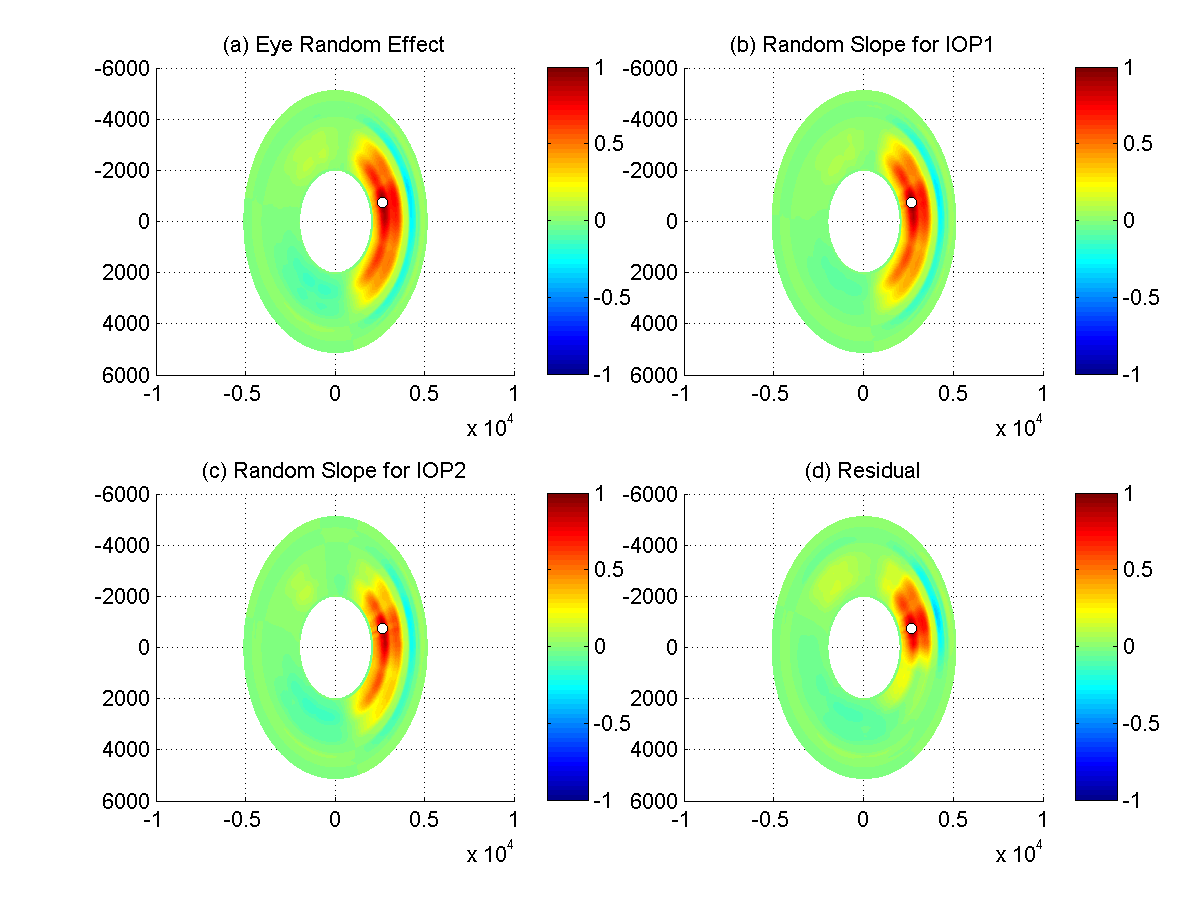}}
	\caption{ \footnotesize Tensor Wavelet, Model with Left vs. Eye Effect, Empirical Bayes Shrinkage}
	\label{fig:sfig1}
\endminipage \hfill
\minipage{0.45\textwidth}
	\centerline{\includegraphics[scale=0.4]{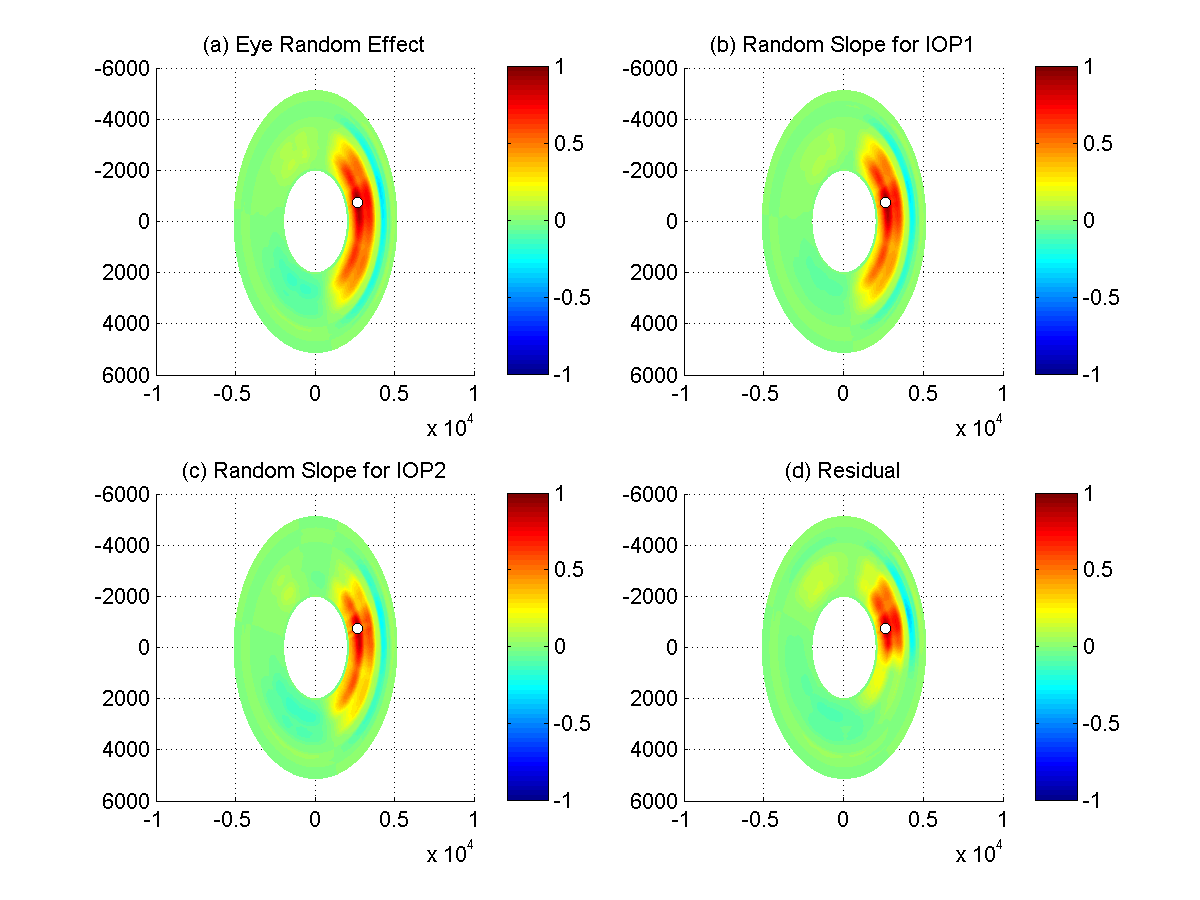}}
	\label{fig:sfig1}
	\caption{ \footnotesize Tensor Wavelet, Main Model, No Shrinkage Prior}
\endminipage
\end{figure}

\begin{figure}
\minipage{0.45\textwidth}
	\centerline{\includegraphics[scale=0.4]{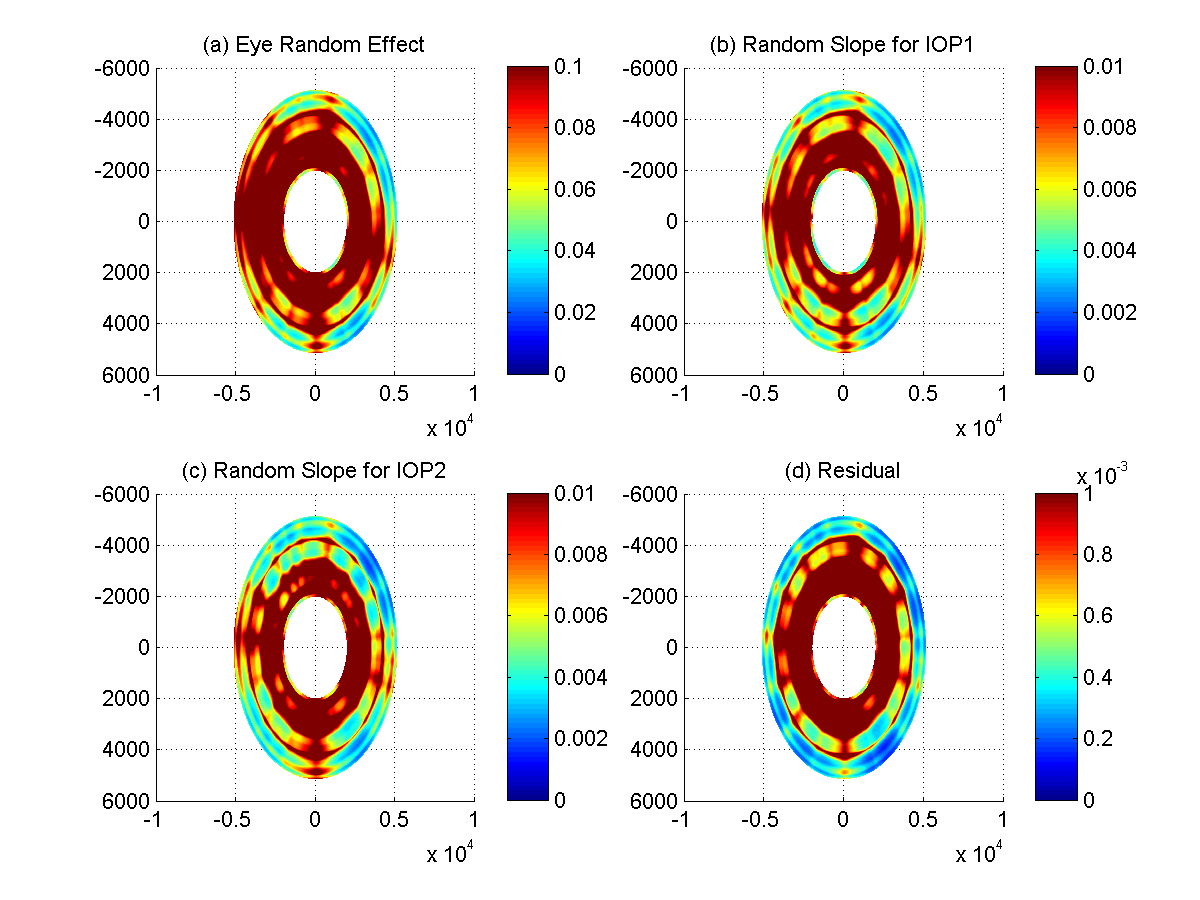}}
	\caption{ \footnotesize Tensor Wavelet, Main Model, Empirical Bayes Shrinkage}
	\label{fig:sfig1}
\endminipage \hfill
\minipage{0.45\textwidth}
	\centerline{\includegraphics[scale=0.4]{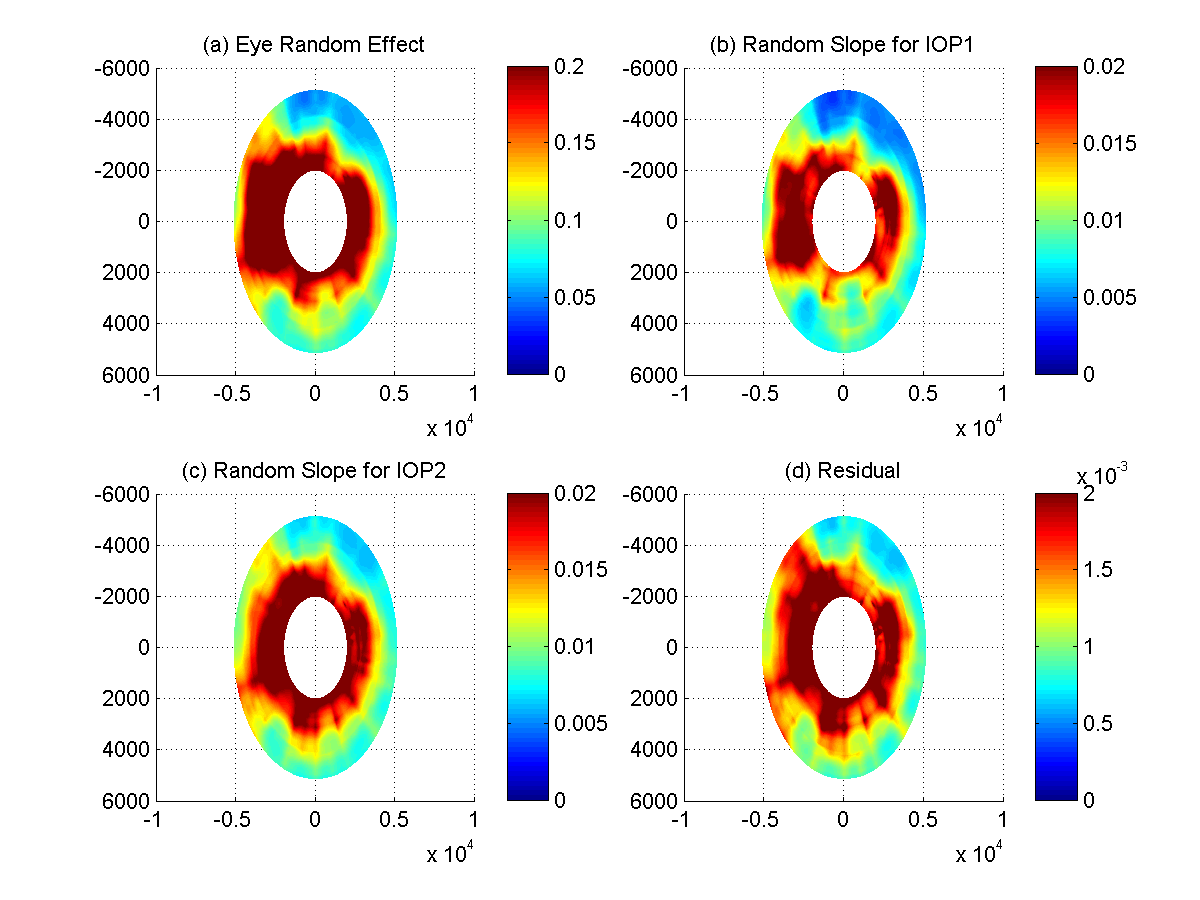}}
	\label{fig:sfig1}
	\caption{ \footnotesize Wavelet-Regularized PC Basis, Main Model}
\endminipage \\
\minipage{0.45\textwidth}
	\centerline{\includegraphics[scale=0.4]{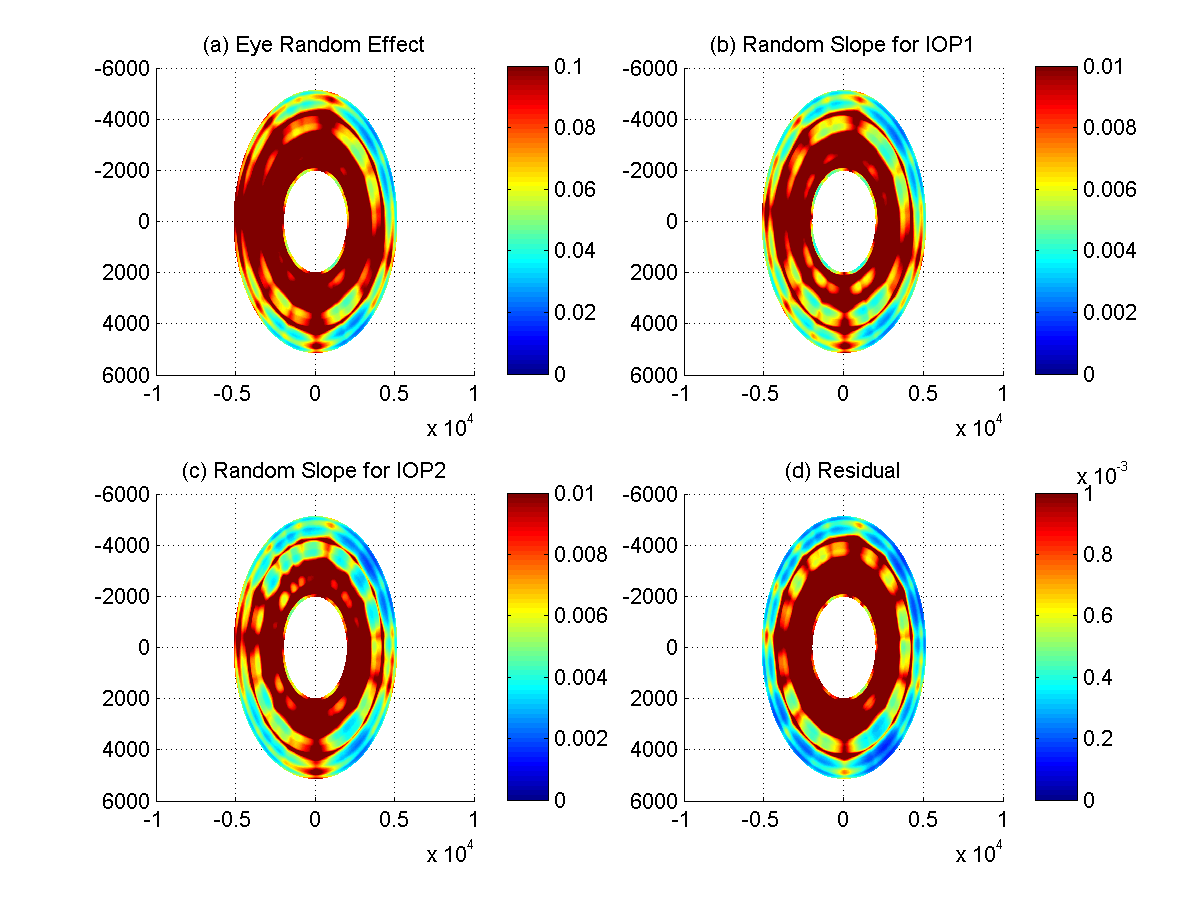}}
	\caption{ \footnotesize Tensor Wavelet, Model with Left vs. Eye Effect, Empirical Bayes Shrinkage}
	\label{fig:sfig1}
\endminipage \hfill
\minipage{0.45\textwidth}
	\centerline{\includegraphics[scale=0.4]{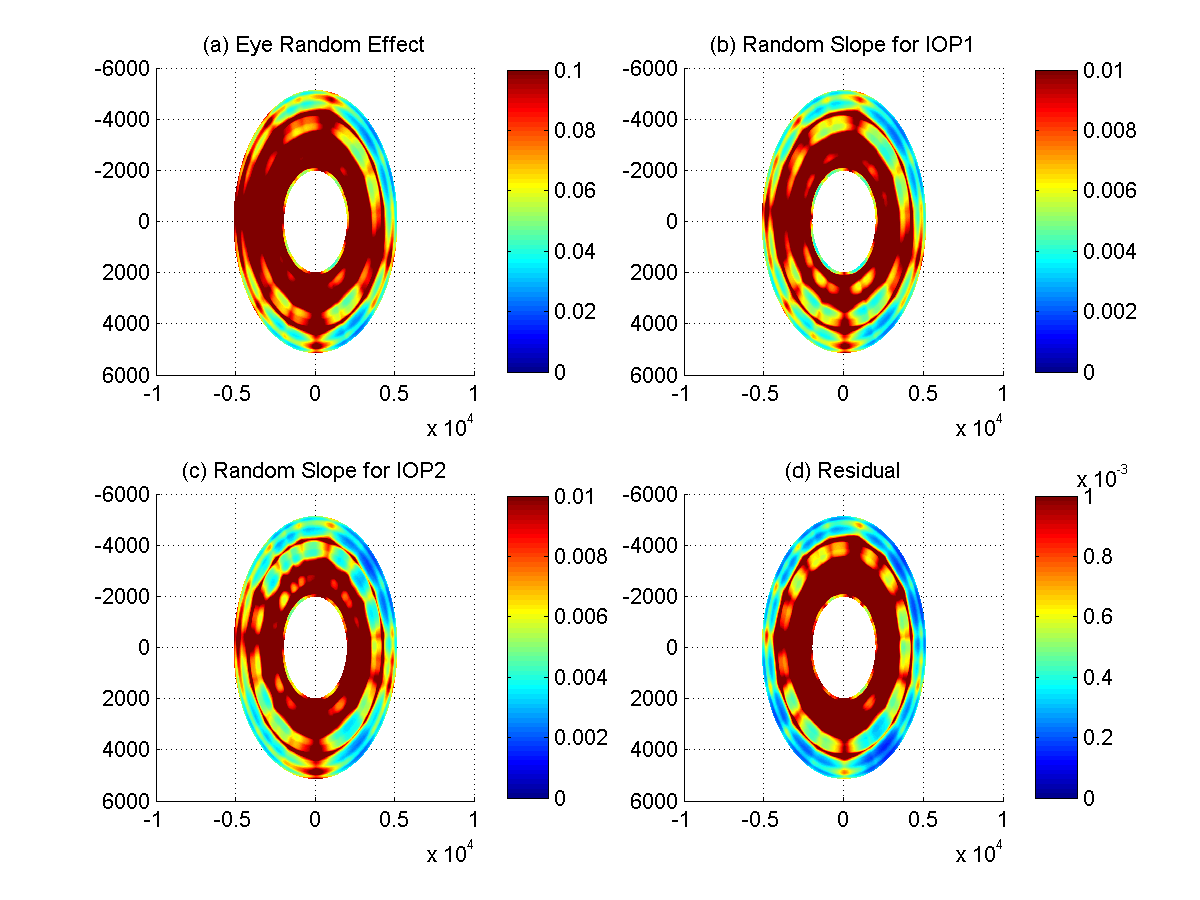}}
	\label{fig:sfig1}
	\caption{ \footnotesize Tensor Wavelet, Main Model, No Shrinkage Prior}
\endminipage
\end{figure}

\newpage
\section*{Other Results from Main Paper}

There are other supplemental plots and movies referenced in the paper and included in the supplement including:
\begin{description}
\item[RawMPScurves.zip:] plots of all raw MPS curves, results after robust filtering to remove outliers, and results after wavelet compression to show that virtually no information is lost by the reduction from $14,400$ observatrions to $269$ basis coefficients.  The following figure contains the raw MPS data for one eye from a subject at the 9 IOP levels, plus plots after outlier removal and compression to show how near-lossless the transform is.

\begin{figure}
	\centerline{\includegraphics[scale=0.8]{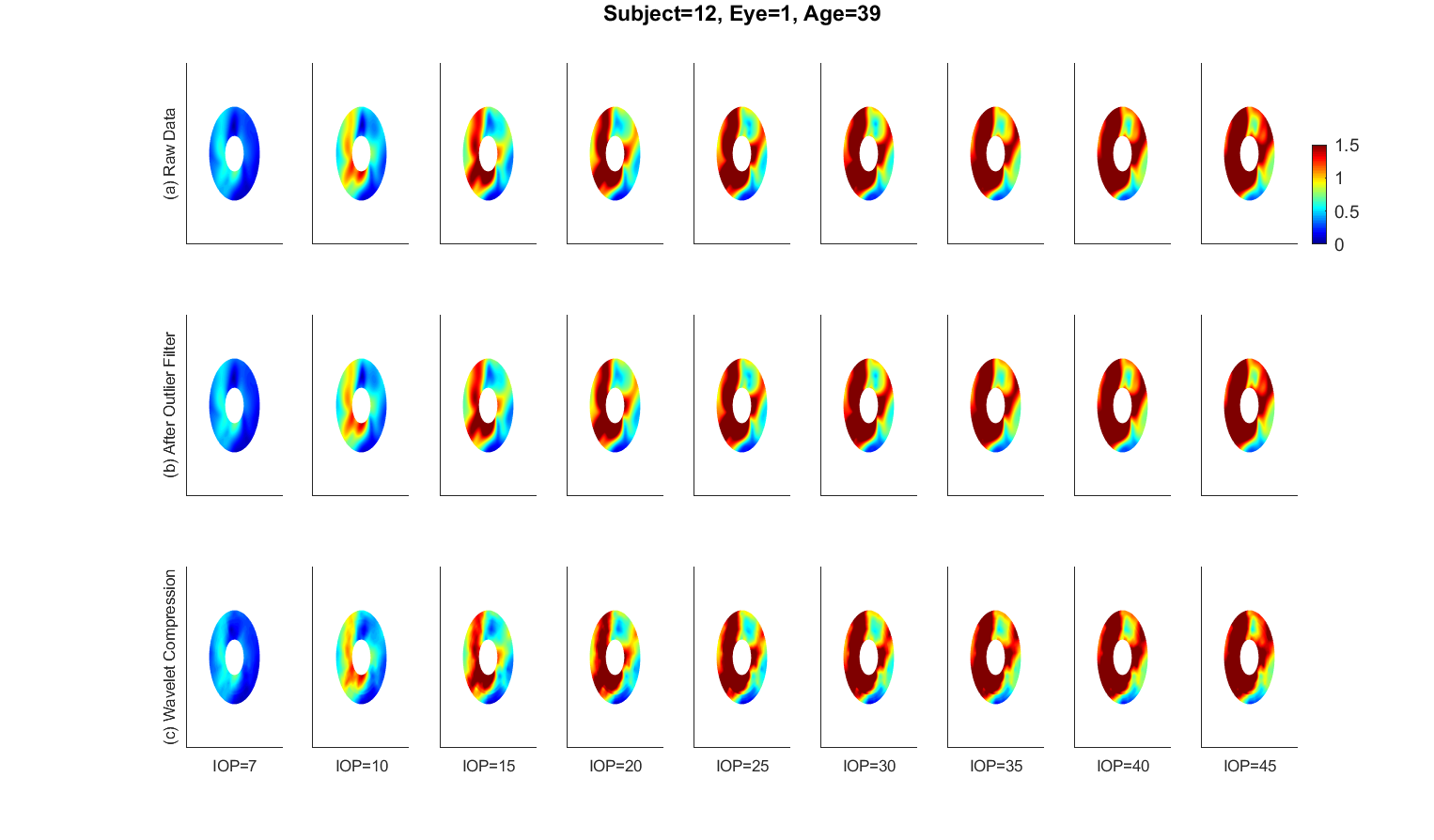}} \label{fig:compression}
	\caption{ \footnotesize Plot of raw data for one eye from 39 year old subject at 9 levels of IOP (top), plus results after outlier removal (middle), and after wavelet compression down to 269 coefficients (bottom), demonstrating the near-lossless nature of the transform.  The plots for all other eyes are in RawMPScurves.zip.}
\end{figure}

\item[MPSvsAge-wave.mp4:] Movie of MPS vs. age for each IOP based on the tensor wavelet basis function.

\item[Combo\_plot.mp4:] Movie of key summary results based on the tensor wavelet basis functions.

\item[Intrafunctional\_correlations.mp4] Movie showing intrafunctional correlations induced by tensor wavelet basis functions.

\item[Intra\_IOP\_corr.mp4] Movie showing interfunctional variance and serial correlation across IOP from same eye based on the tensor wavelet basis functions.
\ 
\item[MPSvsAge-pc.mp4:] Movie of MPS vs. age for each IOP based on the principal component basis functions.

\item[AUCvsAge-pc.mp4:] Movie of AUC vs. age based on the principal component basis functions.

\item[Intrafunctional\_correlations-pc.mp4] Movie showing intrafunctional correlations induced by the principal component basis functions.


\item[MPSvsAge-eye.mp4:] Movie of MPS vs. age for each IOP based on tensor wavelet basis functions and model including left vs. right eye effect.

\item[AUCvsAge-eye.mp4:] Movie of AUC vs. age based on tensor wavelet basis functions and model including left vs. right eye effect.

\item[Intrafunctional\_correlations-eye.mp4] Movie showing intrafunctional correlations induced by the tensor wavelet basis functions and model including left vs. right eye effect


\item[MPSvsAge-nosmooth.mp4:] Movie of MPS vs. age for each IOP based on tensor wavelet basis functions and model with no smoothing ($\pi_{\cdot}=1, \tau_{\cdot}=10^6$).

\item[AUCvsAge-nosmooth.mp4:] Movie of AUC vs. age based on tensor wavelet basis functions and model with no smoothing ($\pi_{\cdot}=1, \tau_{\cdot}=10^6$).

\item[Intrafunctional\_correlations-nosmooth.mp4] Movie showing intrafunctional correlations induced by the tensor wavelet basis functions and model with no smoothing ($\pi_{\cdot}=1, \tau_{\cdot}=10^6$).
\end{description}

\section*{Details of MCMC Convergence Diagnostics}

We used the package {\itshape coda} in R to run the Geweke convergence diagnostics. We looked at three different sets of parameters: the fixed effects coefficients (FE), the nonparametric age coefficients (Age NP), the variance components (VCs), and the combined results for all parameters. For each Markov chain, we test equality of means for the first 25\% of chain and the last 25\% of the chain. Tables \ref{TableDiag1}- \ref{TableDiag4} show the results for the Geweke Z-score quantiles and mean, the quantiles and mean of the respective p-values, the median of the effective sample size (ESS), the proportion of rejection of equality of the means, and the Metropolis-Hastings (M-H) acceptance  probability for the variance components. 
\begin{table}[!h]
\centering
\caption{Geweke convergence diagnostics summaries for the main model presented in the paper.}
\label{TableDiag1}
\begin{tabular}{|c|ccccc|}
\hline
& &FE & Age NP & VCs & Combined  \\
\hline
P-value & Mean & 0.496 & 0.493 & 0.472 & 0.489 \\
 & Q025 & 0.025 & 0.022 & 0.006 & 0.000 \\
 & Q05 & 0.053 & 0.060 & 0.018 & 0.000 \\
 & Median & 0.506 & 0.493 & 0.473 & 0.392 \\
 & Q95 & 0.919 & 0.942 & 0.950 & 0.948 \\
 & Q975 & 0.956 & 0.969 & 0.972 & 0.974 \\
 \hline
Geweke & Mean & 0.046 & -0.006 & 0.040 & 0.007 \\
 & Q025 & -1.924 & -1.951 & -2.058 & -1.983 \\
 & Q05 & -1.465 & -1.724 & -1.752 & -1.726 \\
 & Median & -0.042 & 0.119 & 0.022 & 0.092 \\
 & Q95 & 1.754 & 1.521 & 1.907 & 1.625 \\
 & Q975 & 1.976 & 1.802 & 2.433 & 1.941 \\
 \hline
ESS & Median & 912.414 & 887.777 & 249.245 & 762.605 \\
\hline
\% of Rejections &  & 2.23\% & 4.24\% & 6.39\% & 4.43\% \\
\hline
M-H & Mean &  &  & 0.929 &  \\
\hline
 & Q025 &  &  & 0.877 &  \\
 & Q05 &  &  & 0.911 &  \\
 & Median &  &  & 0.930 &  \\
 & Q95 &  &  & 0.957 &  \\
 & Q975 &  &  & 0.971 & \\
 \hline
\end{tabular}
\end{table}
\begin{table}[!h]
\centering
\caption{Geweke convergence diagnostics summaries for the model including left vs. right eye fixed effect function.}
\label{TableDiag2}
\begin{tabular}{|c|ccccc|}
\hline
& &FE & Age NP & VCs & Combined  \\
\hline
P-value & Mean & 0.514 & 0.487 & 0.473 & 0.487 \\
 & Q025 & 0.024 & 0.026 & 0.008 & 0.000 \\
 & Q05 & 0.057 & 0.047 & 0.023 & 0.000 \\
 & Median & 0.525 & 0.480 & 0.460 & 0.442 \\
 & Q95 & 0.962 & 0.945 & 0.954 & 0.943 \\
 & Q975 & 0.978 & 0.973 & 0.977 & 0.971 \\
 \hline
Geweke & Mean & 0.009 & -0.004 & -0.011 & -0.004 \\
 & Q025 & -1.965 & -1.922 & -2.291 & -1.985 \\
 & Q05 & -1.549 & -1.631 & -1.841 & -1.669 \\
 & Median & 0.033 & -0.045 & -0.002 & -0.028 \\
 & Q95 & 1.494 & 1.780 & 1.835 & 1.777 \\
 & Q975 & 1.887 & 2.085 & 2.236 & 2.108 \\
 \hline
ESS & Median & 865.289 & 899.544 & 253.117 & 820.286 \\
\hline
\% of Rejections & - & 0.032 & 0.055 & 0.074 & 0.056 \\
\hline
M-H & Mean &  &  & 0.929 &  \\
 & Q025 &  &  & 0.888 &  \\
 & Q05 &  &  & 0.910 &  \\
 & Median &  &  & 0.930 &  \\
 & Q95 &  &  & 0.958 &  \\
 & Q975 &  &  & 0.968 & \\
 \hline
\end{tabular}
\end{table}

\begin{table}[!h]
\centering
\caption{Geweke convergence diagnostics summaries for the model  with no regularization/shrinkage.}
\label{TableDiag3}
\begin{tabular}{|c|ccccc|}
\hline
& &FE & Age NP & VCs & Combined  \\
\hline
P-value & Mean & 0.491 & 0.486 & 0.481 & 0.486 \\
\hline
 & Q025 & 0.024 & 0.015 & 0.009 & 0.000 \\
 & Q05 & 0.045 & 0.032 & 0.029 & 0.000 \\
 & Median & 0.482 & 0.483 & 0.494 & 0.389 \\
 & Q95 & 0.960 & 0.953 & 0.945 & 0.913 \\
 & Q975 & 0.982 & 0.975 & 0.969 & 0.956 \\
 \hline
Geweke & Mean & -0.026 & 0.011 & 0.010 & 0.006 \\
 & Q025 & -2.067 & -2.182 & -2.144 & -2.132 \\
 & Q05 & -1.754 & -1.723 & -1.788 & -1.759 \\
 & Median & -0.043 & 0.030 & 0.041 & 0.021 \\
 & Q95 & 1.652 & 1.890 & 1.759 & 1.873 \\
 & Q975 & 1.958 & 2.137 & 2.203 & 2.128 \\
 \hline
ESS & Median & 1,000.000 & 608.125 & 253.413 & 590.288 \\
\hline
\% of Rejections &  & 0.059 & 0.083 & 0.058 & 0.075 \\
\hline
M-H & Mean &  &  & 0.928 &  \\
 & Q025 &  &  & 0.900 &  \\
 & Q05 &  &  & 0.909 &  \\
 & Median &  &  & 0.928 &  \\
 & Q95 &  &  & 0.958 &  \\
 & Q975 &  &  & 0.969 & \\
 \hline
\end{tabular}
\end{table}

\begin{table}[!h]
\centering
\caption{Geweke convergence diagnostics summaries for the  wavelet-regularized principal component' model.}
\label{TableDiag4}
\begin{tabular}{|c|ccccc|}
\hline
& &FE & Age NP & VCs & Combined  \\
\hline
P-value & Mean & 0.466 & 0.426 & 0.454 & 0.437 \\
 & Q025 & 0.020 & 0.000 & 0.009 & 0.000 \\
 & Q05 & 0.027 & 0.015 & 0.021 & 0.000 \\
 & Median & 0.476 & 0.303 & 0.450 & 0.080 \\
 & Q95 & 0.955 & 0.860 & 0.926 & 0.711 \\
 & Q975 & 0.972 & 0.900 & 0.943 & 0.941 \\
 \hline
Geweke & Mean & -0.158 & 0.074 & -0.161 & -0.002 \\
 & Q025 & -2.272 & -2.351 & -2.168 & -2.285 \\
 & Q05 & -2.026 & -1.991 & -1.773 & -1.991 \\
 & Median & -0.098 & 0.203 & -0.307 & 0.077 \\
 & Q95 & 1.528 & 2.110 & 1.674 & 1.932 \\
 & Q975 & 2.029 & 3.488 & 2.506 & 2.592 \\
 \hline
ESS & Median & 1,000.000 & 631.661 & 237.495 & 611.360 \\
\hline
\% of Rejections & - & 0.086 & 0.138 & 0.074 & 0.118 \\
\hline
M-H & Mean &  &  & 0.935 &  \\
 & Q025 &  &  & 0.908 &  \\
 & Q05 &  &  & 0.909 &  \\
 & Median &  &  & 0.930 &  \\
 & Q95 &  &  & 0.970 &  \\
 & Q975 &  &  & 0.979 & \\
 \hline
\end{tabular}
\end{table}

\section*{Simulated Pseudo Data}
We generated pseudo-data from the model (12) in Section 2.5. For each basis coefficient, the model (12) was fitted using the \textit{lme} function in \textit{nlme} R package \citep{nlme}. The estimated parameters were used to generate the simulation data. Once the basis coefficients were generated, they were transformed back to the data space using the inverse of the 2D rectangular wavelet transformation. As a result, the simulation data set has 306 curves and each curve has 14400 functional locations.   The file \textit{Y\_simulated.mat} contains the simulated data.  Plots of simulated pseudo data for 3 subjects and all 9 IOP levels are given in Supplemental Figure 28, and similar plots of real data for 3 subjects are given in Figure 27, and demonstrate that the simulated data look much like real data.

The zip file \textit{EYE\_Toolbox.zip} contains the pseudo data along with all required scripts to perform all of the analyses contained in this paper, plus many of the additional plots and diagnostics that are provided to illustrate its properties.  The file \textit{Pseudo\_data\_analysis.pdf} contains a detailed description of overall procedure to fit the semiparametric functional mixed model and obtain inferential results with a simulation dataset, and can be adapted by users to analyze their own data.

\begin{figure}
\minipage{0.75\textwidth}
	\centerline{\includegraphics[scale=0.5]{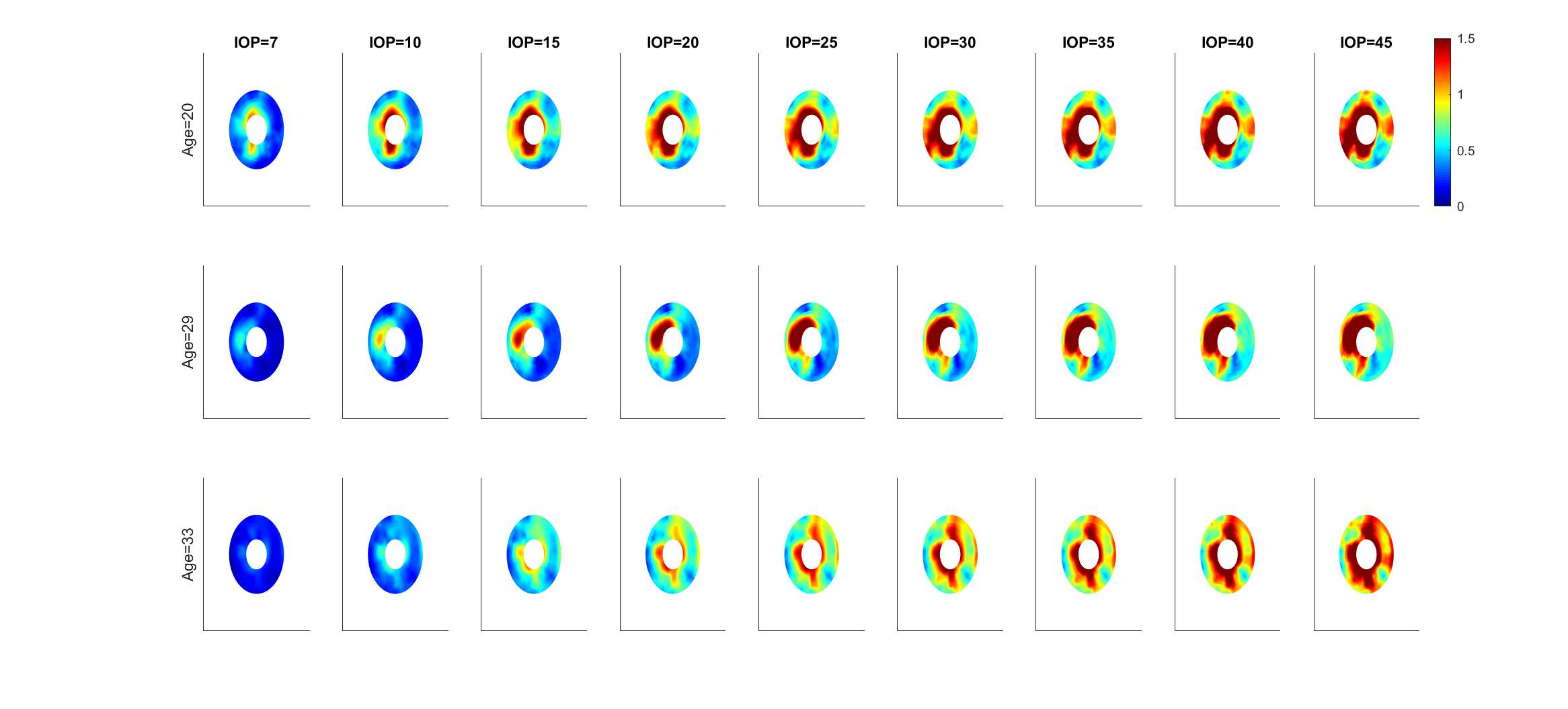}}
	\caption{ \footnotesize Real data for three subjects}
	\label{fig:sfig1}
\endminipage \\
\minipage{0.75\textwidth}
	\centerline{\includegraphics[scale=0.5]{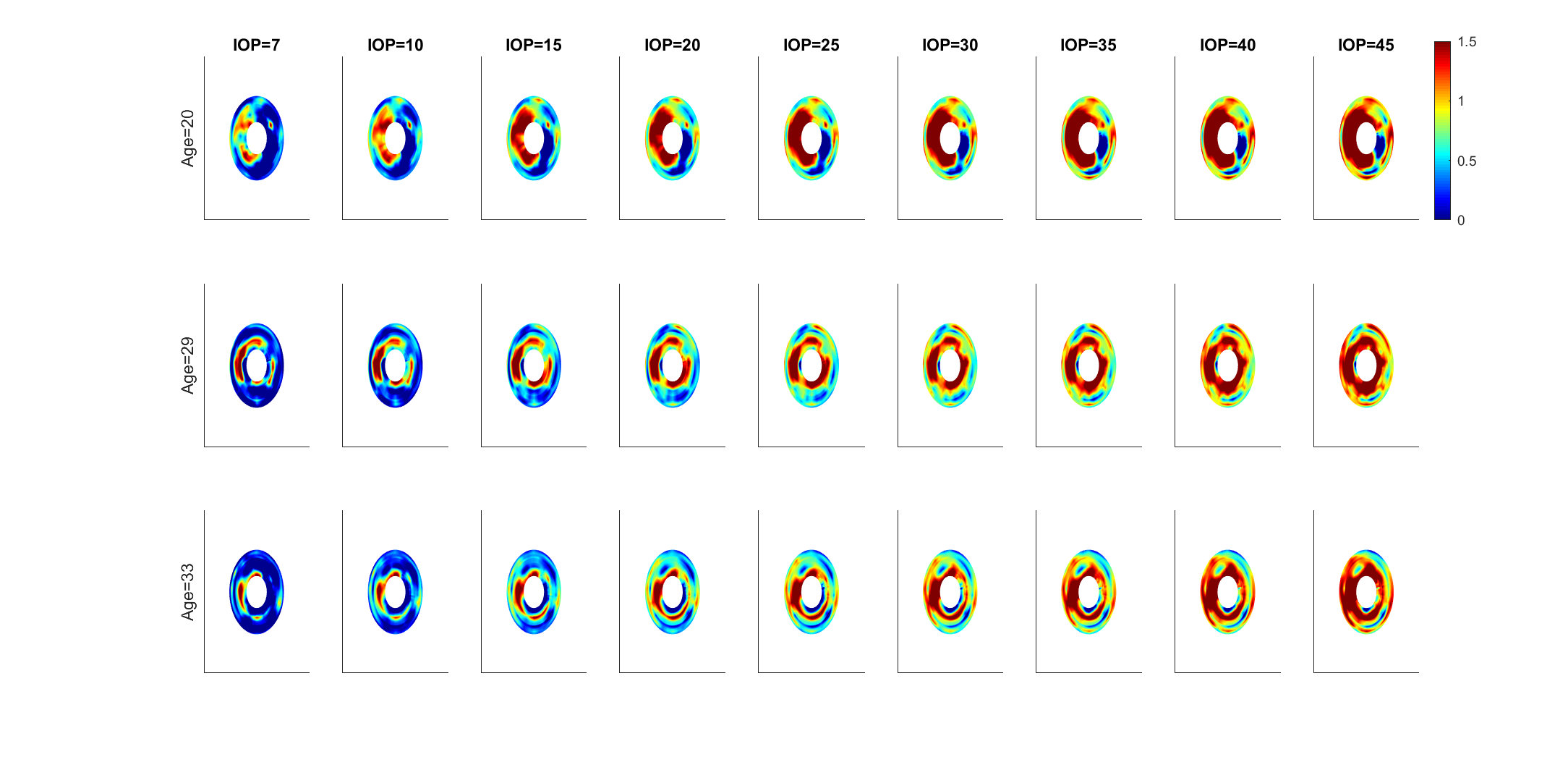}}
	\label{fig:sfig1}
	\caption{ \footnotesize Simulated pseudo data for three simulated subjects}
\endminipage 
\end{figure}

\clearpage
\newpage
\newpage
\bibliographystyle{ECA_jasa}
\bibliography{JASA_ACS}
\end{document}